\documentclass[pdflatex,sn-basic,iicol]{sn-jnl}% Basic Springer Nature Reference Style/Chemistry Reference Style
\usepackage[final]{changes} % final or markup
\setaddedmarkup{\textcolor{teal}{#1}}
\setdeletedmarkup{\textcolor{red}{\sout{#1}}}

\usepackage{graphicx}%
\usepackage{multirow}%
\usepackage{amsmath,amssymb,amsfonts}%
\usepackage{amsthm}%
\usepackage{mathrsfs}%
\usepackage[title]{appendix}%
\usepackage{xcolor}%
\usepackage{textcomp}%
\usepackage{manyfoot}%
\usepackage{booktabs}%
\usepackage{algorithm}%
\usepackage{algorithmicx}%
\usepackage{algpseudocode}%
\usepackage{listings}%

\usepackage[capitalise]{cleveref}       % smart cross-referencing
\usepackage{tikz}
\usepackage{subcaption}
\usepackage[switch]{lineno}
\usepackage{colortbl}

\newcommand{\M}{\bar{M}}
\newcommand{\R}{\mathbb{R}}
\newcommand{\N}{\mathbb{N}}
\newcommand{\E}{\mathbb{E}}

\newcommand{\var}{\text{Var}}

\newcommand{\textmatrix}[1]{\left(\begin{smallmatrix} #1 \end{smallmatrix}\right)}
\newcommand{\bigmatrix}[1]{\begin{pmatrix} #1 \end{pmatrix}}

\theoremstyle{thmstyleone}%
\theoremstyle{thmstyletwo}%

\theoremstyle{thmstylethree}%

\begin{document}

%\title[Mixing in chaotic flows: a sensitivity analysis]{Mixing in chaotic flows: a sensitivity analysis}
\title[Time-varying sensitivity analysis for mixing in chaotic flows: a comparison study]{Time-varying sensitivity analysis for mixing in chaotic flows: a comparison study}


\newbox{\orcid}\sbox{\orcid}{\includegraphics[scale=0.06]{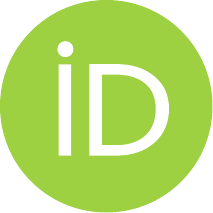}} 
\author[1]{\mbox{\href{https://orcid.org/0000-0003-0111-3835}{\usebox{\orcid}\hspace{1mm}}\fnm{Carla} \sur{Feistner}}}

\author[1,2]{\mbox{\href{https://orcid.org/0000-0003-0210-3874}{\usebox{\orcid}\hspace{1mm}}\fnm{Francesca} \sur{Ziliotto}}}

\author[3]{\mbox{\href{https://orcid.org/0000-0001-6908-6015}{\usebox{\orcid}\hspace{1mm}}\fnm{Barbara} \sur{Wohlmuth}}}

\author[1]{\mbox{\href{https://orcid.org/0000-0003-4850-2037}{\usebox{\orcid}\hspace{1mm}}\fnm{Gabriele} \sur{Chiogna}}}\email{gabriele.chiogna@fau.de}

\affil[1]{Chair of Applied Geology, GeoZentrum Nordbayern, Friedrich-Alexander-Universität Erlangen-Nürnberg, Schlossgarten 5, 91054 Erlangen, Germany}

\affil[2]{Chair of Hydrology and River Basin Management, Technical University of Munich, Arcisstrasse 21, 80333 Munich, Germany}

\affil[3]{Chair of Numerical Mathematics, School of Computation, Information and Technology, Technical University of Munich, Arcisstrasse 21, 80333 Munich, Germany}


\abstract{Engineered injection and extraction systems that create chaotic advection are promising procedures for enhancing mixing between two species. Mixing efficiencies vary considerably, so carefully selecting the design parameters, like pumping rates, well locations, or operation times, is crucial. While numerous studies investigate the conditions required to achieve chaotic flow, sensitivity analyses addressing its impact on mixing have rarely been performed. However, selecting a suitable sensitivity analysis method depends on the underlying system and is often restricted by the computational cost, especially when considering complex, high-dimensional models. Moreover, the most appropriate metric to quantify mixing (e.g., plume area, peak concentration) can also be system-specific. We perform a time-varying sensitivity analysis on the mixing enhancement of two chaotic flow fields with different complexities. The rotated potential mixing (RPM) flow is parametrized using two or four hyperparameters, while the quadrupole flow utilizes $16$ hyperparameters. We compare three global sensitivity analysis methods: Sobol indices, Morris scores, and a modification of the activity scores. We evaluate the temporal evolution of the sensitivity of the design parameters, compare the performance of the three methods, and highlight their potential in analyzing parameter interactions. The analysis of the RPM flow shows comparable sensitivities for all methods. Additionally, our numerical experiments show that Morris is the cheapest method, needing at most four times fewer model evaluations than Sobol to reach convergence. This motivates us to only use the computationally cheaper but as reliable Morris and activity scores on the $16$-dimensional model, yielding again consistent results.}




\maketitle

% full textwidth = 15.9cm -> 6.259843 inch
% one column width = 7.5cm -> 2.95276 inch

\section{Introduction}

%{\color{gray} If you like, you can leave a comment in the PDF in your assigned color. Use your initials like \textbackslash CF\{...\}, \textbackslash FZ\{...\}, \textbackslash BW\{...\} or \textbackslash GC\{...\}.}

%%% Introduction chaotic mixing and applications %%%
Mixing processes in laminar flow regimes are often dominated by molecular diffusion and are hence slow \citep{stone_2001,stremler_2004,rolle_2019}. \added{This is a major limitation} for applications where fast mixing is required, like for chemical reactions in which the reactants need to come into contact, \deleted{this is a major limitation} \citep{neupauer_2014,borgne_2014,bertran_2023}. In such regimes, enhanced mixing can still occur through chaotic advection. The stretching and folding caused by this process generate complex flow paths that increase the interfacial area between the reactive species \citep{aref_1984, aref_1990, liu_2000, stremler_2004}. This enhanced interfacial area leads to sharp concentration gradients, accelerating mixing and hence chemical reaction rates \citep{borgne_2014}.
Applications of chaotic advection to enhance mixing can be found, for example, for in-situ groundwater remediation \citep{mays_2012,zhang_2017,ma_2022,neupauer_2014,ziliotto_,bertran_2023,bagtzoglou_2007,zhang_2009}, microfluidics \citep{aref_2017, ward_2015}, gene expression profiling \citep{stremler_2004}, \deleted{and} the synthesis of biological molecules \citep{teh_2008}\added{, wastewater treatment \citep{bagtzoglou_2006}, biogas production, and organic removal efficiency in Anaerobic hydrogen production systems \citep{huang_2018}}. One can implement chaotic advection using a passive design, where the geometry of the system generates the chaotic flow field \citep{stremler_2004,song_2003,liu_2000}, or actively using external perturbations such as magnetic, pressure, or acoustic disturbances \citep{aref_2017,bagtzoglou_2006} for microstirrers, acoustic mixers, or flow pulsation \citep{ward_2015}. 

%%% EIE systems and their potential to enhance mixing %%%
In Engineered injection and extraction (EIE) systems, mixing efficiency depends on the properties of the flow field, like pumping rates or well locations. 
Mixing can, for example, be limited by Kolmogorov--Arnold--Moser (KAM) islands in the flow, where fluids are trapped around elliptic points in the periodic flow field \citep{lester_2009,chate_1999}. For different configurations of the same EIE system, mixing efficiencies can vary considerably \citep{gubanov_2009,feistner_2024}.

%%% time varying sensitivity analysis %%%
To ensure the effectiveness of these EIE systems for real-world applications, it is essential to assess the uncertainty of the system with respect to its design parameters \citep{ciriello_2013}. Nonetheless, a time-varying sensitivity analysis of mixing for EIE systems has been rarely performed \citep{ziliotto_2025}, and different sensitivity analysis methods have not been tested on these systems.
To assess temporal changes, time-varying sensitivity analysis methods are mandatory and require validation \citep{sarrazin_2016}. Time-varying refers to an analysis of the sensitivity at different times of the model instead of aggregating the overall sensitivity into one single value \citep{pianosi_2016a}.
When using a time-varying approach, it is essential to use a quantitative sensitivity metric, as it allows for comparison of the results among different time-steps.
%%% Differences of sensitivity methods and importance of intercomparison %%%
The choice of an appropriate method for the sensitivity analysis depends heavily on the computational complexity of the problem at hand, the dimensionality of the parameter span, and the properties of the model \citep{pianosi_2016,iooss_2015, chiogna_2024}. A given problem might allow for different sensitivity analysis methods, which, in general, do not yield identical results \citep{constantine_2017}. It is hence advised to apply multiple sensitivity metrics to support the general conclusion of the analysis \citep{pianosi_2016}. However, especially for complex models with a high-dimensional input dimension, this analysis can be computationally expensive, and therefore it is conducted in only a few analyses \citep{sarrazin_2016,dai_2024}. Assessing the computational costs of the sensitivity metrics, we concentrate on the number of Monte-Carlo samples, as the execution of the model is usually the most expensive part of the sensitivity analysis \citep{pianosi_2016}.  

% Sobol
One of the most popular methods for sensitivity analysis is the Sobol index \citep{sobol_2001,saltelli_2010,sobol_1993} used beyond others in \citep{mazziotta_2024,chiogna_2024,ciriello_2013,li_2022,khorashadizadeh_2017}. It is a variance-based method that relates the variance of the output to the variance of the input \citep{saltelli_2010}, assuming that the second moment is sufficient to describe the uncertainty of the model \citep{saltelli_2002a}. 
Approximating the Sobol indices can be done using Monte-Carlo sampling \citep{saltelli_2010}. The computational effort grows with the number of uncertain parameters, requiring $N \cdot (2n + 2)$ model evaluations, where $N$ is the number of samples, and $n$ is the input dimension. If the second-order indices are not needed, the number of evaluations can be reduced to $N \cdot(n + 2)$. High-dimensional models usually require more samples to converge than low-dimensional ones \citep{pianosi_2016}. %The work of \citep{baroni_2014} needs $N = 1024$ sample for a $5$ dimensional model while in \citep{nossent_2011} $N = 12\,000$ samples are used for a $26$ dimensional model. 
Further, the work of \citep{borgonovo_2006, borgonovo_2011, pianosi_2015,pianosi_2016,delloca_2017} shows that Sobol indices may represent output uncertainty incorrectly if the model output is highly skewed or multi-modal.
In cases, however, where Sobol indices are applicable, the results can lead to a good understanding of the model uncertainties.

% Morris
Further, we look at the Morris method \citep{morris_1991,campolongo_2007}, which is also used to perform sensitivity analysis in groundwater remediation in \citep{ma_2022,ziliotto_2025}. \added{This method quantifies sensitivity by the mean of the elementary effects for each input parameter on the model response \citep{morris_1991,campolongo_2007}.} Using Monte-Carlo sampling, the method generates sample trajectories of length $n+1$ by changing each input parameter one at a time. The number of model evaluations also grows linearly with the number of uncertain parameters $N \cdot (n + 1)$ ($N$ number of trajectories). Compared to the Sobol indices, far fewer samples are needed to reach convergence as the works of \citep{ma_2022,ziliotto_2025} both yield realistic results with only $10$ trajectories. 

 % ASM
One more recent sensitivity analysis method is the activity score that can be computed from the results of the active subspace method (ASM) \citep{constantine_2017,bittner_2020,parente_2019a}. The ASM is based on the derivatives of the model output with respect to the uncertain parameters and can lead to low computational costs if gradients are available. If not, we can use a finite difference approach, leading to $N \cdot (n+1)$ samples to create $N$ gradients. 
In addition to the activity scores, the ASM also computes the active direction of the model, which gives insights into the important linear combinations in the model, and we can directly construct a surrogate model \citep{constantine_2013,constantine_2015}. However, following \citep{constantine_2017}, activity scores may cause different results than ANOVA-based methods, like Sobol indices, for highly non-linear systems.

In this work, we assess which sensitivity analysis methods are suitable for systems influenced by a chaotic flow field, considering the dimensionality of the uncertain model input, the computational cost, the significance of the results, and the metric used to quantify mixing. As we are not interested in the sensitivity of the chaotic flow with respect to an initial location, we do not face problems of divergence of sensitivity metrics like \citep{lea_2002,chandramoorthy_2017}. Instead, we focus on the overall effect of chaotic advection to enhance mixing.
We perform a time-varying sensitivity analysis on two chaotic flow fields with different input dimensionality
\deleted{: the RPM flow (Lester et al. 2009) utilizing two and four input parameters that control the rotation angles and the time between rotations of one source and one sink, and the quadrupole flow (Mays and Neupauer 2012) with 16 input parameters representing pumping rates, well locations and hydraulic conductivities.}
\added{. The RPM flow \citep{lester_2009,metcalfe_2010} utilizes two and four input parameters that control the rotation angles and the time between rotations of one source and one sink. While the quadrupole flow \citep{mays_2012} represents a modification to the pulsating dipole in \citep{jones_1988}, adding periodic reorientations of the well locations to enhance mixing. Its $16$ input parameters describe pumping rates, well locations, and hydraulic conductivities.} Both flow fields find applications for groundwater remediation \citep{bertran_2023,cho_2019,neupauer_2014,ziliotto_2025}. We use total Sobol indices, Morris scores, and a modified version of the activity scores from ASM that reuses the Morris samples to reduce computational costs. As the Morris and the activity scores are both qualitative methods, we introduce a scaling such that the results get a quantitative interpretation in terms of percentage contribution to the overall variability. 
For the RPM flow, we quantify mixing through the fraction of the domain covered by solute after a specific time, while for the quadrupole flow, we consider the maximal solute concentration.
We compare the results of the different sensitivity analysis methods on both models and show how they are used to assess sensitivity and parameter interactions. Finally, we analyze the convergence rates of the sensitivity metrics.%, showing that although all metrics converge with $\mathcal{O}(N^{-\frac{1}{2}})$ there is a significant offset from Morris compared to Sobol. This offset makes the Morris method the cheapest of the three methods, requiring four times fewer samples than Sobol to achieve the same accuracy. The sensitivity for the RPM flow shows similar results for the total Sobol index compared to the Morris scores and the activity scores from ASM. Together with the convergence analysis, this motivates the usage of only Morris and ASM for the quadrupole flow, where we again find very similar results for the Morris scores and the activity scores.

Our paper is structured as follows: In \replaced{}{the methodology }\cref{sec:Meth}, we start by introducing the chaotic flow systems that we use for our sensitivity analysis, the RPM flow in \cref{sec:Meth:RPM} and the quadrupole flow in \cref{sec:Meth:quad}. We then introduce the sensitivity analysis methods in \cref{sec:Meth:Sens}, Sobol indices in \cref{sec:Meth:Sobol}, Morris scores in \cref{sec:Meth:Morris}, and activity scores in \cref{sec:Meth:ASM}. We show the results of our numerical experiments in \cref{sec:Results}, starting with the two versions of the RPM flow in \cref{sec:Results:RPM1,sec:Results:RPM2} and then continuing with the quadrupole flow in \cref{sec:Results:quad}. Finally, we conclude our work in \cref{sec:Conclusion}.

\section{Methodology}\label{sec:Meth}

\subsection{Chaotic flow fields}\label{sec:Meth:flow}

\subsubsection{The RPM flow, a low dimensional problem}\label{sec:Meth:RPM}

\paragraph{Model setup}
We consider the two-dimensional rotated potential mixing (RPM) flow that was introduced in \citep{lester_2009}. The flow is produced by using one source and one sink that lie opposite each other on the edge of a unit circle, which serves as the domain for the flow. Both wells operate simultaneously, while fluid that is pumped out of the domain at the sink is reinjected instantaneously at the source. After operating the source and the sink for time $\tau$, the location of the wells is rotated instantaneously by the angle $\Theta$ around the origin. The rotation makes the flow unsteady, which is needed to generate chaotic advection \citep{ottino_1989,stremler_2004,aref_1990,zhang_2009}. Our quantity of interest is $\M$, the fraction of the domain covered by particles initially placed in a tiny region at the top of the domain. As we do not consider diffusion, $\M$ is in theory constant in time. Due to our discretization, we add numerical diffusion when evaluating $\M$, leading to a monotonically increasing $\M$ over time. 
This choice is motivated in \citep{feistner_2024} where the computation of $\M$ on a non-diffusive RPM flow is a good indicator for the effectiveness of chaotic mixing in the same configuration of the RPM flow with diffusion. For $t \rightarrow \infty$, the particles occupy the entire chaotic region of the domain, leaving only the KAM islands particle-free. For large $t$, the value of $\M$ hence approximates the size of the chaotic region.

\begin{figure}[tb!]
	\centering
	\begin{subfigure}[t]{3.6cm}
		\centering
		\resizebox{3.6cm}{!}{
		\begin{tikzpicture}
			% Define colors
			\definecolor{bluecircle}{rgb}{0.2,0.6,1}
			\definecolor{bluecircle1}{rgb}{0.4,0.7,1}
			\definecolor{bluecircle2}{rgb}{0.6,0.8,1}
			\definecolor{redcircle}{rgb}{1,0.2,0.2}
			\definecolor{redcircle1}{rgb}{1,0.4,0.4}
			\definecolor{redcircle2}{rgb}{1,0.6,0.6}
		
			% Draw the unit circle
			\draw[ultra thick, black] (0,0) circle(1.3);
		
			% Connect opposite dots with gray lines (drawn first to go behind colored dots)
			\draw[semithick, gray] (0,1.3) -- (0,-1.3); % Connect top and bottom
			\draw[semithick, gray] ({1.3*cos(135)},{1.3*sin(135)}) -- ({1.3*cos(315)},{1.3*sin(315)}); % Connect rotated positions 1
			\draw[semithick, gray] ({1.3*cos(180)},{1.3*sin(180)}) -- ({1.3*cos(0)},{1.3*sin(0)}); % Connect rotated positions 2
		
			% Draw the blue and red dots on the unit circle
			\fill[bluecircle2] (0,1.3) circle (0.1); % Blue dot at the top
			\node at (0.125,1.55) {\(\tau\)};
			\node at (-0.125,1.6) {\includegraphics[scale=0.02]{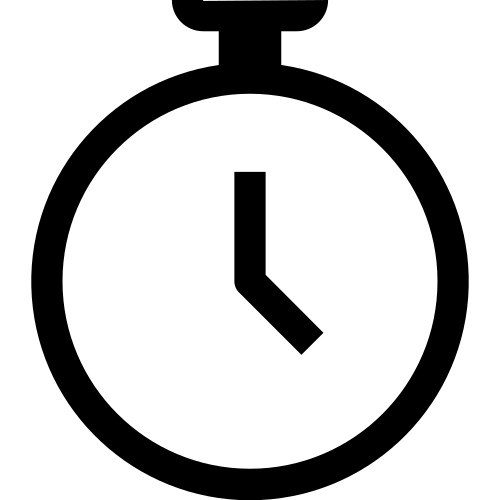}};
			\fill[redcircle2] (0,-1.3) circle (0.1); % Red dot at the bottom
			\node at (0.125,-1.65) {\(\tau\)};
			\node at (-0.125,-1.6) {\includegraphics[scale=0.02]{Fig1_clock}};
		
			% Draw rotated positions 1
			\fill[bluecircle1] ({1.3*cos(135)},{1.3*sin(135)}) circle (0.1); % Blue rotated by theta
			\node at (-1.1,1.15) {\(\tau\)};
			\node at (-1.35,1.2) {\includegraphics[scale=0.02]{Fig1_clock}};
			\fill[redcircle1] ({1.3*cos(315)},{1.3*sin(315)}) circle (0.1); % Red rotated by theta
			\node at (1.4,-1.25) {\(\tau\)};
			\node at (1.15,-1.2) {\includegraphics[scale=0.02]{Fig1_clock}};
		
			% Draw rotated positions 2
			\fill[bluecircle] ({1.3*cos(180)},{1.3*sin(180)}) circle (0.1); % Blue rotated by theta
			\node at (-1.55,0) {\(\tau\)};
			\node at (-1.8,0.05) {\includegraphics[scale=0.02]{Fig1_clock}};
			\fill[redcircle] ({1.3*cos(0)},{1.3*sin(0)}) circle (0.1); % Red rotated by theta
			\node at (1.85,0) {\(\tau\)};
			\node at (1.6,0.05) {\includegraphics[scale=0.02]{Fig1_clock}};
		
			% Draw arrows indicating rotation 1
			\draw[->, ultra thick, black] ({1.3*cos(95)},{1.3*sin(95)}) arc[start angle=95, end angle=133, radius=1.3];
			\draw[->, ultra thick, black] ({1.3*cos(275)},{1.3*sin(275)}) arc[start angle=275, end angle=313, radius=1.3];
		
			% Draw arrows indicating rotation 2
			\draw[->, ultra thick, black] ({1.3*cos(140)},{1.3*sin(140)}) arc[start angle=140, end angle=178, radius=1.3];
			\draw[->, ultra thick, black] ({1.3*cos(320)},{1.3*sin(320)}) arc[start angle=320, end angle=358, radius=1.3];
		
			% Indicate the angle theta at the center for the first rotation
			\draw[->, semithick, black] ({0.9*cos(90)},{0.9*sin(90)}) arc[start angle=90, end angle=135, radius=0.9]; 
			\node at (-0.225,0.6) {\(\Theta\)};
			\draw[->, semithick, black] ({0.9*cos(270)},{0.9*sin(270)}) arc[start angle=270, end angle=315, radius=0.9]; 
			\node at (0.225,-0.6) {\(\Theta\)};
		
			% Indicate the angle theta at the center for the second rotation
			\draw[->, semithick, black] ({0.9*cos(135)},{0.9*sin(135)}) arc[start angle=135, end angle=180, radius=0.9]; 
			\node at (-0.6,0.225) {\(\Theta\)};
			\draw[->, semithick, black] ({0.9*cos(315)},{0.9*sin(315)}) arc[start angle=315, end angle=360, radius=0.9]; 
			\node at (0.6,-0.225) {\(\Theta\)};
		\end{tikzpicture}
		}

		\caption{Influence of the two hyperparameters $\Theta$ and $\tau$ on the RPM flow in the non-randomized case. Other than for the randomized case, the parameters stay fixed over the time interval $[0, t_\text{max}]$.}
		\label{fig:setup_fixed}
	\end{subfigure}
    \hfill
	\begin{subfigure}[t]{3.6cm}
		\centering
		\resizebox{3.6cm}{!}{
		\begin{tikzpicture}
			% Define colors
			\definecolor{bluecircle}{rgb}{0.2,0.6,1}
			\definecolor{bluecircle1}{rgb}{0.4,0.7,1}
			\definecolor{bluecircle2}{rgb}{0.6,0.8,1}
			\definecolor{redcircle}{rgb}{1,0.2,0.2}
			\definecolor{redcircle1}{rgb}{1,0.4,0.4}
			\definecolor{redcircle2}{rgb}{1,0.6,0.6}
		
			% Draw the unit circle
			\draw[ultra thick, black] (0,0) circle(1.3);
		
			% Connect opposite dots with gray lines (drawn first to go behind colored dots)
			\draw[semithick, gray] (0,1.3) -- (0,-1.3); % Connect top and bottom
			\draw[semithick, gray] ({1.3*cos(125)},{1.3*sin(125)}) -- ({1.3*cos(305)},{1.3*sin(305)}); % Connect rotated positions 1
			\draw[semithick, gray] ({1.3*cos(185)},{1.3*sin(185)}) -- ({1.3*cos(5)},{1.3*sin(5)}); % Connect rotated positions 2
		
			% Draw the blue and red dots on the unit circle
			\fill[bluecircle2] (0,1.3) circle (0.1); % Blue dot at the top
			\node at (0.15,1.55) {\(\tau_1\)};
			\node at (-0.15,1.6) {\includegraphics[scale=0.02]{Fig1_clock}};
			\fill[redcircle2] (0,-1.3) circle (0.1); % Red dot at the bottom
			\node at (0.15,-1.65) {\(\tau_1\)};
			\node at (-0.15,-1.6) {\includegraphics[scale=0.02]{Fig1_clock}};
		
			% Draw rotated positions 1
			\fill[bluecircle1] ({1.3*cos(125)},{1.3*sin(125)}) circle (0.1); % Blue rotated by theta
			\node at (-1.0,1.2) {\(\tau_2\)};
			\node at (-1.3,1.25) {\includegraphics[scale=0.02]{Fig1_clock}};
			\fill[redcircle1] ({1.3*cos(305)},{1.3*sin(305)}) circle (0.1); % Red rotated by theta
			\node at (1.3,-1.25) {\(\tau_2\)};
			\node at (1.0,-1.2) {\includegraphics[scale=0.02]{Fig1_clock}};
		
			% Draw rotated positions 2
			\fill[bluecircle] ({1.3*cos(185)},{1.3*sin(185)}) circle (0.1); % Blue rotated by theta
			\node at (-1.6,-0.2) {\(\tau_3\)};
			\node at (-1.9,-0.15) {\includegraphics[scale=0.02]{Fig1_clock}};
			\fill[redcircle] ({1.3*cos(5)},{1.3*sin(5)}) circle (0.1); % Red rotated by theta
			\node at (1.9,0.15) {\(\tau_3\)};
			\node at (1.6,0.2) {\includegraphics[scale=0.02]{Fig1_clock}};
		
			% Draw arrows indicating rotation 1
			\draw[->, ultra thick, black] ({1.3*cos(95)},{1.3*sin(95)}) arc[start angle=95, end angle=123, radius=1.3];
			\draw[->, ultra thick, black] ({1.3*cos(275)},{1.3*sin(275)}) arc[start angle=275, end angle=303, radius=1.3];
		
			% Draw arrows indicating rotation 2
			\draw[->, ultra thick, black] ({1.3*cos(140)},{1.3*sin(140)}) arc[start angle=140, end angle=183, radius=1.3];
			\draw[->, ultra thick, black] ({1.3*cos(320)},{1.3*sin(320)}) arc[start angle=320, end angle=363, radius=1.3];
		
			% Indicate the angle theta at the center for the first rotation
			\draw[->, semithick, black] ({0.9*cos(90)},{0.9*sin(90)}) arc[start angle=90, end angle=125, radius=0.9]; 
			\node at (-0.19,0.6) {\(\Theta_1\)};
			\draw[->, semithick, black] ({0.9*cos(270)},{0.9*sin(270)}) arc[start angle=270, end angle=305, radius=0.9]; 
			\node at (0.22,-0.6) {\(\Theta_1\)};
		
			% Indicate the angle theta at the center for the second rotation
			\draw[->, semithick, black] ({0.9*cos(125)},{0.9*sin(125)}) arc[start angle=125, end angle=185, radius=0.9]; 
			\node at (-0.55,0.25) {\(\Theta_2\)};
			\draw[->, semithick, black] ({0.9*cos(305)},{0.9*sin(305)}) arc[start angle=305, end angle=365, radius=0.9]; 
			\node at (0.55,-0.25) {\(\Theta_2\)};
		\end{tikzpicture}
		}
		
		\caption{In the randomized case, the parameters $\Theta_i$ and $\tau_i$ vary over time. They are sampled uniformly from the intervals $\left[\bar{\Theta} - \frac{\Theta_r}{2}, \bar{\Theta} + \frac{\Theta_r}{2}\right]$ and $\left[\bar{\tau} - \frac{\tau_r}{2}, \bar{\tau} + \frac{\tau_r}{2}\right]$ that depend on four hyperparameters $\bar{\Theta}$, $\Theta_r$, $\bar{\tau}$ and $\tau_r$.}
		\label{fig:setup_random}
	\end{subfigure}
			
	\caption{Visualization for the source (blue) and sink (red) location for the RPM flow for the non-randomized and the randomized case.}
	\label{fig:RPMsetup}
\end{figure}

\paragraph{Uncertain parameters}
By construction, the RPM flow is parametrized by two flow parameters, $\Theta$ and $\tau$. We give a graphical representation of the system in \cref{fig:setup_fixed}. In addition to the standard RPM flow, we investigate a setup in which we modify the RPM flow by introducing randomization. Therefore, for each rotation the angle $\Theta_i$ is sampled from the uniform distribution $\mathcal{U}(\bar{\Theta} - \frac{\Theta_r}{2}, \bar{\Theta} + \frac{\Theta_r}{2})$, similar we sample $\tau_i$ from $\mathcal{U}(\bar{\tau} - \frac{\tau_r}{2}, \bar{\tau} + \frac{\tau_r}{2})$. The flow is hence described using four parameters: the means and the deviations of the respective randomization intervals. We visualize the randomized version of the RPM flow in \cref{fig:setup_random}. To avoid confusion with the flow parameters $\Theta_i$ and $\tau_i$, we will refer to $\Theta$ and $\tau$ of the non-randomized system and to $\bar{\Theta}$, $\Theta_r$, $\bar{\tau}$ and $\tau_r$ of the randomized system as hyperparameters. During our numerical experiments, we consider both versions of the flow. We give the ranges of the hyperparameters in \cref{tab:parameter_ranges}.

\begin{table*}[tb!]
	\renewcommand{\arraystretch}{1.2}
	\centering
	\begin{subtable}[t]{\textwidth}
		\centering
		\begin{tabular}{|c|c|c|c|c|c|c|}
			\hline
            \rowcolor[gray]{0.9} \multicolumn{7}{|l|}{Non-randomized RPM flow, large intervals} \\ \hline
			$\Theta$ & $\tau$ & \multicolumn{5}{|c|}{} \\ \hline
			$[0, \pi]$ & $[0.001, 1]$ & \multicolumn{5}{|c|}{} \\ \hline
            \rowcolor[gray]{0.9} \multicolumn{7}{|l|}{Non-randomized RPM flow, small intervals} \\ \hline
			$\Theta$ & $\tau$ & \multicolumn{5}{|c|}{} \\ \hline
			$[0.4\pi, 0.6\pi]$ & $[0.4, 0.6]$ & \multicolumn{5}{|c|}{} \\ \hline \multicolumn{7}{c}{}\\[-1.2em] \hline
            \rowcolor[gray]{0.9} \multicolumn{7}{|l|}{Randomized RPM flow, large intervals} \\ \hline
			$\bar{\Theta}$ & $\Theta_r$ & $\bar{\tau}$ & $\tau_r$ & \multicolumn{3}{|c|}{} \\ \hline
			$[0, \pi]$ & $[0, 0.2\pi]$ & $[0.1, 1]$ & $[0, 0.2]$ & \multicolumn{3}{|c|}{} \\ \hline
            \rowcolor[gray]{0.9} \multicolumn{7}{|l|}{Randomized RPM flow, small intervals} \\ \hline
			$\bar{\Theta}$ & $\Theta_r$ & $\bar{\tau}$ & $\tau_r$ & \multicolumn{3}{|c|}{} \\ \hline
			$[0.45\pi, 0.55\pi]$ & $[0, 0.1\pi]$ & $[0.45, 0.55]$ & $[0, 0.1]$ & \multicolumn{3}{|c|}{} \\ \hline \multicolumn{7}{c}{}\\[-1.2em] \hline
            \rowcolor[gray]{0.9} \multicolumn{7}{|l|}{Quadruple flow} \\ \hline
            $log(k_1)$ & $log(k_2)$ & $log(k_3)$ & $log(k_4)$ & $q_{n,s,e,w}$ & $r_{n,s,e,w}$ & $\theta_{n,s,e,w}$ \\ \hline
			$[-0.122, 0.354]$ & $[0.354, 0.697]$ & $[0.698, 1.042]$ & $[1.042, 1.539]$ & $[0.95, 1.05]$ & $[0, 5]$ & $[0, 2\pi]$ \\ \hline
		\end{tabular}
	\end{subtable}

    \vspace{0.2cm}
	\caption{Hyperparameter ranges for the sensitivity analysis. We sample the hyperparameters for each experiment according to the uniform distribution from the respective intervals.}
	\label{tab:parameter_ranges}
\end{table*}

\paragraph{Flow equations}
The Lagrangian form of the RPM flow with source at $(0, 1)^T$ and the sink at $(0, -1)^T$ can be described using the Hamiltonian system \citep{sposito_1998,durst_2022}
\begin{align}\label{eq:PSS_hamiltonian_system}
    \frac{dx}{dt} = \frac{\partial \psi}{\partial y}, \quad
    \frac{dy}{dt} = -\frac{\partial \psi}{\partial x}, \quad
    \bigmatrix{x(0)\\y(0)} = \bigmatrix{x_0\\y_0}
\end{align}
with Lagrange stream function \citep{lester_2009}
\begin{align}\label{eq:RPM_psi}
    \psi(x, y) = \tan^{-1} \left(\frac{2x}{1-x^2-y^2}\right).
\end{align}
\cite{lester_2009} derive the analytical solution to \eqref{eq:PSS_hamiltonian_system} with stream function $\psi$ in \eqref{eq:RPM_psi}. For the convenience of the author, we briefly describe the process in the following. We focus first on the positive side of the $x$-axis (i.e., $x > 0$) and represent the particle position in terms of the angle to the origin $\theta \in (-\pi/2, \pi/2)$ and its streamline $\psi \in (0, \pi/2]$. This parametrization is convenient, as $\psi$ is constant in time. The advection time of a particle along its streamline, until it reaches $\theta = 0$, was established in \citep{lester_2009} as
\begin{align}\label{eq:RPM_t_adv}
    \begin{split}
        t_\text{adv}(\theta, \psi)& = \csc^2(\psi) \\
        \times\Biggl\{ &\cot(\psi) \arctan \left[ \frac{\sin(\theta) \cot(\psi)}{\sqrt{1 + \cos^2(\theta) \cot^2(\psi)}} \right] \\
        &+ \sin(\theta) \sqrt{1 + \cos^2(\theta) \cot^2(\psi)} \\
        &- |\cot(\psi)| (\theta + \cos(\theta) \sin(\theta)) \Biggr\}.
    \end{split}
\end{align}
This formula yields a negative value for particles downstream of $\theta = 0$. Using \eqref{eq:RPM_t_adv}, we can compute the residence time of a particle inside the domain from injection to extraction as
\begin{align}\label{eq:RPM_t_res}
 T_{\text{res}}(\psi) = \left(t_{\text{adv}}\left(\frac{\pi}{2}, \psi\right) - t_{\text{adv}}\left(-\frac{\pi}{2}, \psi\right)\right).
\end{align}
Combining \eqref{eq:RPM_t_adv} and \eqref{eq:RPM_t_res}, we compute the angle $\theta(t)$ of a particle $(\theta_0, \psi_0)$ after time $t$ as
\begin{align}\label{eq:RPM_formular_advection}
    \begin{split}
        t&_{\text{adv}}(\theta(t), \psi_0) + \frac{T_{\text{res}}(\psi_0)}{2} = \\
        &\left(t_{\text{adv}}(\theta_0, \psi_0) + \frac{T_{\text{res}}(\psi_0)}{2} - t\right) \bmod T_{\text{res}}(\psi_0).
    \end{split}
\end{align}
The modulo operator $\bmod$ is needed to account for the instantaneous reinjection of a particle.
We cannot specify the solution of \eqref{eq:RPM_formular_advection} analytically, hence we use a linear interpolation of the inverse $\theta(t_\text{adv}, \psi)$ on a $100 \times 100$ grid. Together, this yields the flow $\Psi_+^t$ for all particles with $x > 0$. For $x < 0$, we use the reflection symmetry of $\psi$ with respect to the $y$-axis, yielding the flow $\Psi_-^{t}\textmatrix{x\\y} = \textmatrix{-1&0\\0&1}\Psi_+^{t}\textmatrix{-x\\y}$. 

For $x = 0$, the usage of $(\theta, \psi)$ coordinates is insufficient. Alternatively, we directly solve \eqref{eq:PSS_hamiltonian_system}
\begin{align}\label{eq:RPM_1D_ODE}
    \begin{split}
	\dfrac{dx}{dt}=&\dfrac{4xy}{(y^2+x^2-1)^2+4x^2}\\
	\dfrac{dy}{dt}=&\dfrac{-2(x^2-y^2+1)}{(x^2+y^2)^2+2x^2-2y^2+1}\\
	\bigmatrix{x(0)\\y(0)} =& \bigmatrix{0\\y_0}
    \end{split}
\end{align}
using the definition of $\psi$ in \eqref{eq:RPM_psi}.
Given that $x(0) = 0$ we find $\frac{dx}{dt}|_{t=0} = 0$ and hence $x(t) = 0$ for all $t \in \R$. This simplifies \eqref{eq:RPM_1D_ODE} to 
\begin{align}\label{eq:RPM_1D_ODE_simplified}
	x(t) = 0 , \qquad
	\dfrac{d y}{d t} = \dfrac{2}{y^2-1} , \qquad
	y(0) = y_0.
\end{align} 
We solve \eqref{eq:RPM_1D_ODE_simplified} for $y$ by using the separation of variables, leading to
\begin{align}\label{eq:RPM_1D_ODE_simplified_solution}
	t_{\text{adv}}(y) = \frac{y}{2} - \frac{y^3}{6}, \qquad T_{\text{res}} = \frac{2}{3}.
\end{align}
We can use $t_{\text{adv}}$ from \eqref{eq:RPM_1D_ODE_simplified_solution} in \eqref{eq:RPM_formular_advection} where $y$ takes the place of $\theta$ to compute the flow $\Psi_0^t$ for all particles with $x = 0$.
Combining the three flows, we can define the flow $\Psi^t$ for the complete domain.

To solve the equations when the source and the sink are rotated after time $\tau_i$, we introduce the rotation operator $R[\Theta]:\R^2\rightarrow\R^2$ that rotates the particles by the angle $\Theta$. The operators $R_k^+$ and $R_k^-:\R^2\rightarrow\R^2$ are defined accordingly by $R_k^+=R[\sum_{j=1}^k\Theta_j]$ and $R_k^-=R[-\sum_{j=1}^k\Theta_j]$ with $R_0^+ = R_0^- = R[0]$. Using $R^-_{i}R^+_{i-1} = R[-\Theta_{i}]$, the position of a particle $\textmatrix{x\\y}$ after time $t_k=\sum_{j=1}^k \tau_j$ is given by
\begin{align}\label{eq:RPM_flow}
\begin{split}
	\bigmatrix{x(t_k)\\y(t_k)} 
	&= \prod_{j=1}^{k} R^+_{j-1}\Psi^{\tau_j} R^-_{j-1} \bigmatrix{x_0\\y_0}\\
	&=R[\Theta_n] \prod_{j=1}^k R[-\Theta_j]\Psi^{\tau_j} \bigmatrix{x_0\\y_0}.
\end{split}
\end{align}

\paragraph{Numerical model}
We use the analytical solution of the RPM flow in \eqref{eq:RPM_flow} to perform Lagrangian particle tracking on $100\,000$ particles that start at the source. As two particles with the exact same initial location follow the same trajectory, we sample the initial particle locations according to $(0, 1)^T + \epsilon$ with $\epsilon \sim (\mathcal{N}(0, 10^{-5}))^2$ to ensure that the particles are initially concentrated in the same tiny region inside the domain without having the same initial coordinates. Due to this definition, particles may be initialized outside the flow domain. In this case, we resample the initialization until the particle ends up inside the unit circle. 
%In doing so, we ensure that the particles are initially concentrated in the same tiny region inside the domain without having the same initial coordinates. 
We evolve these particles until $t_\text{max} = 20$. Using the resulting particle distribution for different times $t$, we approximate the fraction of the domain covered by particles $\M$ by dividing the domain into cells of size $0.02^2$ and counting the number of cells that contain at least one particle. We use a time step of $\tau$ for the particle tracking to reduce the computational complexity. Hence, we compute $\bar{M}$ only for times $t \in \{n\tau:n\in\N\}$. We use linear interpolation to get an approximation of $\bar{M}$ for all other times.\added{ The choice of $t_\text{max} = 20$ was found to be sufficient to get close to the steady state of $\bar{M}$ for most configurations (see supplementary material, Figs. S10, S11 and S12), hence we do not expect large changes of the sensitivity metrics for $t > t_\text{max}$.}

\subsubsection{The quadrupole flow, a high dimensional problem}\label{sec:Meth:quad}

% \begin{figure*}[tb!]
% 	\centering
% 	\begin{subfigure}[t]{8cm}
% 		\includegraphics[width=8cm]{fig/Quadrupole/setup_boundaries.png}
% 		\caption{Model domain}
% 		\label{fig:Quad_setup_boundaries}
% 	\end{subfigure}
	
% 	\begin{subfigure}[t]{8.3cm}
% 		\includegraphics[width=8.3cm]{fig/Quadrupole/setup_wells.png}
% 		\caption{Flow hyperparameters concerning the uncertainty of the location of the four wells.}
% 		\label{fig:Quad_setup_wells}
% 	\end{subfigure}
%     \hfill
% 	\begin{subfigure}[t]{7.45cm}
% 		\includegraphics[width=7.45cm]{fig/Quadrupole/setup_conductivity.png}
% 		\caption{Heterogeneous hydraulic conductivity field. \CF{Is this the plot with the correct conductivity field (field $2$)?}}
% 		\label{fig:Quad_setup_conductivity}
% 	\end{subfigure}

% 	\caption{Description of the Quadrupole model. \Cref{fig:Quad_setup_boundaries} shows the model domain with the four wells (white diamonds), \cref{fig:Quad_setup_wells} shows the uncertainty in the well location, and \cref{fig:Quad_setup_conductivity} shows the heterogeneous conductivity field with uncertainty in the four conductivity values $k_1$ to $k_4$. The parameter $L$ represents the distance between each well and the center of the domain and is equal to $25$ m. \protect\citep{ziliotto_2025}}
% 	\label{fig:Quad_setup}
% \end{figure*}

\begin{figure}[tb!]
	\centering
	\begin{subfigure}[t]{7.5cm}
		\includegraphics[width=7.5cm]{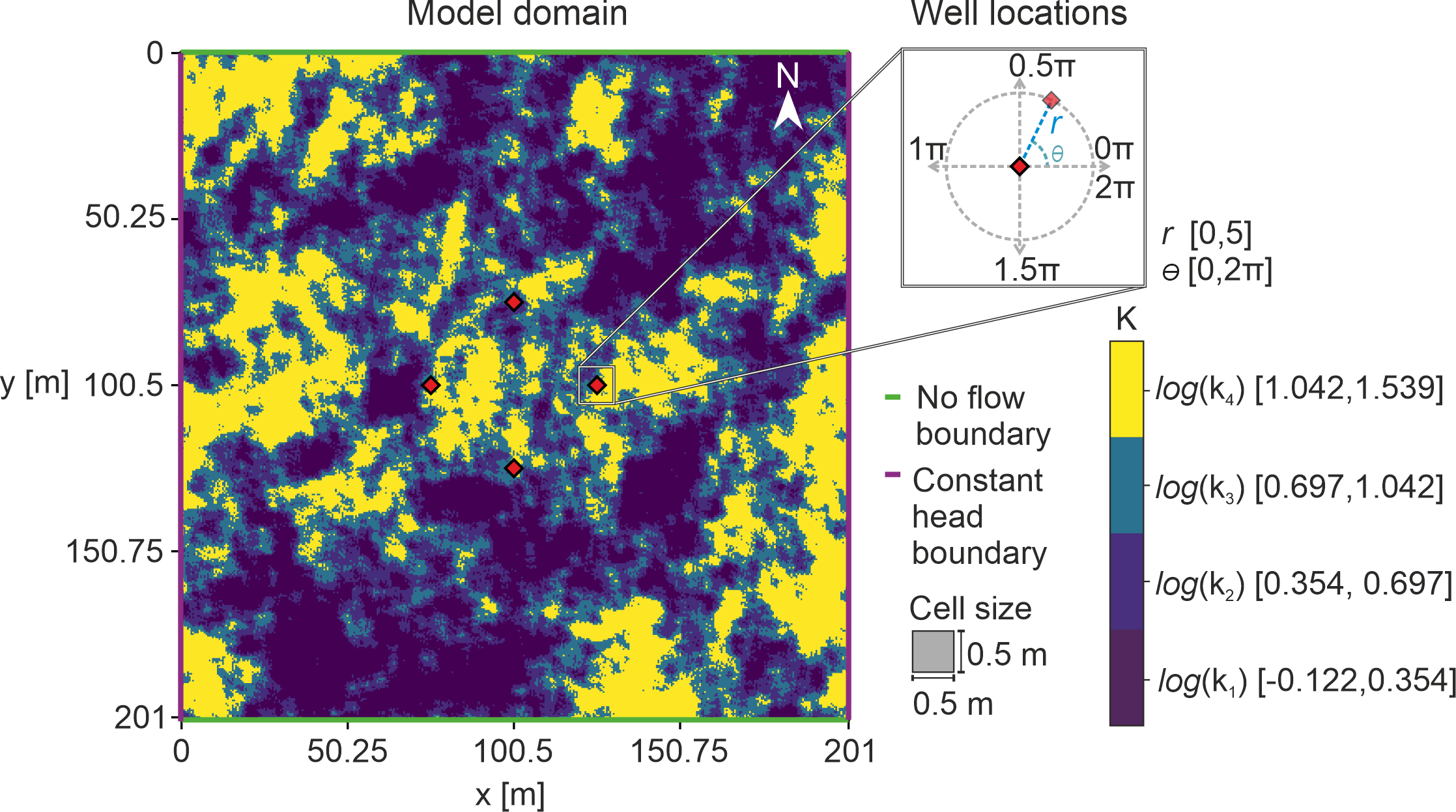}
	\end{subfigure}
	
	\caption{Visualization of the Quadrupole model, including the boundary condition and the conductivity field. The locations of the wells depend on the hyperparameter for $r$ and $\theta$ as depicted in the top right corner.}
	\label{fig:Quad_setup}
\end{figure}

\paragraph{Model setup}
As second model, we consider the quadrupole flow introduced in \citep{mays_2012}. The flow is produced by four wells located in a diamond shape around the center of the domain that represents a two-dimensional confined aquifer (see \cref{fig:Quad_setup}). The wells are operated according to the extraction and injection protocol of \citep{mays_2012}, where the four wells inject and extract water one at a time at different pumping rates. In the ideal case, with constant hydraulic conductivity, the sequence is designed such that the solute, which is initially placed in the center of the domain, is not extracted. In our experiments, extractions appear due to our modifications to the well locations and the hydraulic conductivity field. If solute gets extracted, it leaves the domain and will not be reinjected. Due to this solute extraction, the fraction of the domain covered by solute $\M$ is an improper metric to quantify mixing in this setup. While $\M$ mostly increases with time, it decreases whenever the system experiences solute extraction (see supplementary material, Fig. S1). The results with this metric would hence be highly affected by the solute extraction. 
We hence use the maximal solute concentration at the end of each of the $12$ stress periods as our quantity of interest, which is also known to be a good indicator for mixing efficiency \citep{chiogna_2017}. 
We highlight that there are large differences between the RPM flow and the quadrupole flow. While the RPM flow follows a more synthetic approach, using a non-diffusive system and applying assumptions on the reinjection protocol, the quadrupole is more physical. A detailed discussion of these differences can be found in \citep{lester_2013a}.
%The work of \citep{basiliohazas_2023} demonstrates that there is a linear relationship between the reactor ratio and the maximal solute concentration; hence, we expect a similar relationship also to $\M$.

\paragraph{Uncertain parameters} 
Our analysis reproduces the work of \citep{ziliotto_2025}. The flow is parametrized by $16$ parameters. The first $8$ parameters describe the position of the four wells. We determine the exact well location by perturbing the ideal well locations with polar coordinates $(r, \theta)$ sampled uniformly from their respective intervals. We show this process in \cref{fig:Quad_setup}. Also, the pumping rates of the four wells underlie some uncertainty. We introduce a multiplier $q$ for each well that introduces a $\pm 5$ \% uncertainty to the pumping rates defined in \citep{mays_2012}. Finally, we use an uncertain hydraulic conductivity field that specifies the distribution of four materials in the aquifer while each material covers $25$ \% of the domain area (see \cref{fig:Quad_setup}). The corresponding hydraulic conductivity values $k_1$, $k_2$, $k_3$, and $k_4$ are sampled according to the log uniform distribution. The exact sampling intervals for the $16$ uncertain parameters are given in \cref{tab:parameter_ranges}.

\paragraph{Groundwater flow and transport equations}
We solve the groundwater flow and transport problem using an Eulerian framework. For the two-dimensional groundwater flow in an isotropic confined aquifer, we use the equation \citep{fetter_2001,bear_1988}
\begin{align}\label{eq:groundw_flow}
 S \frac{\partial h}{\partial t} = \nabla \cdot (T \nabla h) + W, \quad h(x, 0) = h_0(x)
\end{align}
with hydraulic head $h$ [\replaced{$L$}{-}], storage coefficient $S$ [-], transmissivity tensor $T = Kb$ [$\frac{L^2}{T}$] with hydraulic conductivity tensor $K$ [$\frac{L}{T}$] and thickness of the aquifer $b$ [$L$] and source and sink term $W$ [$\frac{L}{T}$]. The transport of a conservative solute due to the flow is given in \citep{fetter_2001,bear_1988} by
\begin{align}\label{eq:groundw_transport}
    \frac{\partial c}{\partial t} = - \nabla \cdot (v c) + \nabla \cdot (D \nabla c), \quad c(x, 0) = c_0(x)
\end{align}
with solute concentration $c$ [$\frac{M}{L^3}$] and hydrodynamic dispersion tensor $D$ [$\frac{L^2}{T}$]. The velocity vector $v = -\frac{K}{n_e} \nabla h$ [$\frac{L}{T}$] is given by Darcy's law and depends on the effective porosity $n_e$ [-] and the solution of \eqref{eq:groundw_flow} \citep{fetter_2001}.

\paragraph{Numerical model}
We simulate the engineered injection and extraction (EIE) system using MODFLOW-2005 \citep{arlenw.harbaugh_2017}, to solve the groundwater flow and MT3D-USGS \citep{tonkin_2016}, to solve the transport of a conservative solute. For the simulation of the wells, we use the well package of \citep{arlenw.harbaugh_2017}. The model consists of a square domain with extension $201$ m $\times$ $201$ m and a grid size $\Delta x = \Delta y = 0.5$ m. For the boundaries, we apply a constant head on the east and west sides and no flow on the north and south sides. We generate the random hydraulic conductivity field using the field generator \citep{chiang_2005}, with $\mu_{\log(K)} = 0.7$, $\sigma_{\log(K)}^2 = 0.25$, and correlation length $\lambda = 10$ m. For the values in the lowest quartile, we assign $k_1$, in the second-lowest quartile $k_2$, in the second-highest quartile $k_3$, and in the highest quartile $k_4$. The total duration of the extraction and injection sequence is $t_{\text{max}} = 75$ days, which is divided into $12$ stress periods of $6.25$ days. We save the concentration distribution of the solute after each stress period. The initial normalized concentration of the plume $c_0(x)$ occupies an $11\text{ m} \times 11\text{ m}$ square in the center of the domain, and we use a hydrodynamic dispersion tensor $D = 4 \cdot 10^{-2}$ m$^2$/day. A more detailed summary of the parameters of the quadrupole system is given in the supplementary material Table S1 as well as in \citep{ziliotto_2025}.

\subsection{Sensitivity analysis}\label{sec:Meth:Sens}

Sensitivity analysis is a powerful tool to understand the influence of input parameters on the output of a model. It can be used to identify the most important parameters, to reduce the dimensionality of the problem, and to improve the model's performance. In this work, we use three different methods for sensitivity analysis. We estimate how sensitive the mixing efficiency of two chaotic flow systems introduced in \cref{sec:Meth:RPM} and \cref{sec:Meth:quad} depends on the flow parametrization. Given a parametrization $x=(x_1, ..., x_n) \in \R^n$, we compute the mixing efficiency using $f:\R^n \rightarrow \R$. For the RPM flow, $f(x)$ is given by the area covered by particles at time $t$, and for the quadrupole flow, we use the peak concentration at the end of each stress period.

\subsubsection{Sobol indices}\label{sec:Meth:Sobol}

The Sobol index is a variance-based approach \citep{sobol_2001,saltelli_2010,sobol_1993}, derived on the ANOVA representation \citep{archer_1997,jansen_1999,sobol_2001,sobol_2003,sobol_2005}, of $f$. 
Details are given in the appendix \cref{app:Meth:Sobol}.
For the model $f:\R^n \Rightarrow \R$ we compute the total variance $V$
and the conditional variances $V_i$ and $V_{i,j}$, that describe the output variance due to a single or pairs of input parameters, by
\begin{align*}
    V &= \var(f(x))\\
    V_{i} &= \var(\E[f(x)|x_i])\\
    V_{i,j} &= \var(\E[f(x)|x_i, x_j]) - V_i - V_j,
\end{align*}
\citep{saltelli_2010,sobol_1993,homma_1996}. 
Using this, we can define the first- and second-order Sobol indices as
\begin{align}\label{eq:S_i}
 S_{i} := \frac{V_{i}}{V}, \qquad S_{i,j} := \frac{V_{i,j}}{V}.
\end{align}
Also the computation of higher-order interactions via $S_{i_1, ..., i_s} := \frac{V_{i_1, ..., i_s}}{V}$
is possible \citep{saltelli_2002a, homma_1996,sobol_1993,sobol_2001}. It holds that 
\begin{align}\label{eq:Sobol_sum}
	\sum_i S_i + \sum_i \sum_{j>i} S_{i,j} + ... + S_{1, 2, ..., k} = 1.
\end{align}
We also use the total Sobol index with respect to the parameter $x_i$ introduced in \citep{homma_1996,saltelli_2002}
\begin{align*}
	\begin{split}
		S_{Ti} &:= \frac{\E[\var(f(x)|\{x_j:j\neq i\})]}{V} \\
		&= 1 - \frac{\var(\E[f(x)|\{x_j:j\neq i\}])}{V}.
	\end{split}
\end{align*}
It measures the contribution of the parameter $x_i$ considering its first-order effect and all interactions with other parameters \citep{pianosi_2016}.
Following \citep{sobol_2001,archer_1997}, the total Sobol index can also be written in terms of the classical Sobol indices as
\begin{align}\label{eq:S_ti_sum}
    \begin{split}
        S_{Ti} &= \sum_{s=1}^{n} \sum_{\substack{j_1<...<j_s, \\ \exists k: j_k=i}} S_{j_1, ..., j_s} %\\
        %&= 1 - \sum_{s=1}^{n-1} \sum_{\substack{j_1<...<j_s, \\ j_k\neq i\, \forall k}} S_{j_1, ..., j_s}.
    \end{split}
\end{align}

For the numerical estimation of the Sobol indices, one can use Monte-Carlo sampling \citep{sobol_2001,saltelli_2010,saltelli_2002,sobol_1993}. In general, estimating Sobol indices of higher order is computationally expensive \citep{saltelli_2002}. Therefore, we restrict our analysis to first-order, second-order, and total Sobol indices. Our numerical experiments use the Python package SALib \citep{herman_2017,iwanaga_2022}. We use Saltelli Sampling and create the sample matrices $A$ and $B \in \R^{N\times n}$, with $N$ number of samples and $n$ input dimension of the model. The $i$'th column of the matrices contains input samples for $x_i$. We define $AB\in \R^{2N\times n}$ as the concatenation of the two matrices $A$ and $B$. Further, we use the matrices $A_B^{(i)}$ where all columns are taken from $A$, but the $i$'th column is replaced by the $i$'th column of $B$. In the same way, we also create $B_A^{(i)}$. The implementation of the first-order Sobol index follows \citep{saltelli_2010}. The estimators for $V_i$ and $V$ for computing $S_i$ in \eqref{eq:S_i} are given by 
\begin{align}\label{eq:V_i}
\begin{split}
 V_i &= \frac{1}{N} \sum_{j=1}^N f(A)_j (f(A_B^{(i)})_j - f(B)_j) \\
 V &= \frac{1}{2N} \sum_{j=1}^{2N} \left(f(AB)_j - \frac{1}{2N}\sum_{k=1}^{2N} f(AB)_k \right)^2.
\end{split}
\end{align}
For the second-order Sobol index, we use the estimator given in \citep{saltelli_2002}
\begin{align}\label{eq:V_ij}
 V_{i,j} = \frac{1}{N} \sum_{j=1}^N f(B_A^{(i)})_j f(A_B^{(j)})_j.
\end{align}
Finally, the estimator for the total Sobol index is given in \citep{saltelli_2010,jansen_1999} as
\begin{align}\label{eq:total_sobol}
 S_{Ti} = \frac{\frac{1}{2N} \sum_{j=1}^N (f(A)_j - f(A_B^{(i)})_j)^2}{V}.
\end{align}
The estimators in \eqref{eq:V_i} and \eqref{eq:total_sobol} are variations of a standard mean estimator for which the rate of convergence is in $O(N^{-1/2})$ \citep{iooss_2015}. We hence expect a convergence rate of $\frac{1}{2}$.

\subsubsection{Morris method}\label{sec:Meth:Morris}

The second sensitivity analysis method under consideration is the Morris method, introduced in \citep{morris_1991,campolongo_2007}. For this method, the input parameters $x$ are sampled as a set of trajectories from the unit hypercube $[0, 1]^n$. From this distribution, they are later transformed into their actual distribution. Each element of the trajectory is bound to a discrete grid of $p$ levels ($p$ even number), allowing for $x_i \in \{0, \frac{1}{p-1}, \frac{2}{p-1}, ..., 1\}$. Starting from a random sample $x \in \R^n$, the trajectory is created by changing one parameter at a time until all $n$ parameters have been changed once. Thereby, the step size for this change is given by $\Delta = \pm\frac{p}{2(p-1)}$, to ensure a symmetric treatment of the input parameters \citep{morris_1991}, and avoid focusing on local behaviors \citep{pianosi_2016}. We use the sign in front of $\Delta$ to guarantee that the new sample remains inside the unit hypercube $[0, 1]^n$. 
A trajectory is hence given by $\{x, x + \Delta e_j, x + \Delta e_j + \Delta e_k, ...\}$, with $e_j$ being the unit vector with a one entry at position $j$. 

From a trajectory, we compute the elementary effect of each input factor $x_i$ by
\begin{align}\label{eq:elementary_effect}
 d_i(x) = \frac{f(x_1, ..., x_{i-1}, x_i + \Delta, x_{i+1}, ..., x_n) - f(x)}{\Delta}.
\end{align}
The values of these elementary effects can be interpreted as local measures of sensitivity at point $x$ in direction $x_i$. Estimating global sensitivity requires approximating the first and second moments of the elementary effects, which we approximate by creating $N$ trajectories \citep{morris_1991,campolongo_2007}. The first and second moments are given by 
\begin{align*}
    \mu_i = \frac{1}{N} \sum_{j=1}^N d_i^{(j)}, \quad
    \sigma_i = \sqrt{\frac{1}{N-1} \sum_{j=1}^N (d_i^{(j)} - \mu_i)^2}
\end{align*}
where $d^{(j)} \in \R^n$ is the vector of the elementary effects $d_i^{(j)}$ that correspond to the $j$-th trajectory. To avoid the problem of non-monotonic models where elementary effects may have different signs, \citep{campolongo_2007} introduces a refined version by estimating the mean over the absolute value of the elementary effects
\begin{align}\label{eq:mu_i}
    \mu_i^* = \frac{1}{N} \sum_{j=1}^N |d_i^{(j)}|.
\end{align}
Again, we see that the sensitivity metric is in the form of a mean estimator, and hence, we expect a convergence rate of $\frac{1}{2}$ with respect to the number of samples $N$. 

Comparing \eqref{eq:mu_i} with the formula for approximating the conditional variances $V_i$ and $V_{i,j}$ in \eqref{eq:V_i} and \eqref{eq:V_ij}, we observe that the quantities have different units. Given that $f(x)$ has the unit $[F]$, then $V_i$ and $V_{i,j}$ are given in terms of $[F^2]$ while $\mu_i^*$ has unit $[F]$. The sensitivity ranking remains unaffected by this; however, if one compares the actual values of the two metrics, this can have a significant influence. Therefore, whenever we consider the actual values of the Morris scores we use
\begin{align}\label{eq:mu_i2}
    \mu_i^{2*} = \left(\frac{1}{N} \sum_{j=1}^N |d_i^{(j)}|\right)^2.
\end{align}

The outputs of this method are $n$ values of $\mu_i^*$, $\mu_i^{2*}$ and $\sigma_i$, each corresponding to one input parameter. Unlike the Sobol indices, where the indices sum up to one in \eqref{eq:Sobol_sum}, the magnitude of $\mu_i^*$, $\mu_i^{2*}$ and $\sigma_i$ depend on the overall variability of the model. As we aim to compare the sensitivity scores for different times in our numerical experiment, we normalize the values of $\mu_i^*$, $\mu_i^{2*}$, and $\sigma_i$ such that the metric gets quantitative \citep{pianosi_2016}
%\begin{align*}
%	\begin{split}
%		\hat{\mu}_i^* = 100 \cdot\frac{\mu_i^*}{\sum_{j=1}^n \mu_j^*},& \qquad
%		\hat{\sigma}_i = 100 \cdot\frac{\sigma_i}{\sum_{j=1}^n \mu_j^*}, \\
%		\hat{\mu}_i^{2*} = 100 &\cdot\frac{\mu_i^{2*}}{\sum_{j=1}^n \mu_j^{2*}}.
%	\end{split}
%\end{align*} 
\begin{align}\label{eq:morris_norm}
	\begin{split}
		\hat{\mu}_i^* = \frac{100\cdot\mu_i^*}{\sum_{j} \mu_j^*}, \quad
		\hat{\sigma}_i = \frac{100\cdot\sigma_i}{\sum_{j} \mu_j^*}, \quad
		\hat{\mu}_i^{2*} = \frac{100\cdot\mu_i^{2*}}{\sum_{j} \mu_j^{2*}}.
	\end{split}
\end{align} 
The normalized values can be interpreted as a percentage contribution to the sensitivity. 
A large value of $\hat{\mu}_i^*$ indicates a high sensitivity of the model output with respect to the input parameter $x_i$. If additionally $\hat{\sigma}_i$ is small, i.e., $\frac{\hat{\sigma}_i}{\hat{\mu}_i^*} \leq 0.1$, we can assume that the model output depends almost linearly on $x_i$. Higher values of $\hat{\sigma}_i$ indicate the presence of interactions and/or non-linearities in the model. For $\frac{\hat{\sigma}_i}{\hat{\mu}_i^*} \in [0.1, 1]$, we assume monotonous or almost monotonous influence and for $\frac{\hat{\sigma}_i}{\hat{\mu}_i^*} \geq 1$ highly non-monotonous non-linearities or interactions \citep{garciasanchez_2014,merchan-rivera_2022, richieri_2024, pianosi_2016}.

\subsubsection{Active subspace method}\label{sec:Meth:ASM}

Although, the active subspace method (ASM) is mostly used for its ability to identify the most important directions in the input parameter space of the model and reduce a model's dimensionality \citep{constantine_2013,constantine_2015,parente_2019,parente_2019a}, it can also be used to compute a global sensitivity metric, the activity scores \citep{constantine_2017,erdal_2019,erdal_2020,bittner_2020}. 
The variation of the model within its input parameter space $[0, 1]^n$ is given by the covariance matrix \citep{constantine_2015}
\begin{align}\label{eq:covariance}
 C = \int_{[0,1]^n} \nabla f(x) \nabla f(x)^T \, dx = W\Lambda W^T.
\end{align}
As $C$ is symmetric, positive-semidefinite, there exists an eigenvalue decomposition with an orthogonal matrix of eigenvectors $W$ and a diagonal matrix of eigenvalues in decreasing order $\Lambda = \text{diag}(\lambda_1, ..., \lambda_n)$. 
An estimation of $C$ in \eqref{eq:covariance} can be computed using Monte-Carlo sampling \citep{constantine_2017}. Given $N$ samples $x_j \in [0, 1]^n$, we set
\begin{align}\label{eq:covariance_approx}
    \hat{C} = \frac{1}{N} \sum_{j=1}^N \nabla \hat{f}(x_j) \nabla \hat{f}(x_j)^T.
\end{align} 
To lower the cost when computing multiple sensitivity measures simultaneously, we use the elementary effects of the Morris method and set $\nabla\hat{f}_i(x) = d_i(x)$, where $d_i(x)$ is the elementary effect of the Morris method in \eqref{eq:elementary_effect}.
For each eigenpair $(\lambda_i, w_i)$ of $C$, it holds that
\begin{align*}
	\lambda_i = w_i^T C w_i = \int_{[0,1]^n} (\nabla f(x)^T w_i)^2 \, dx,
\end{align*}
\citep{constantine_2017}.
We can hence interpret $\lambda_i$ as the importance of the direction $w_i$ in the input parameter space on the model output. 

Starting from the eigenvalue decomposition of the covariance matrix, the authors of \citep{constantine_2017} define the activity score by 
\begin{align}\label{eq:alpha_i}
	\alpha_i(k) = \sum_{j=1}^k \lambda_j w_{ij}^2, \qquad k\leq n.
\end{align}
The value of $k$ is typically chosen to be the location of the biggest gap between the eigenvalues in log scale, i.e., $\log(\lambda_k) \gg \log(\lambda_{k+1})$ \citep{constantine_2015}. In the context of surrogate modeling, the first $k$ dimensions, corresponding to $w_1, ..., w_k$, are the active dimensions that impact the model output most. The remaining $n-k$ dimensions are the inactive dimensions \citep{constantine_2017}. In cases without a significant eigenvalue gap, $k = n$ is reasonable. Although $\alpha_i(n)$ also includes the directions of less importance, these directions are weighted by a smaller eigenvalue and contribute less to the sensitivity value. During our experiments, we estimate the sensitivity of the input parameters at different times. Since we find a different location for the gap depending on the time step considered, we use $k = n$ for all times to ensure a consistent approach. It is shown in \citep{constantine_2017} that, in this case, the activity score is equivalent to the derivative-based global sensitivity measure
\begin{align}\label{eq:deriv_based_sens}
	\nu_i = \frac{1}{N} \sum_{j=1}^{N} \left(\frac{\partial f(x_j)}{\partial x_i}\right)^2.
\end{align}
For applications where the surrogate model is not of interest, this is advantageous, as the evaluation does not require the eigendecomposition of the covariance matrix, but only the sum of the squared derivatives in \eqref{eq:deriv_based_sens}.
It also highlights that for our implementation with $\frac{\partial f(x_j)}{\partial x_i} = d_i(x_j)$, the activity score can be seen as a modification of $\mu_i^*$ from the Morris method, using the mean of the square elementary effects in place of the mean of the absolute elementary effects. Analyzing the units of $\alpha_i$ based on \eqref{eq:alpha_i}, we obtain $[F^2]$ alike for the Sobol indices. The values are hence comparable without adding a transformation. 

As for the Morris method, we normalize the activity scores as 
\begin{align}\label{eq:asm_norm}
    \hat{\alpha}_i = 100 \cdot \frac{\alpha_i}{\sum_{j=1}^n \alpha_j}
\end{align}
to make the metric quantitative \citep{pianosi_2016}, ensuring comparability of the results at different times. We compute the activity scores from the eigenvalues of the covariance matrix in \eqref{eq:covariance_approx}. As $\hat{C}$ converges to $C$ with order $\frac{1}{2}$, we expect a convergence rate of $\frac{1}{2}$ for the activity scores $\hat{\alpha}$.

The signs of the eigenvector entries corresponding to the largest eigenvalues are also interesting in analyzing the relationships between input and output. If the sign is the same, a simultaneous increase of both hyperparameters results in the largest output change; if the sign is different, we need to decrease the value of one hyperparameter and increase the other \citep{constantine_2017}.

When constructing a surrogate model, we need the first $k$ eigenvectors $W_k \in \R^{n \times k}$ of the covariance matrix. We can then reuse the samples $(x_i, f(x_i))$ from the approximation of $\hat{C}$ and compute the projection $y_i = W_k^T x_i$. The surrogate model is given by a regression surface $G:\R^k \rightarrow \R$ of the pairs $(y_i, f(x_i))$ with $G(y_i) \approx f(x_i)$ \citep{constantine_2015}. We use a polynomial fit of order $m$ to construct $G$ using the Python package Scikit-learn \citep{pedregosa_2011}.

\section{Results and Discussion}\label{sec:Results}

During our numerical experiments, we analyze the sensitivity of the two flow systems introduced in \cref{sec:Meth:flow}. We compare the different sensitivity metrics by taking the dimensionality of the problem into account. For the comparison, we use the total Sobol index $S_T$, the Morris index $\hat{\mu}^{2*}$, and the activity score $\hat{\alpha}$\added{, as all three quantities represent the total-order influence of one hyperparameter on the model output}. After demonstrating that the results for the RPM flow are consistent over all three methods, we look at the higher-dimensional quadrupole flow. Here, we only consider the Morris method and ASM to reduce the computational cost. We motivate this choice by the convergence behavior of the metrics for the RPM flow, where, for reaching the same relative accuracy, Morris scores require significantly fewer model evaluations compared to Sobol indices.

\subsection{Sensitivity analysis on low dimensional problems}\label{sec:Results:RPM}

In our first experiment, we analyze the sensitivity of the two versions of the RPM flow. We regulate the non-randomized versions by varying the two hyperparameters $\Theta$ and $\tau$ and the randomized version with the hyperparameters $\bar{\Theta}$, $\Theta_r$, $\bar{\tau}$ and $\tau_r$. For the sampling intervals, we choose one configuration that allows for a large range of values for the flow parameters and one configuration with a more restricted choice of flow parameters, to investigate the effects of the prior information on the parameters. We give the exact intervals in \cref{tab:parameter_ranges}. For each configuration, we compute all three sensitivity metrics defined in \cref{sec:Meth:Sens}. To do so, we create Saltelli samples for the first, second and total-order Sobol indices \eqref{eq:S_i}, \eqref{eq:total_sobol} and Morris samples for the Morris scores \eqref{eq:mu_i} and the activity scores \eqref{eq:alpha_i}, with $N = 4096$. Our convergence analysis demonstrates that this sample size is large enough to ensure that all methods converge.

We assess the result of our sensitivity analysis using two approaches.
We start by ranking the parameters according to their sensitivity scores. The ranks are often used to present the sensitivity analysis results in the literature \citep{pianosi_2016,sarrazin_2016,ziliotto_,richieri_2024,mazziotta_2024} and allow for an easier comparison between the different methods. 
To better understand the sensitivity, we also look at the exact values of the sensitivity metrics for each parameter. Doing so, we use a transformation on the Morris score to obtain $\mu^{2*}$ as presented in \eqref{eq:mu_i2} and normalize the Morris scores and activity scores using \eqref{eq:morris_norm} and \eqref{eq:asm_norm} to allow for inter-comparison between the different time steps.

\subsubsection{Sensitivity analysis for the non-randomized RPM flow}\label{sec:Results:RPM1}

\begin{figure*}[tb!]
	\centering
	\begin{subfigure}[t]{7.95cm}
		\includegraphics[width=7.95cm]{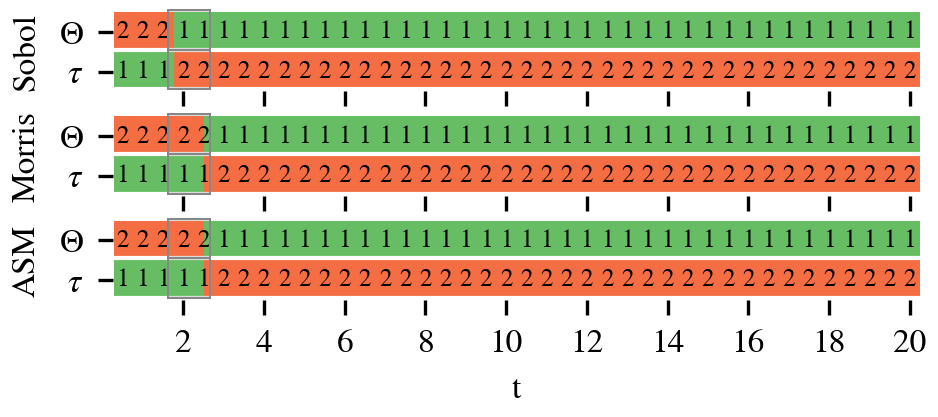}
		\caption{Ranking for the large intervals experiment}
		\label{fig:sSens_ranking_bigintervals}
	\end{subfigure}
	\hfill
	\begin{subfigure}[t]{7.95cm}
		\includegraphics[width=7.95cm]{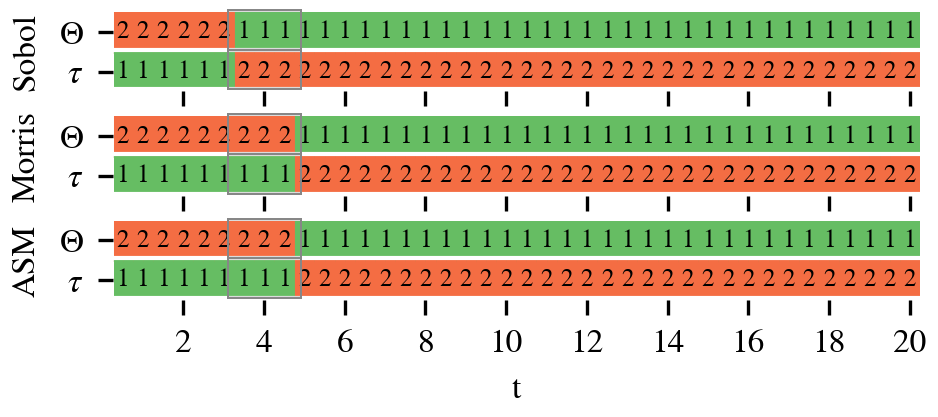}
		\caption{Ranking for the small intervals experiment}
		\label{fig:sSens_ranking_smallintervals}
	\end{subfigure}

	\caption{Sensitivity rankings for the non-randomized RPM flow with $\Theta \in [0, \pi]$ and $\tau \in [0, 1]$ in \cref{fig:sSens_ranking_bigintervals} and $\Theta \in [0.4\pi, 0.6\pi]$ and $\tau \in [0.4, 0.6]$ in \cref{fig:sSens_ranking_smallintervals} over time $t$. The lower the score, the higher the parameter is ranked in sensitivity. Differences between the rankings of the different methods are highlighted with gray boxes.}
	\label{fig:sSens_ranking}
\end{figure*}

Starting with the sensitivity analysis for the non-randomized RPM flow, we show the rankings in \cref{fig:sSens_ranking}. For both experiments, with large and small hyperparameter intervals, $\tau$ \replaced{seems to be}{is} the most sensitive hyperparameter for small times. Hence, at the beginning of the experiment, the number of performed rotations is more important to the result than the rotation angle. At time $t>1.5$, there is a switch, and $\Theta$ gets more sensitive. Thereby, the switch happens simultaneously for the Morris method and the ASM ($t \approx 2.5$ for the large interval experiment and $t \approx 4.75$ for the small interval experiment). Although the switch for the total Sobol index appears earlier ($t \approx 1.75$ for the large interval experiment and $t \approx 3.25$ for the small interval experiment), we conclude that the results over the three methods are very consistent.

\begin{figure*}[tb!]
	\centering
	\begin{subfigure}[t]{7.95cm}
		\includegraphics[width=7.95cm]{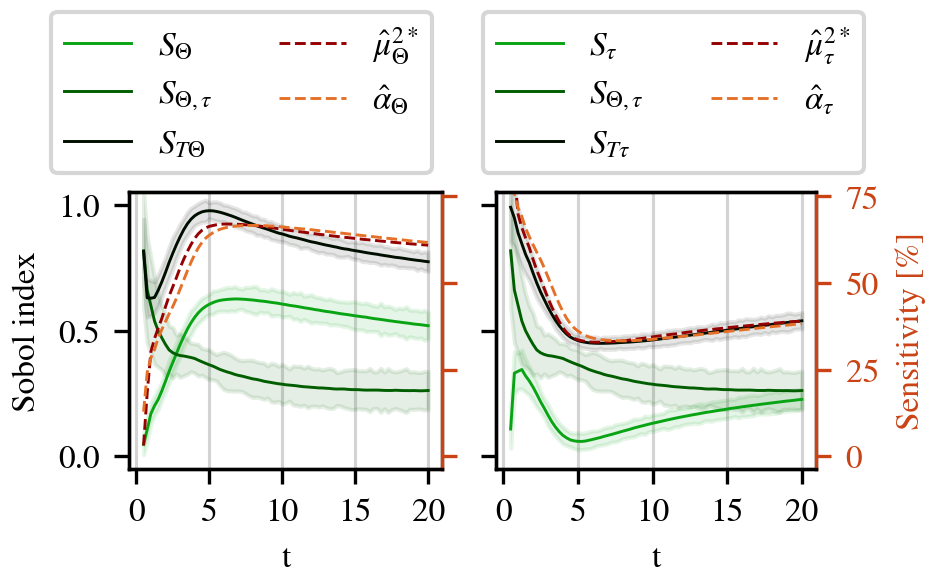}
		\caption{Sensitivity of $\Theta$ (left) and $\tau$ (right) in the large intervals experiment}
		\label{fig:sSens_bigintervals}
	\end{subfigure}
	\hfill
	\begin{subfigure}[t]{7.95cm}
		\includegraphics[width=7.95cm]{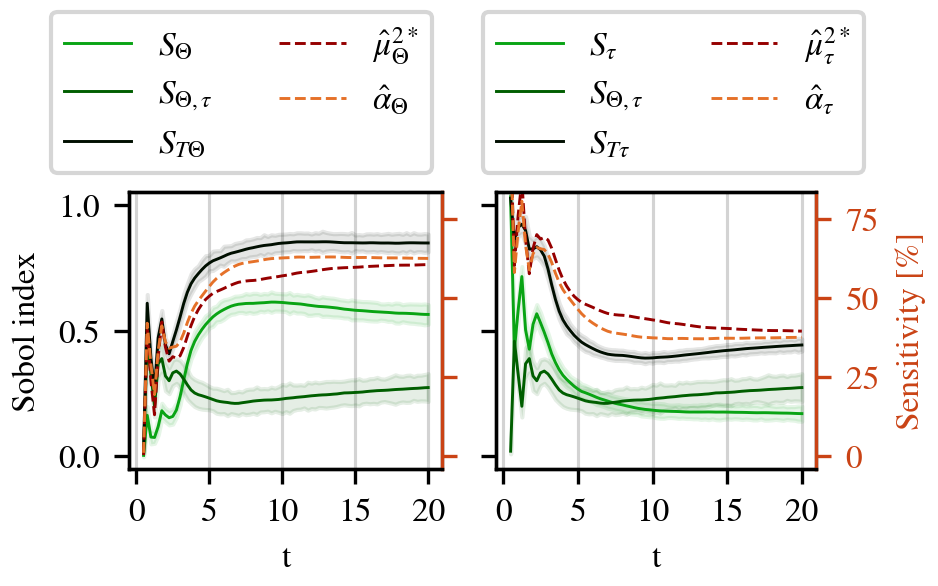}
		\caption{Sensitivity of $\Theta$ (left) and $\tau$ (right) in the small intervals experiment}
		\label{fig:sSens_smallintervals}
	\end{subfigure}

	\caption{Results of the sensitivity analysis for the non-randomized RPM flow with $\Theta \in [0, \pi]$ and $\tau \in [0, 1]$ in \cref{fig:sSens_bigintervals} and $\Theta \in [0.4\pi, 0.6\pi]$ and $\tau \in [0.4, 0.6]$ in \cref{fig:sSens_smallintervals} over time $t$. We show the first, second, and total-order Sobol indices $S$ with respect to the left y-axis with solid lines and the Morris activity scores $\hat{\mu}^{2*}$ as well as the active subspace activity scores $\hat{\alpha}$ with respect to the right y-axis with dashed lines. For the results of Sobol we also show a $95~\%$ confidence interval computed using bootstrapping.}
	\label{fig:sSens}
\end{figure*}

To better understand the sensitivity, we extend our analysis and show the sensitivity metrics' values in \cref{fig:sSens}. As the curves for the total Sobol index $S_T$, the Morris index $\hat{\mu}^{2*}$, and the activity score $\hat{\alpha}$ show the same behavior, the analysis reinforces that the results for different sensitivity metrics reflect the same underlying sensitivity for both experiments. However, considering the exact values instead of the ranking reveals additional information. For the experiment with the large hyperparameter intervals, we find a minimum in the sensitivity of $\tau$ and a maximum in the sensitivity of $\Theta$ at $t \approx 5$. We notice the \replaced{variable}{unsteady} line for the sensitivity scores at small times for the small hyperparameter intervals experiment in \cref{fig:sSens_smallintervals}. This behaviour can be explained by the fact that the overall output variability is very small at the beginning of the experiment, and hence \added{although $\tau$ reaches higher sensitivity scores, }none of the hyperparameters is dominant. Due to the normalization of the sensitivity metrics, we cannot yield zero sensitivity for all hyperparameters. We show the output variability in the supplementary material, Fig. S2b.

\begin{figure*}[tb!]
	\centering
	\begin{subfigure}[t]{5.4cm}
		\includegraphics[width=5.4cm]{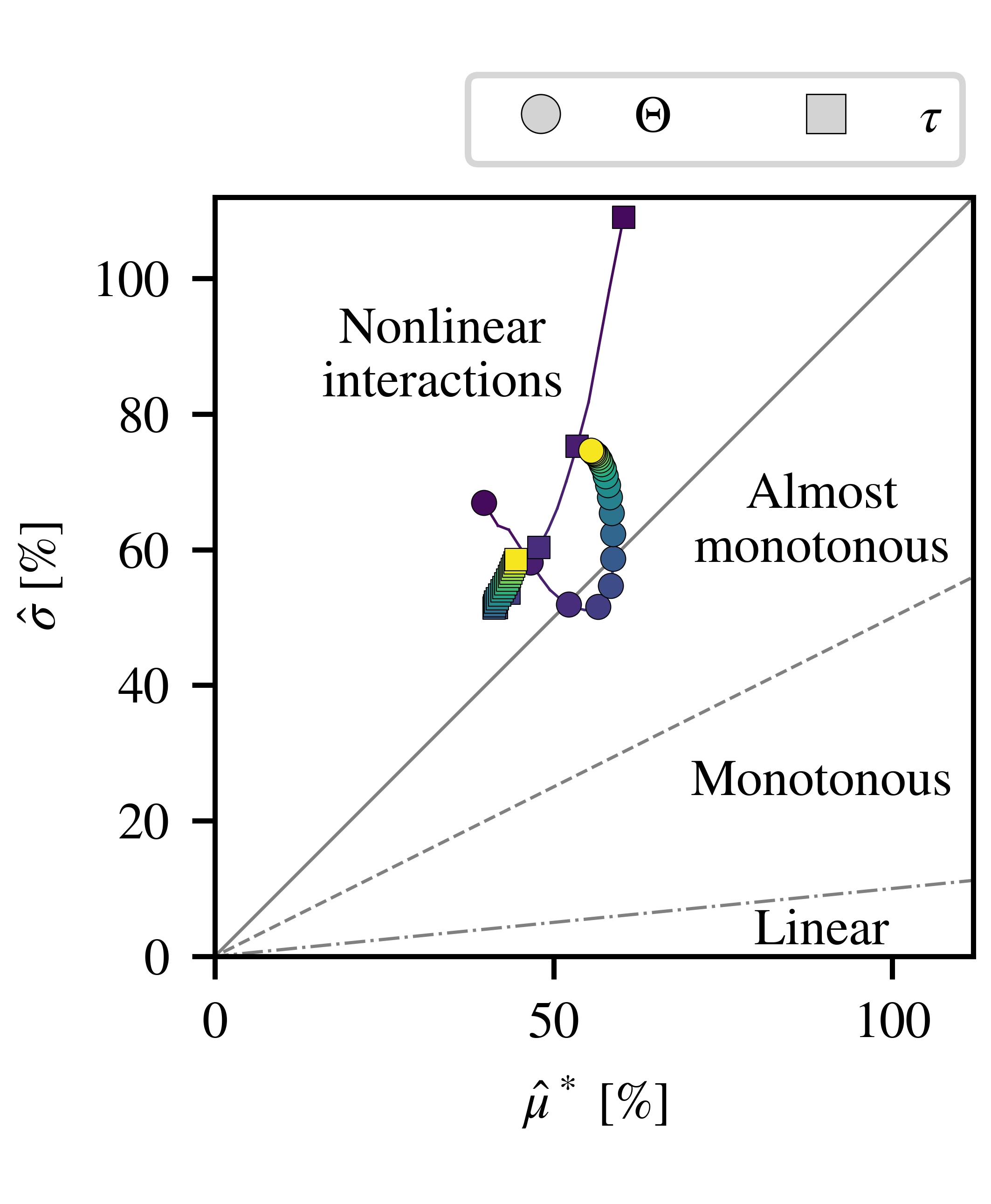}
		\caption{Large intervals experiment}
		\label{fig:sSens_relationship_big}
	\end{subfigure}
    ~
	\begin{subfigure}[t]{5.4cm}
		\includegraphics[width=5.4cm]{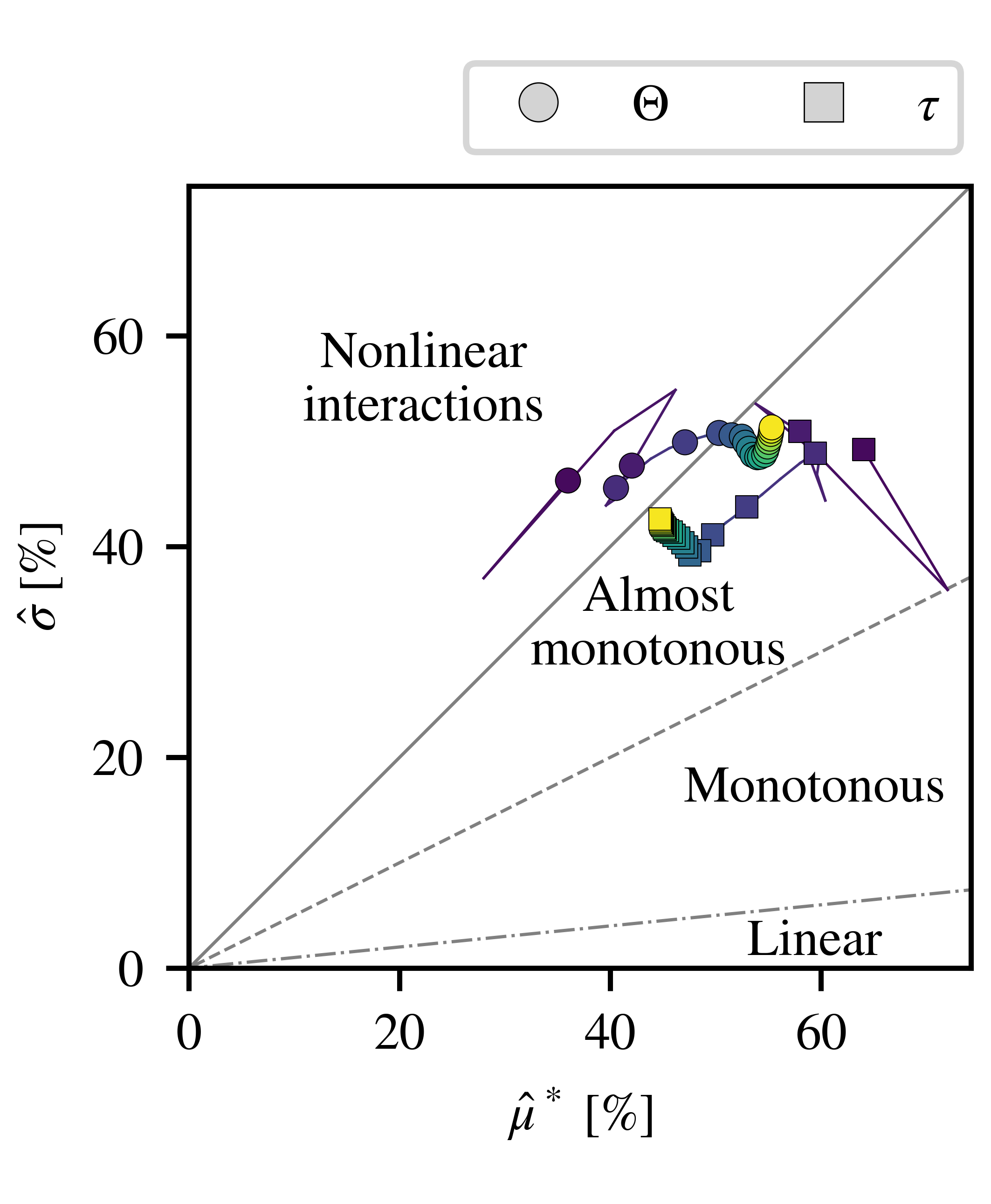}
		\caption{Small intervals experiment}
		\label{fig:sSens_relationship_small}
	\end{subfigure}
    ~
	\begin{subfigure}[t]{1.2cm}
		\includegraphics[width=1.2cm]{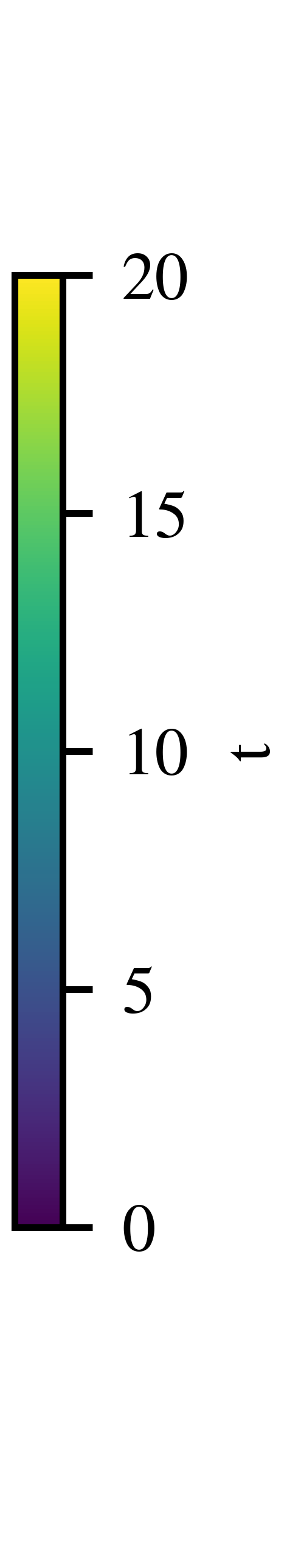}
	\end{subfigure}

	\caption{Relationship between $\hat{\mu}^*$ and $\hat{\sigma}$ over time $t \geq 1$ for the non-randomized RPM flow with $\Theta \in [0, \pi]$ and $\tau \in [0, 1]$ in \cref{fig:sSens_relationship_big} and $\Theta \in [0.4\pi, 0.6\pi]$ and $\tau \in [0.4, 0.6]$ in \cref{fig:sSens_relationship_small}. The solid line indicates $\frac{\hat{\sigma}}{\hat{\mu}^*} = 1$, the dashed line indicates $\frac{\hat{\sigma}}{\hat{\mu}^*} = 0.5$ and the dash-dotted line indicates $\frac{\hat{\sigma}}{\hat{\mu}^*} = 0.1$.}
	\label{fig:sSens_relationship}
\end{figure*}

Using the three different methods, we can also analyze hyperparameter interactions. For the Sobol method, we compute the first- and second-order Sobol indices. As we consider a two-dimensional model where $A_B^{(1)} = B_A^{(2)}$ and $A_B^{(2)} = B_A^{(1)}$, this comes with no additional cost, and the results give a detailed impression about hyperparameter interactions. 
The first-order Sobol index shows the influence of one hyperparameter without any interactions, while the second-order index represents the interaction between exactly two hyperparameters. We show all first and second-order Sobol indices in \cref{fig:sSens}. We find differences comparing the experiment with large and small hyperparameter intervals. For large hyperparameter intervals, there is a strong interaction between the two hyperparameters for $t < 5$ that decreases over time. For the small hyperparameter intervals, after we exit the time frame where the output variability is too small to draw reasonable conclusions on sensitivity at $t \approx 4$, the interaction stays more or less constant throughout the experiment. 
Using the Morris method, we consider $\frac{{\hat\sigma}}{\hat{\mu}^*}$ to analyze the type of interactions inside the model. \Cref{fig:sSens_relationship} displays the relationship between $\hat{\mu}^*$ and $\hat{\sigma}$ over time for the two experiments. We see nonlinear interactions for the experiment with large hyperparameter intervals, while for the small hyperparameter intervals, the interactions are at the transition between nonlinear and almost monotonous. 
Finally, one can also observe relationships between hyperparameters using ASM by considering signs of the eigenvector entries corresponding to the largest eigenvalue \citep{constantine_2017}. 
In the small hyperparameter intervals experiment, the first eigenvector has the same sign for both hyperparameters over the whole experiment, indicating a positive correlation. We cannot find any indication of a linear relationship for the large hyperparameter intervals, as most of the time, one of the eigenvector entries is close to zero. We give the eigenvectors in the supplementary material, Fig. S7.

An interpretation of the sensitivity analysis tells us that to enhance mixing in the RPM flow, controlling the rotation angle $\Theta$ is important, as long as we do not have to operate the system only for a short time. \added{We highlight that an arbitrary rotation angle $\Theta$ might not be applicable for all applications. In groundwater remediation, for example, this may lead to a new pair of wells being drilled after each rotation of source and sink, which is practically infeasible, while a modification of $\tau$ could be implemented more easily.}
Interestingly, a more constrained hyperparameter interval does not allow for identifying a leading parameter controlling mixing at short times. For longer times, the size of the parameter interval does not influence the ranking of the parameters controlling mixing.

% \begin{figure}[tb!]
% 	\centering
% 	\begin{subfigure}[t]{8.6cm}
% 		\includegraphics[width=8.6cm]{fig/sSens/A_T[0. 1.]pi_tau[0. 1.]_num5000_saltellisamples4096/SobolMorris_ranking2a.png}
% 	\end{subfigure}

% 	\caption{Sensitivity rankings for the non-randomized RPM flow with $\Theta \in [0, \pi]$ and $\tau \in [0, 1]$ over time $t$. The lower the score the higher the parameter is ranked in terms of sensitivity.}
% 	\label{fig:sSens_ranking_bigintervals}
% \end{figure}

% \begin{figure}[tb!]
% 	\centering
% 	\begin{subfigure}[t]{8.6cm}
% 		\includegraphics[width=8.6cm]{fig/sSens/A_T[0.4 0.6]pi_tau[0.4 0.6]_num5000_saltellisamples4096/SobolMorris_ranking2a.png}
% 	\end{subfigure}

% 	\caption{Sensitivity rankings for the non-randomized RPM flow with $\Theta \in [0.4\pi, 0.6\pi]$ and $\tau \in [0.4, 0.6]$ over time $t$. The lower the score the higher the parameter is ranked in terms of sensitivity.}
% 	\label{fig:sSens_ranking_smallintervals}
% \end{figure}

\subsubsection{Sensitivity analysis for the randomized RPM flow}\label{sec:Results:RPM2}

\begin{figure}[tb!]
	\centering
	\begin{subfigure}[t]{7.5cm}
		\includegraphics[width=7.5cm]{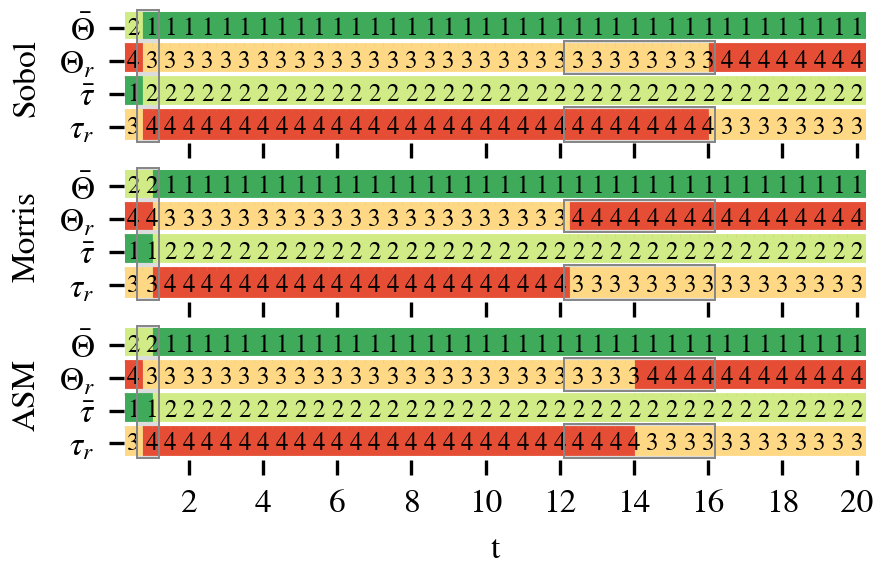}
	\end{subfigure}

	\caption{Sensitivity rankings for the randomized RPM flow with $\bar{\Theta} \in [0, \pi]$, $\Theta_r \in [0, 0.2\pi]$, $\bar{\tau} \in [0.1, 1]$ and $\tau_r \in [0, 0.2]$ over time $t$. The lower the score, the higher the parameter is ranked in sensitivity. Differences between the rankings of the different methods are highlighted with gray boxes.}
	\label{fig:Sens_ranking_bigintervals}
\end{figure}

\begin{figure*}[tb!]

	\begin{subfigure}[t]{1cm}
		\includegraphics[width=1cm]{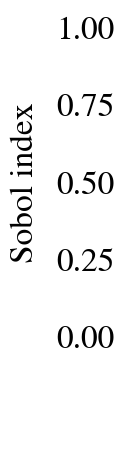}
	\end{subfigure}
	\begin{subfigure}[t]{3.4cm}
		\includegraphics[width=3.4cm]{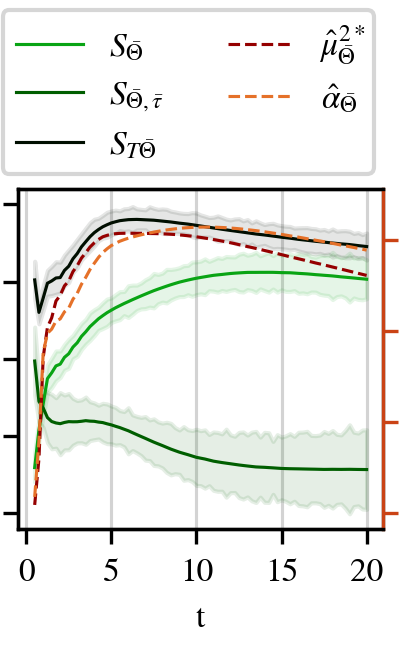}
		\caption{Sensitivity of $\bar{\Theta}$}
	\end{subfigure}
	\begin{subfigure}[t]{3.4cm}
		\includegraphics[width=3.4cm]{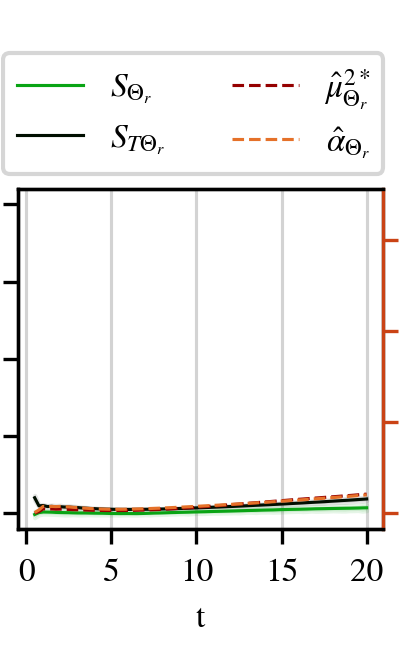}
		\caption{Sensitivity of $\Theta_r$}
	\end{subfigure}
	\begin{subfigure}[t]{3.4cm}
		\includegraphics[width=3.4cm]{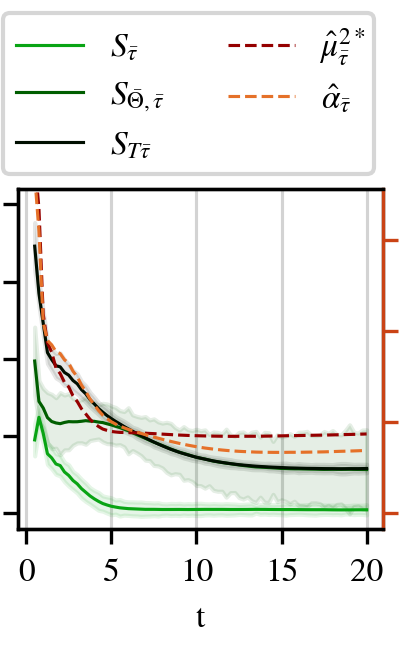}
		\caption{Sensitivity of $\bar{\tau}$}
	\end{subfigure}
	\begin{subfigure}[t]{3.4cm}
		\includegraphics[width=3.4cm]{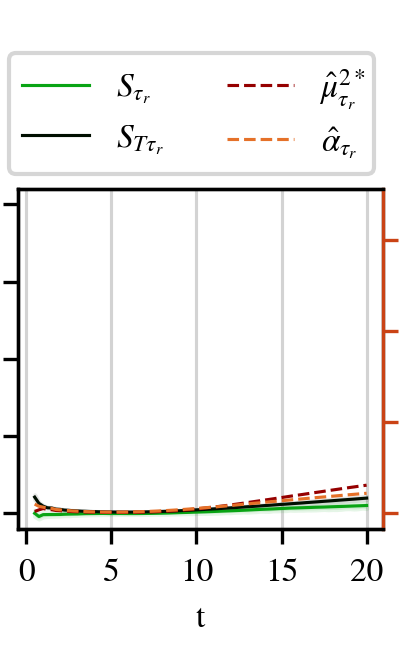}
		\caption{Sensitivity of $\tau_r$}
	\end{subfigure}
	\begin{subfigure}[t]{0.8cm}
		\includegraphics[width=0.8cm]{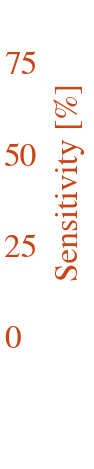}
	\end{subfigure}

	\caption{Results of the sensitivity analysis for the randomized RPM flow with $\bar{\Theta} \in [0, \pi]$, $\Theta_r \in [0, 0.2\pi]$, $\bar{\tau} \in [0.1, 1]$ and $\tau_r \in [0, 0.2]$ over time $t$. We show the first, second, and total-order Sobol indices $S$  with respect to the left y-axis with solid lines, and the Morris activity scores $\hat{\mu}^{2*}$ as well as the active subspace activity scores $\hat{\alpha}$ with respect to the right y-axis with dashed lines. For the results of Sobol we also show a $95~\%$ confidence interval computed using bootstrapping.}
	\label{fig:Sens_bigintervals}
\end{figure*}

Looking at the sensitivity analysis of the randomized RPM flow, we start with the experiment with large hyperparameter intervals. The ranking in \cref{fig:Sens_ranking_bigintervals} shows the same trends for all three methods. After the large switch at $t \approx 1$, there is a second switch of the ranking between $\Theta_r$ and $\tau_r$. The timing for this switch is different over all three methods ($t \approx 16$ for the Sobol indices, $t \approx 12.25$ for Morris, and $t \approx 14$ for ASM). This inconsistency is due to the values of the sensitivity metrics, which are very close to each other as shown in \cref{fig:Sens_bigintervals}. We observe that the sensitivities of $\Theta_r$ and $\tau_r$ converge towards similar values, which makes locating the exact time of the jump very difficult. The analysis of \cref{fig:Sens_bigintervals} also reveals that the metrics yield very similar values. This could only be achieved due to the transformation of the Morris scores in \eqref{eq:mu_i2}. The modification due to the transformation is not reflected in the sensitivity ranking, as the transformation is strictly monotonically increasing. 
\replaced{Overall}{All in all}, we find that the deviations of the randomization $\Theta_r$ and $\tau_r$ are less sensitive than the means $\bar{\Theta}$ and $\bar{\tau}$. This result are due to the choice of the larger sampling intervals for the means than for the deviations. The sensitivity of the means is similar to the sensitivity of the respective hyperparameters in the non-randomized case, while the sensitivity of the deviations is nearly zero but increases with time. We explain these sensitivity increases of the deviation parameters as they enable particles to enter the KAM islands of the RPM flow. These KAM islands are non-mixing regions that arise in the non-randomized RPM flow; the randomization slowly deconstructs these islands such that the particles can cover the entire domain. As the effect of the mean values is much larger than the effect of the deviation parameters, it is important to characterize the mean of the random rotations in order to control mixing.

\begin{figure*}[tb!]
	\centering
	\begin{subfigure}[t]{5.4cm}
		\includegraphics[width=5.4cm]{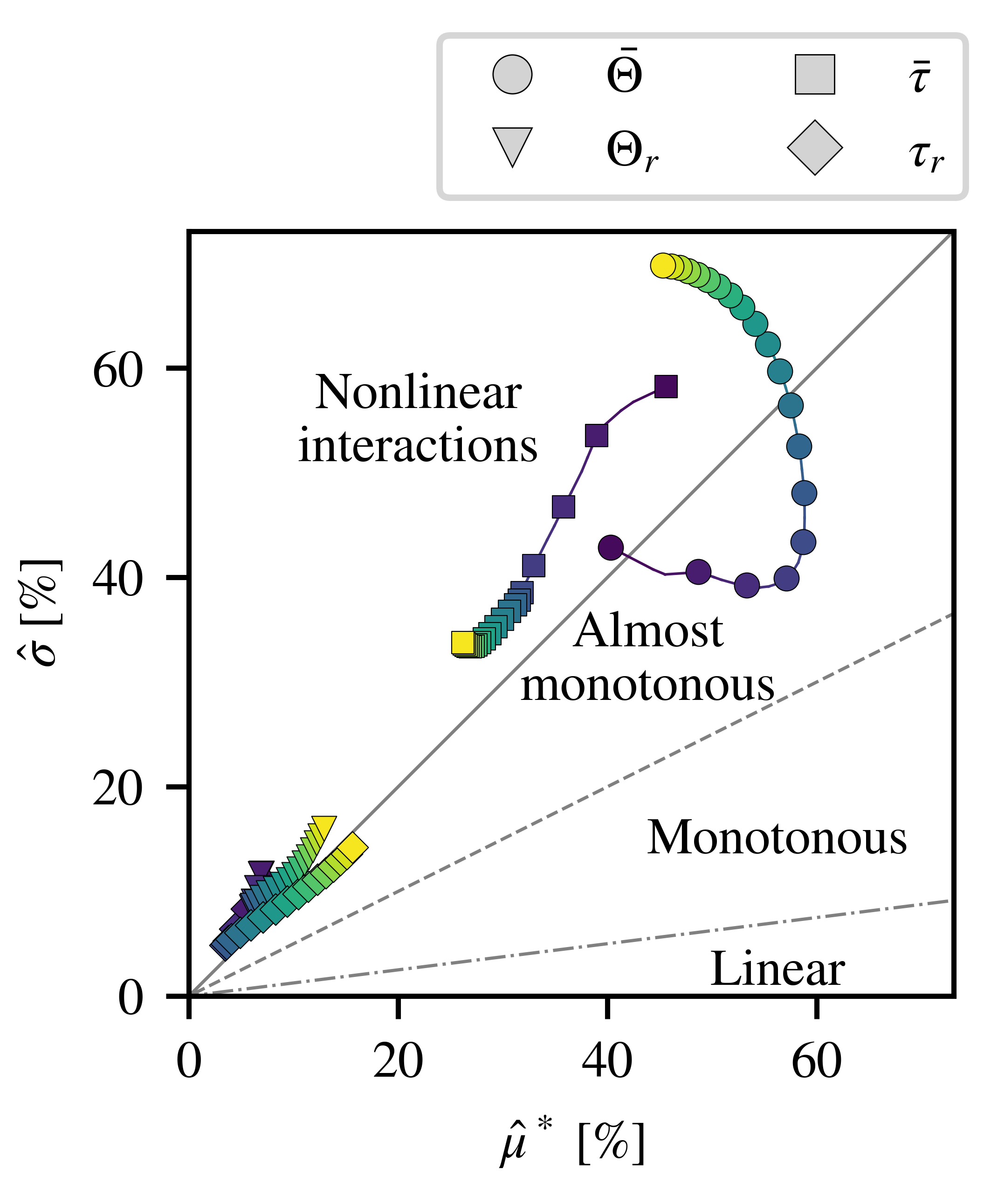}
		\caption{Large intervals experiment}
		\label{fig:Sens_relationship_big}
	\end{subfigure}
	\begin{subfigure}[t]{5.4cm}
		\includegraphics[width=5.4cm]{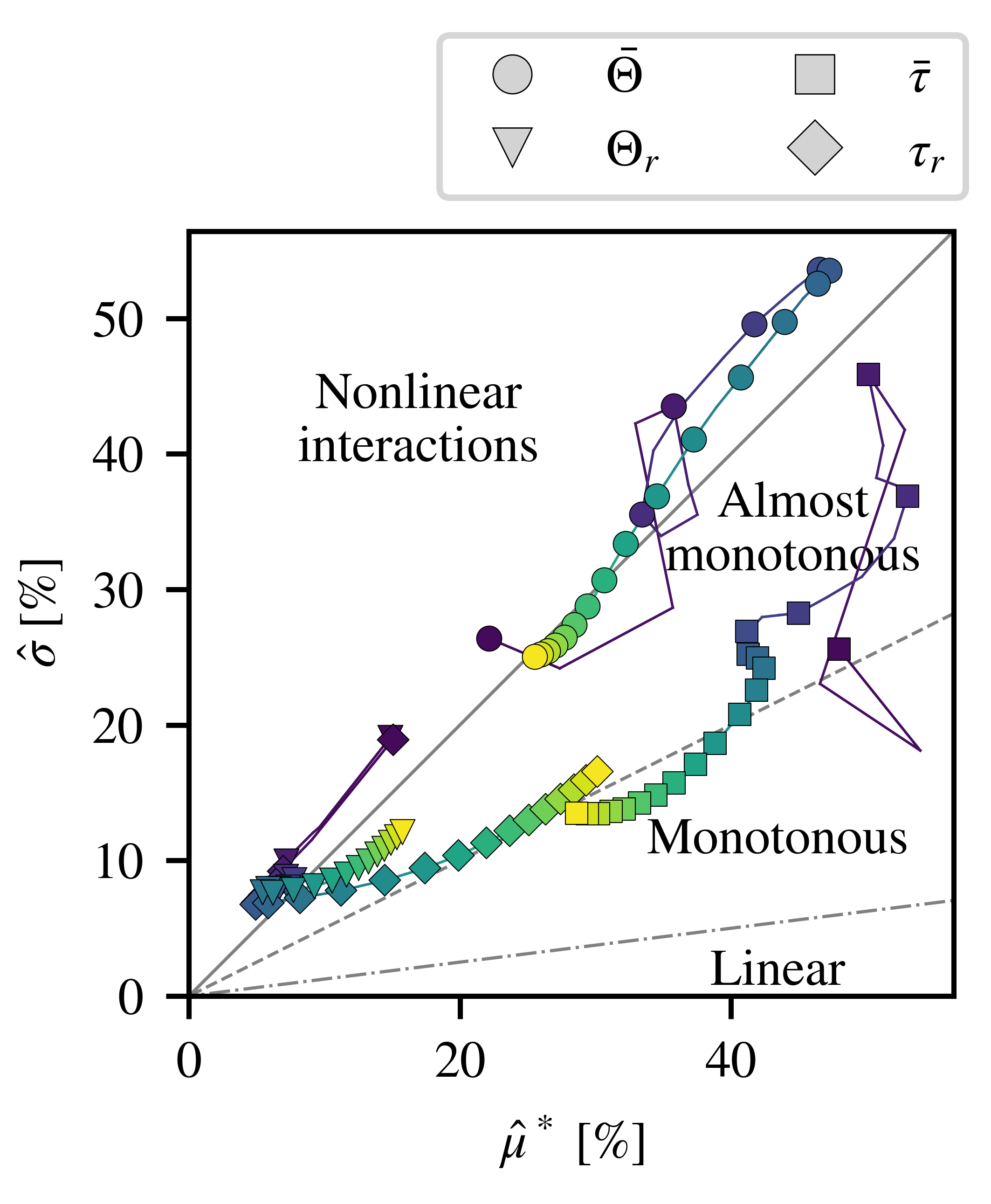}
		\caption{Small intervals experiment}
		\label{fig:Sens_relationship_small}
	\end{subfigure}
	\begin{subfigure}[t]{1.2cm}
		\includegraphics[width=1.2cm]{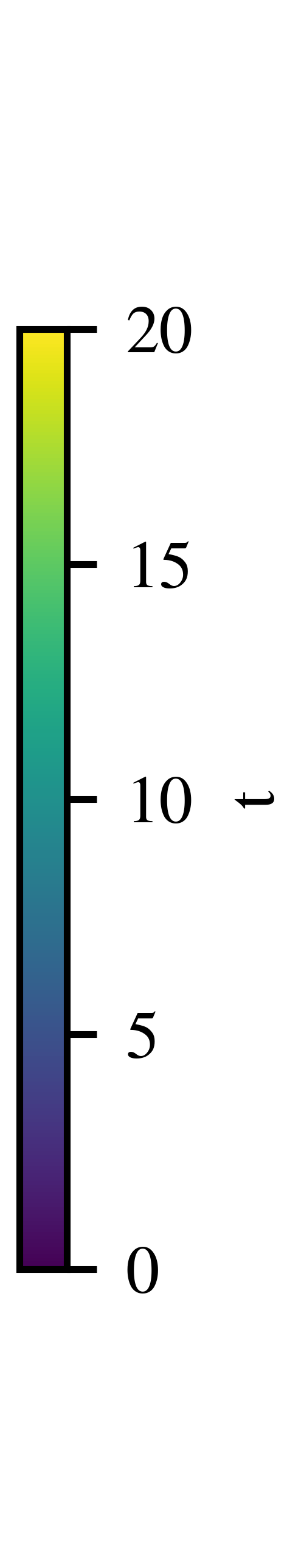}
	\end{subfigure}

	\caption{Relationship between $\hat{\mu}^*$ and $\hat{\sigma}$ over time $t \geq 1$ for the randomized RPM flow with $\bar{\Theta} \in [0, \pi]$, $\Theta_r \in [0, 0.2\pi]$, $\bar{\tau} \in [0.1, 1]$ and $\tau_r \in [0, 0.2]$ in \cref{fig:Sens_relationship_big} and $\bar{\Theta} \in [0.45, 0.55\pi]$, $\Theta_r \in [0, 0.1\pi]$, $\bar{\tau} \in [0.45, 0.55]$ and $\tau_r \in [0, 0.1]$ in \cref{fig:Sens_relationship_small}. The solid line indicates $\frac{\hat{\sigma}}{\hat{\mu}^*} = 1$, the dashed line indicates $\frac{\hat{\sigma}}{\hat{\mu}^*} = 0.5$ and the dash-dotted line indicates $\frac{\hat{\sigma}}{\hat{\mu}^*} = 0.1$.}
	\label{fig:Sens_relationship}
\end{figure*}

This time, analyzing interactions using Sobol indices comes with an additional cost of $N \cdot n$ samples to create $f(B_A^{(i)})|_{i=1}^n$.
We only show the second-order Sobol index $S_{\bar{\Theta},\bar{\tau}}$ in \cref{fig:Sens_bigintervals}, as all other second-order indices yield values very close to zero. There is, hence, only one significant interaction that appears between the two mean parameters. As for the non-randomized case, the interaction is strong for $t < 5$, but decreases over time. Using the finding in \eqref{eq:S_ti_sum} and $S_{T\bar{\tau}} \approx S_{\bar{\Theta},\bar{\tau}}$, we highlight that the total sensitivity of $\bar{\tau}$ is solely due to its interaction with $\bar{\Theta}$. As the total Sobol index is the sum of all lower-order Sobol indices involving the respective hyperparameter, we can conclude that we have nearly no higher-order interactions.
For the Morris method, the analysis of \cref{fig:Sens_relationship_big} demonstrates that the interactions are again mostly nonlinear. Looking at the largest eigenvector of ASM, we find that most of the time, the two means and the two deviations are positively correlated. In contrast, we find a negative correlation between the deviations and the means. The eigenvector entries are given in the supplementary material, Fig. S8a.

\begin{figure}[tb!]
	\centering
	\begin{subfigure}[t]{7.5cm}
		\includegraphics[width=7.5cm]{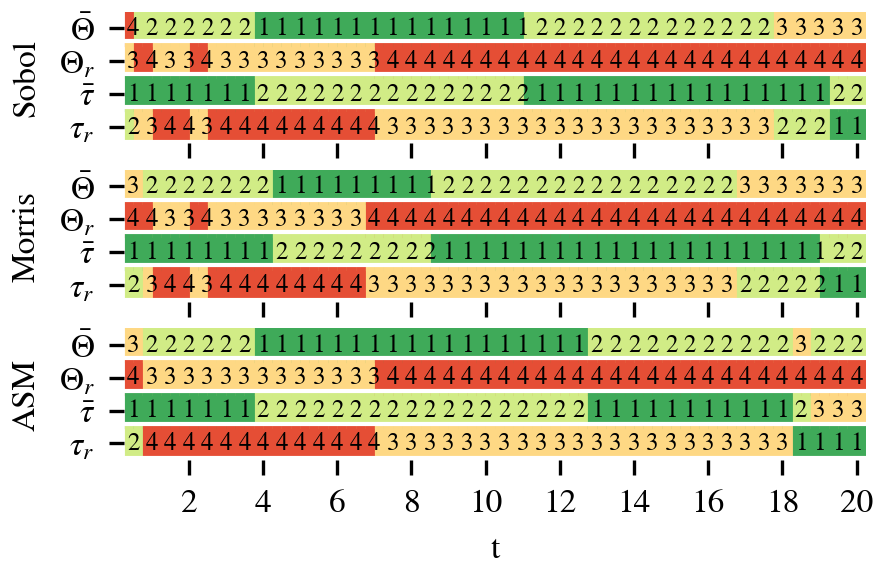}
	\end{subfigure}

	\caption{Sensitivity rankings for the randomized RPM flow with $\bar{\Theta} \in [0.45, 0.55\pi]$, $\Theta_r \in [0, 0.1\pi]$, $\bar{\tau} \in [0.45, 0.55]$ and $\tau_r \in [0, 0.1]$ over time $t$. The lower the score, the higher the parameter is ranked in terms of sensitivity.}
	\label{fig:Sens_ranking_smallintervals}
\end{figure}

\begin{figure*}[tb!]
	\centering

	\begin{subfigure}[t]{1cm}
		\includegraphics[width=1cm]{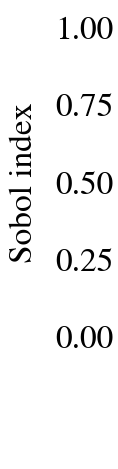}
	\end{subfigure}
	\begin{subfigure}[t]{3.4cm}
		\includegraphics[width=3.4cm]{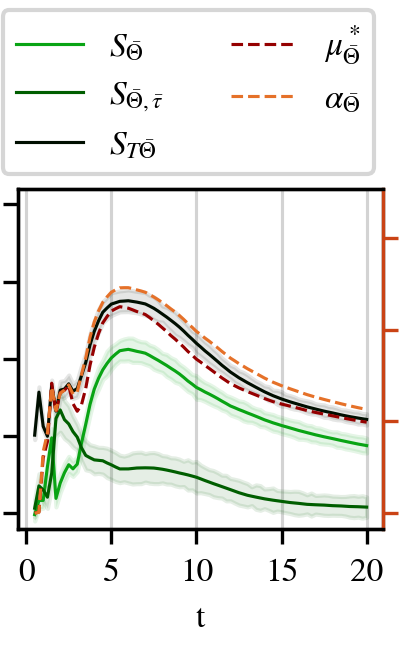}
		\caption{Sensitivity of $\bar{\Theta}$}
	\end{subfigure}
	\hfill
	\begin{subfigure}[t]{3.4cm}
		\includegraphics[width=3.4cm]{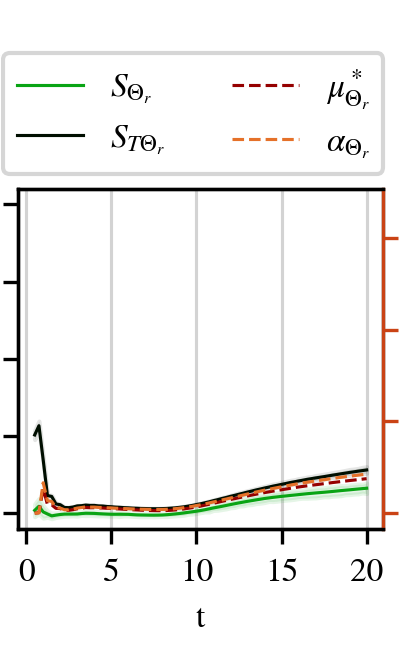}
		\caption{Sensitivity of $\Theta_r$}
	\end{subfigure}
	\hfill
	\begin{subfigure}[t]{3.4cm}
		\includegraphics[width=3.4cm]{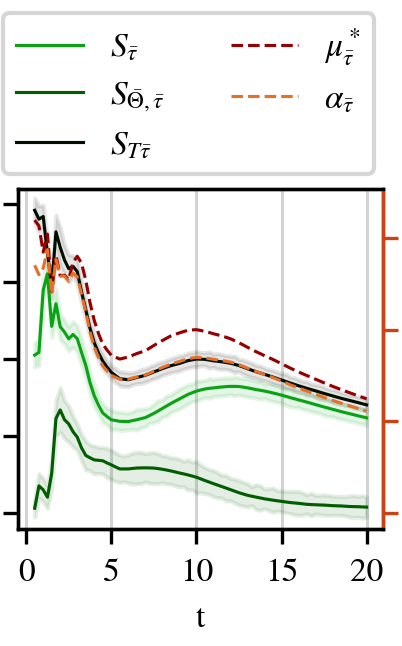}
		\caption{Sensitivity of $\bar{\tau}$}
	\end{subfigure}
	\hfill
	\begin{subfigure}[t]{3.4cm}
		\includegraphics[width=3.4cm]{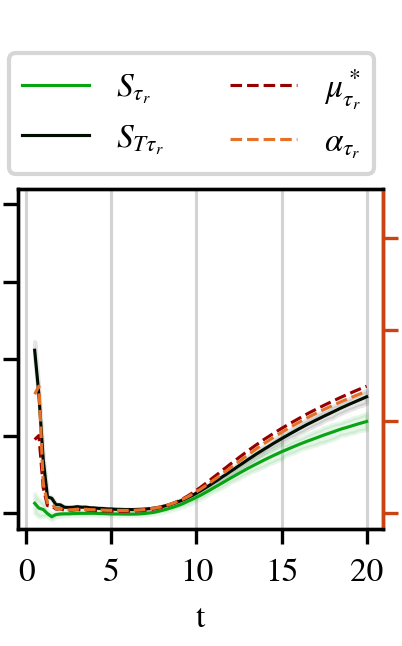}
		\caption{Sensitivity of $\tau_r$}
	\end{subfigure}
	\begin{subfigure}[t]{0.8cm}
		\includegraphics[width=0.8cm]{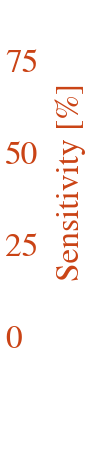}
	\end{subfigure}

	\caption{Results of the sensitivity analysis for the randomized RPM flow with $\bar{\Theta} \in [0.45, 0.55\pi]$, $\Theta_r \in [0, 0.1\pi]$, $\bar{\tau} \in [0.45, 0.55]$ and $\tau_r \in [0, 0.1]$ over time $t$. We show the first, second, and total-order Sobol indices $S$  with respect to the left y-axis with solid lines, and the Morris activity scores $\hat{\mu}^{2*}$ as well as the active subspace activity scores $\hat{\alpha}$ with respect to the right y-axis with dashed lines. For the results of Sobol we also show a $95~\%$ confidence interval computed using bootstrapping.}
	\label{fig:Sens_smallintervals}
\end{figure*}

The situation is similar for the small hyperparameter intervals. Although the ranking of the three methods in \cref{fig:Sens_ranking_smallintervals} seems less consistent, we find a pattern that all three methods follow upon close inspection. Most of the time, the deviation parameters $\Theta_r$ and $\tau_r$ are the least sensitive until the sensitivity of $\tau_r$ reaches rank one at the end of the experiment at $t \approx 18.5$. Most of the time, the mean values $\bar{\Theta}$ and $\bar{\tau}$ are the more sensitive parameters. Thereby, there is a time frame for $t$ between $4$ and $13$ where $\bar{\Theta}$ has rank one. For the remaining time, $\bar{\tau}$ is the most sensitive parameter. Due to the increase in the sensitivity of $\tau_r$ at the end of the experiment, the ranking of the mean values drops by one.
In the randomized RPM flow, a more informed prior (i.e., a smaller range of the parameters) has a larger impact on the sensitivity analysis than the standard RPM flow. This means that randomly operating the system increases the importance of the duration of the strikes. 
Our claim that all three methods yield consistent results is further supported in \cref{fig:Sens_smallintervals}, where the values of $S_T$, $\mu^{2*}$, and $\alpha$ follow a very similar path over the whole experiment.

We \replaced{observe}{realize} many switches in the ranking for small times in \cref{fig:Sens_ranking_smallintervals}. This behavior is again due to the \replaced{variable}{unsteady} behavior of the sensitivity metrics at small times in \cref{fig:Sens_smallintervals}, which is due to the low output variability of the model. We show the output variability in the supplementary material, Fig. S3b. Looking at later times in \cref{fig:Sens_smallintervals}, we observe why the switches in the ranking again appear at slightly different times, as the sensitivity metrics for $\bar{\Theta}$, $\bar{\tau}$ and $\tau_r$ yield very similar values towards the end of the experiment.

Analyzing the parameter interactions with Sobol, we only find interactions between $\bar{\Theta}$ and $\bar{\tau}$. Therefore, we only show this index in \cref{fig:Sens_smallintervals}. For $t < 3$, this index reveals that the sensitivity of $\bar{\Theta}$ is mainly composed of the second-order interaction with $\bar{\tau}$. For $t > 3$, the interaction decreases until the second-order index reaches a value close to zero, while the first-order index yields larger values. A comparison of the first-order and total Sobol indices indicates that the total sensitivity is mainly composed of the first-order influence of the parameters and less due to interactions.
The analysis of $\frac{\hat{\sigma}_i}{\hat{\mu}_i^*}$ in \cref{fig:Sens_relationship_small} shows that the interactions between hyperparameters are between nonlinear and almost monotonous for $\bar{\Theta}$ and $\Theta_r$ and between almost monotonous and monotonous for $\bar{\tau}$ and $\tau_r$. 
Considering the first eigenvector of ASM presents the same results as for the large intervals experiment. The two means and the two deviations are positively correlated, while the means and deviations are negatively correlated. For the exact values of the eigenvector entries, we refer to the supplementary material, Fig. S8b.

\added{From an application perspective, the results of the sensitivity analysis indicate that, depending on the time horizon available for the mixing process, different hyperparameters are relevant for generating optimal mixing. Therefore, when considering a short time horizon (e.g., $t_\text{max} = 5$) the choice of $\Theta_r$ and $\tau_r$ does not affect the expected mixing enhancement and can therefore be disregarded. As mentioned for the non-randomized configurations, controlling $\bar{\Theta}$ and $\Theta_r$ might be infeasible for some applications, where the locations of source and sink can not be chosen arbitrarily; therefore, the modification of $\bar{\tau}$ and $\tau_r$ seems much more reasonable. Given that at least one of the two parameters is classified as sensitive throughout the whole experimental time, we can still control our results using these two parameters.}

\begin{figure*}[tb!]
	\centering
	\begin{subfigure}[t]{1cm}
		% in gimp the width of this figure is 1.058 cm
		\includegraphics[width=1cm]{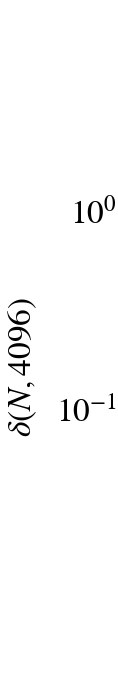}
	\end{subfigure}
	\begin{subfigure}[t]{3.5cm}
		\includegraphics[width=3.5cm]{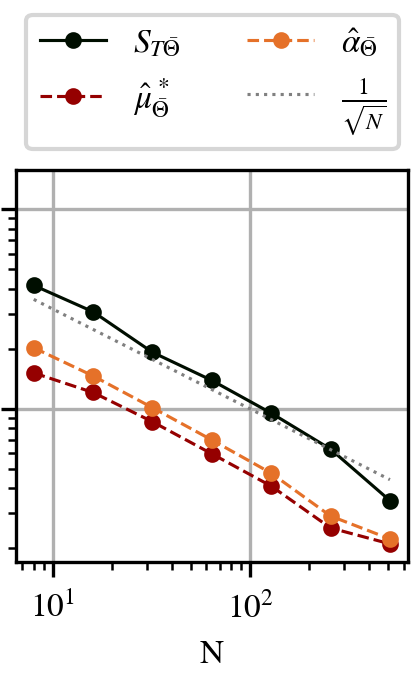}
		\caption{Convergence for $\bar{\Theta}$}
	\end{subfigure}
	\hfill
	\begin{subfigure}[t]{3.5cm}
		\includegraphics[width=3.5cm]{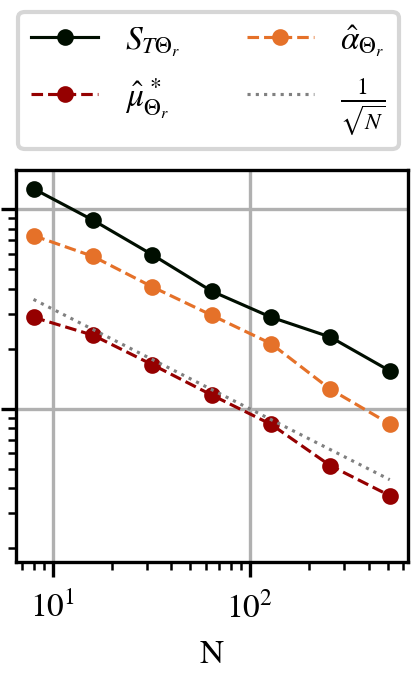}
		\caption{Convergence for $\Theta_r$}
	\end{subfigure}
	\hfill
	\begin{subfigure}[t]{3.5cm}
		\includegraphics[width=3.5cm]{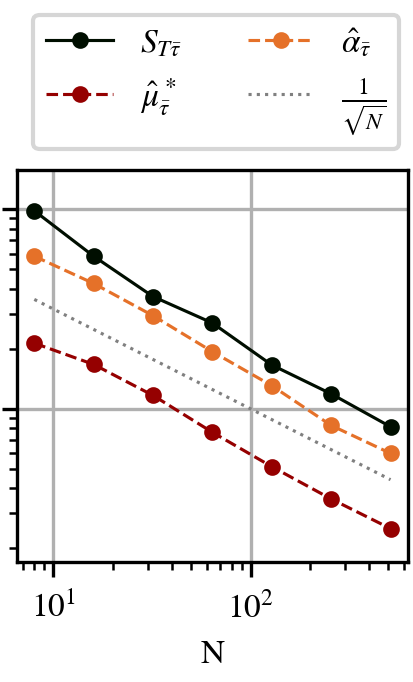}
		\caption{Convergence for $\bar{\tau}$}
	\end{subfigure}
	\hfill
	\begin{subfigure}[t]{3.5cm}
		\includegraphics[width=3.5cm]{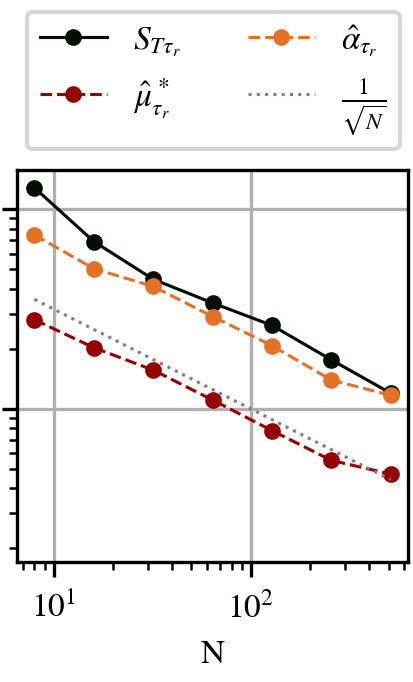}
		\caption{Convergence for $\tau_r$}
	\end{subfigure}
	\hspace{0.72cm}

	\caption{Mean relative error estimate $\delta(N, 4096)$ over batch size $N$ for the different sensitivity scores for the randomized RPM flow with $\bar{\Theta} \in [0, \pi]$, $\Theta_r \in [0, 0.2\pi]$, $\bar{\tau} \in [0.1, 1]$ and $\tau_r \in [0, 0.2]$. }
	\label{fig:Sens_convergence}
\end{figure*}

We analyze the convergence to better understand the quality of the sensitivity metrics. We perform this analysis on the large intervals experiment of the randomized RPM flow. To do so, we compute the metrics for mini-batches of size $N \leq 512$ and compare the results to those with $N = 4096$. A relative error estimate $\delta(N, 4096)$ approximates the mean relative error between the mini-batch results with batch size $N$ and the result with $4096$ samples. \Cref{fig:Sens_convergence} shows the relative error estimate for different mini-batch sizes. All sensitivity metrics converge with order $\frac{1}{2}$, as expected. The same rate was also found in \citep{constantine_2017} for the activity scores and total Sobol indices. Be aware that using the same value of $N$ leads to $N(n+2)$ samples for the Sobol indices and $N(n+1)$ samples for the Morris method and ASM. Although using more samples, the error of the Sobol indices is larger than the error of the two other methods. We find the highest accuracy for Morris, which needs at most $\frac{1}{4}$ of the samples to yield the same relative error as the total Sobol index.
The results highlight that the Morris method is most efficient for this higher-dimensional problem, where the analysis is limited by the cost of computing the required Monte-Carlo samples. \added{This finding is also supported in the literature \citep{pianosi_2016, yang_2011}. }Additionally, we can reuse the samples to compute the activity scores of the ASM, giving us more insights into the model without additional computational costs. For a more detailed analysis, we show the convergence also for both randomized RPM flow experiments, including also the first- and second-order Sobol indices, in the supplementary material, Figs. S5 and S6.

\begin{figure}[tb!]
    % prior = 4.3cm
	\centering
	\begin{subfigure}[t]{3.75cm}
		\includegraphics[width=3.75cm]{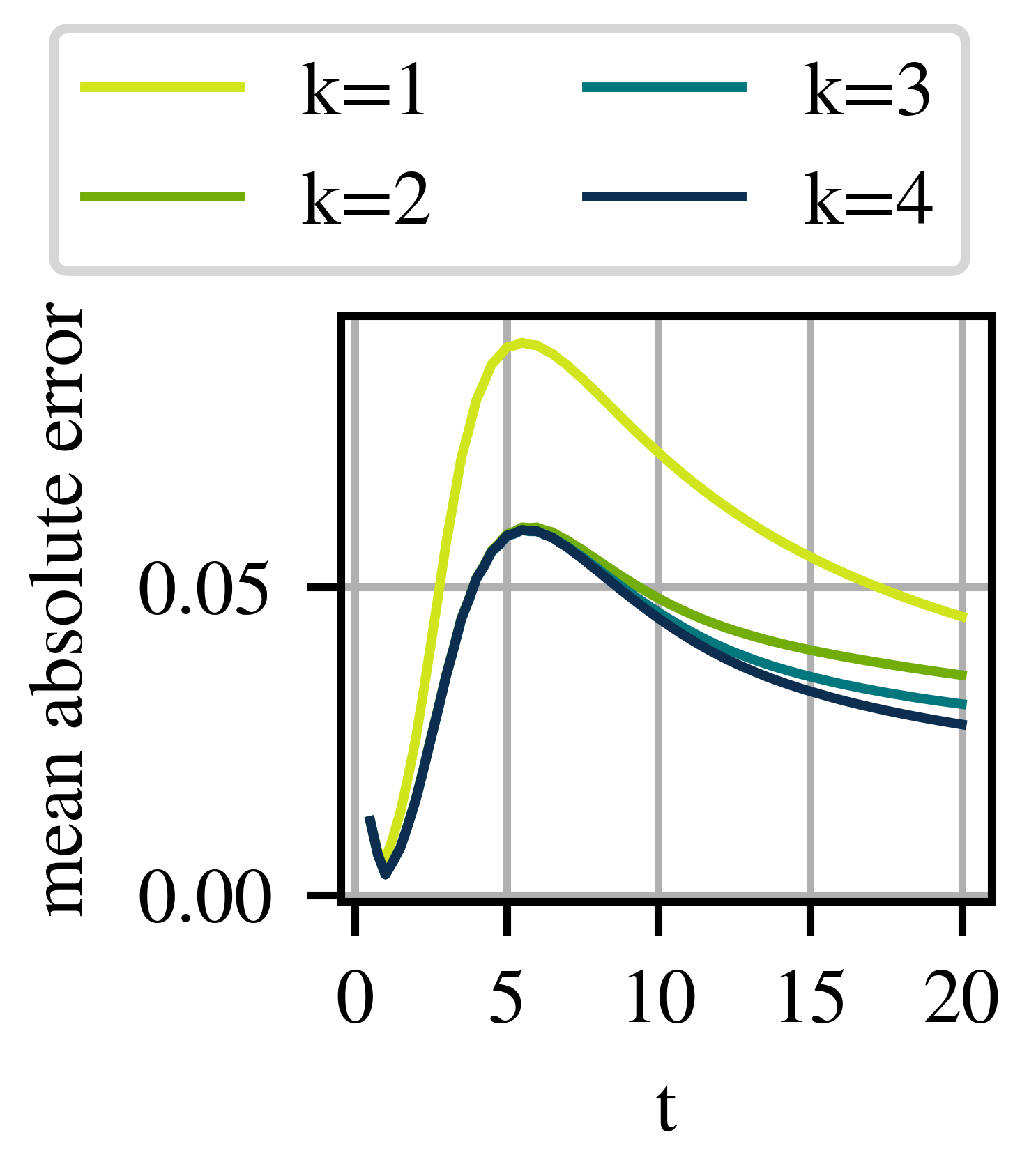}
		\caption{Large intervals experiment}
		\label{fig:Sens_surrugate_MAE_big}
	\end{subfigure}
	\begin{subfigure}[t]{3.75cm}
		\includegraphics[width=3.75cm]{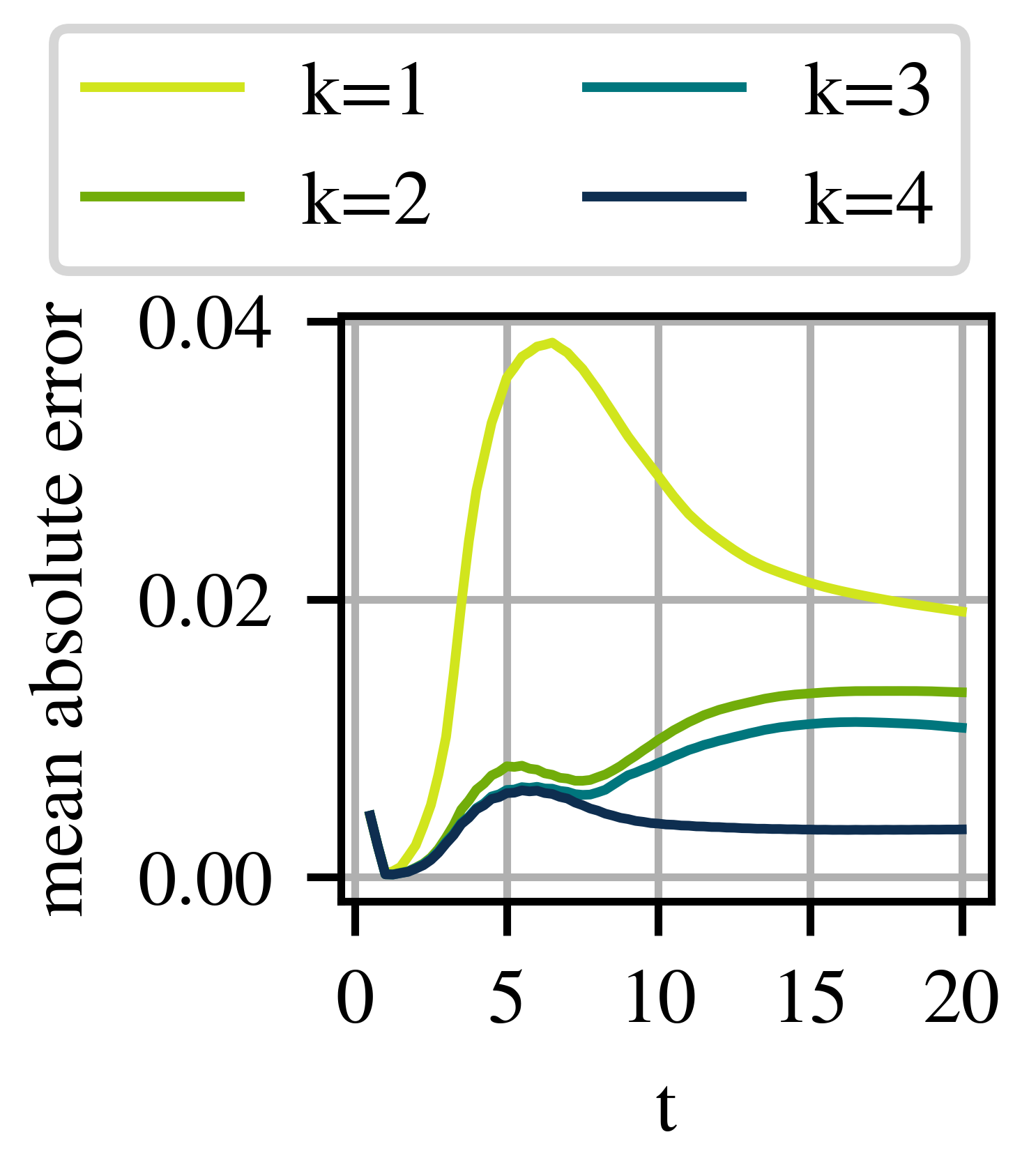}
		\caption{Small intervals experiment}
		\label{fig:Sens_surrugate_MAE_small}
	\end{subfigure}

	\caption{Mean absolute error of the surrogate model with input dimension $k \in \{1, 2, 3, 4\}$ created by ASM over time $t$ for the randomized RPM flow. The surrogate model is a polynomial with maximal order $5$.}
	\label{fig:Sens_surrugate_MAE}
\end{figure}

As described in \cref{sec:Meth:ASM}, ASM enables us to construct a surrogate model. We showcase this process using a polynomial fit of order $5$ and input dimensions $k \in \{1, 2, 3, 4\}$. The surrogate is constructed solely using the Morris samples. We use the Saltelli samples to compute the mean absolute error of the surrogate in \cref{fig:Sens_surrugate_MAE}. The results show that two input dimensions are sufficient to approximate the model output well. We notice that the mean absolute error of the surrogate is maximal for times with a high output variability in Fig. S3. One can increase the accuracy further by fine-tuning the polynomial order of the surrogate. 

\subsection{Sensitivity analysis on high dimensional problems}\label{sec:Results:quad}
 
In our next experiment, we analyze the sensitivity of the quadrupole model from \cref{sec:Meth:quad}. This model is significantly higher-dimensional than the previous flow systems, having $16$ hyperparameters. Due to the higher computational cost of the Monte-Carlo sampling, we restrict the analysis and only compute the Morris sensitivity and ASM activity scores. Our results on the RPM flow show that these sensitivity metrics yield consistent results with the Sobol indices while requiring fewer samples to converge. For our analysis, the choice of $N = 10$ is sufficient for the screening of the sensitive parameters. A detailed analysis of the results of the sensitivity analysis with Morris is given in \citep{ziliotto_2025}; we therefore concentrate more on the comparison of the Morris and the ASM method. Further, due to the complexity of the model and the high dimensional input, a surrogate model as presented in \cref{sec:Meth:ASM} cannot be applied without additional sampling.

\begin{figure}[tb!]
	\centering
	\begin{subfigure}[t]{7.5cm}
		\includegraphics[width=7.5cm]{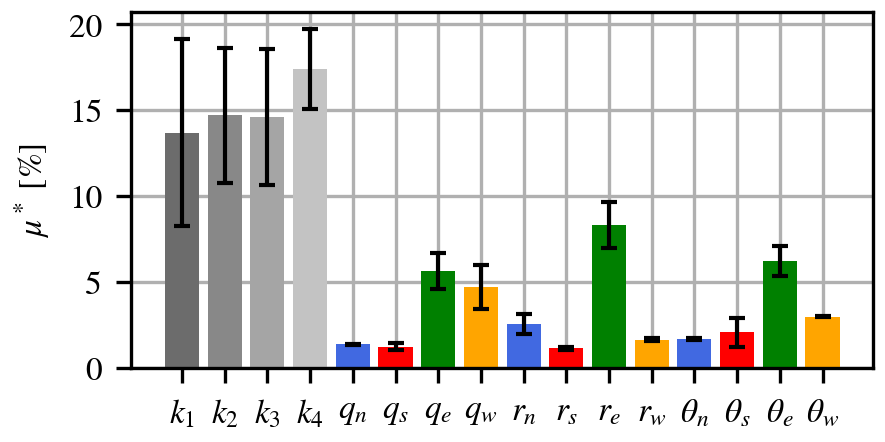}
	\end{subfigure}

	\caption{Morris sensitivity scores $\mu^*$ after the last stress period. The height of the bar represents the sensitivity of each hyperparameter, respectively, while the black line highlights its $95$ \% confidence interval.}
	\label{fig:Quad_sens_conf}
\end{figure}

We start our analysis by showing the results for the sensitivity analysis of Morris after the last stress period using $N = 10$ trajectories in \cref{fig:Quad_sens_conf}. Despite the low number of samples, we can draw some interesting conclusions from these results. While it is generally enough for practical applications related to model calibration and system design to perform a sensitivity screening, i.e., identification of the group of most sensitive parameters, we can observe that the ranking is not fully converged yet, and small variations could be expected. Despite the exact values can vary significantly within the respective confidence intervals, we should consider the interaction among the parameters and hence an overlap of the confidence intervals of two parameters, does not necessarily mean that the two parameters could switch their ranking. Conclusions based on the exact values of the sensitivity index on the contrary need to be considered with special care. Such analysis is often not needed in practical applications focusing on inverse modelling and optimization, however it may be valuable in order to get insights about how dominant the sensitive hyperparameters are compared to the less sensitive ones. 
In the work of \citep{sarrazin_2016}, the authors use the size of the $95$ \% confidence interval as a convergence criterion for the sensitivity analysis. They state that the method is converged if the largest confidence interval has a size below $5$. Knowing that Morris converges with a rate of $\frac{1}{2}$, we expect to get all confidence intervals below size five with $N = 48$ sample trajectories. So, the convergence for the sensitivity indices comes at a significant computational cost, that may not be justified by the purpose of the study.

\begin{figure*}[tb!]
	\centering
	\begin{subfigure}[t]{7.95cm}
		\includegraphics[width=7.95cm]{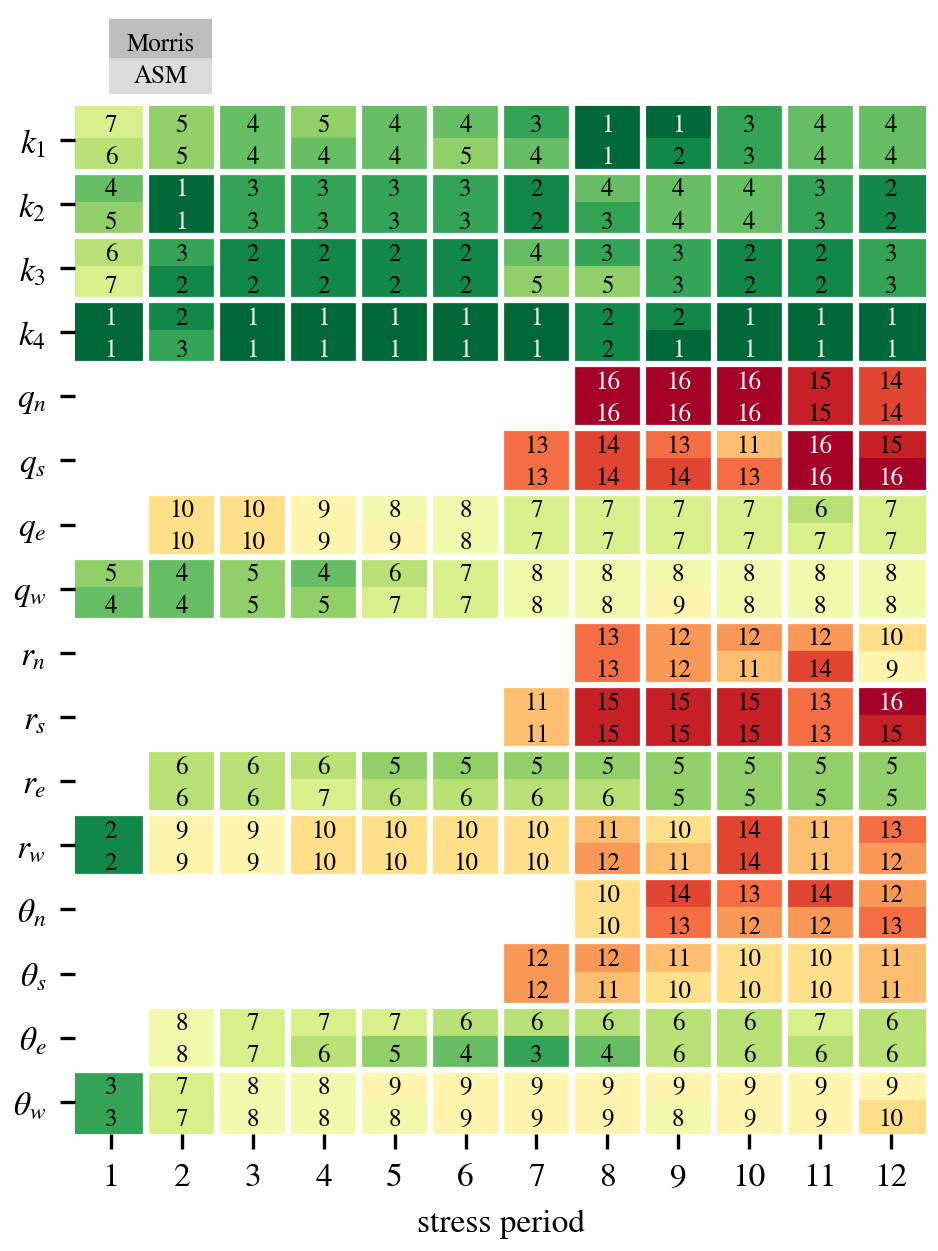}
		\caption{Sensitivity ranking of the Quadrupole model.}
		\label{fig:Quad_ranking}
	\end{subfigure}
	\hfill
	\begin{subfigure}[t]{7.95cm}
		\includegraphics[width=7.95cm]{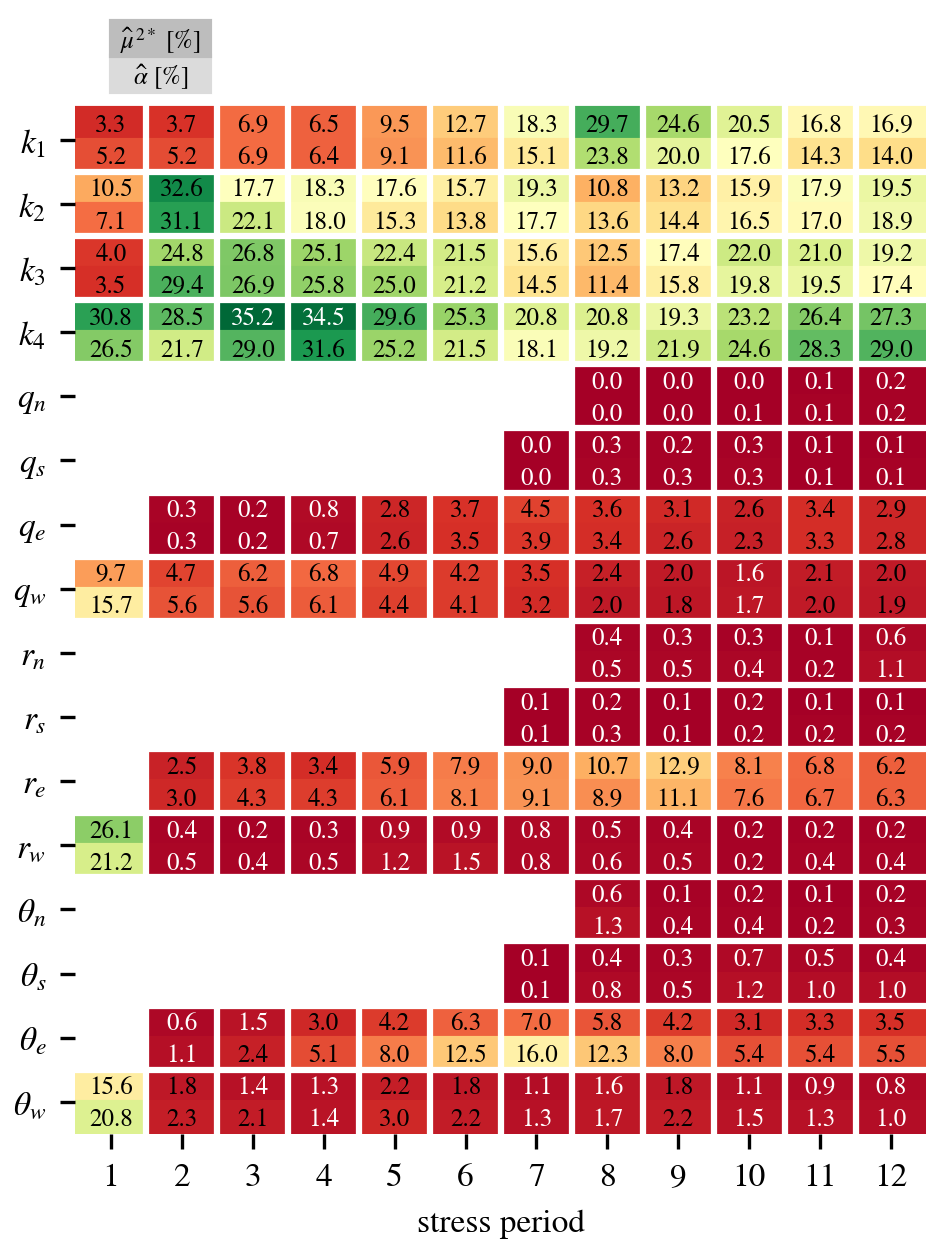}
		\caption{Values of the sensitivity metrics of the Quadrupole model.}
		\label{fig:Quad_values}
	\end{subfigure}

	\caption{Results of the sensitivity analysis for the quadrupole model over the stress periods (each period lasts $6.25$ days). The upper half of each square represents $\hat{\mu}^{2*}$ from the Morris method, and the lower half is the ASM's activity score $\hat{\alpha}$. \Cref{fig:Quad_ranking} shows the sensitivity ranking and \cref{fig:Quad_values} shows the values of the sensitivity metrics. The lower the ranking and the higher the value, the more sensitive the hyperparameter is. }
	\label{fig:Quad_sensitivity}
\end{figure*}

We show the complete results of the sensitivity ranking of Morris and ASM in \cref{fig:Quad_ranking}. The cells with no values belong to wells activated at a later stress period, like the north well that starts pumping in stress period seven. The ranking of the two methods is very similar. We observe the same trends, like the high sensitivity of the conductivity values $k_1$, $k_2$, $k_3$, and $k_4$, followed by the parameters related to the east well location $r_e$ and $\theta_e$. 
We only find minor ranking differences between the two methods. The maximal deviation is three for the hyperparameter $\theta_e$ and stress period $7$. Other than that, we have seven deviations of two and $41$ of one. The remaining $101$ instances match exactly.

Next, we look at the exact values of the sensitivity metrics in \cref{fig:Quad_values}.
As discussed previously, the values require more samples to reach convergence, and hence the results need to be observed more carefully. 
Additionally, we highlight that the two methods are unlikely to obtain the same values, as they use the same data but different formulas to compute the sensitivity metrics. 
Looking at \cref{fig:Quad_values}, we find that both methods (after using the transformation in the Morris scores in \eqref{eq:mu_i2}) yield very similar sensitivity scores. \replaced{Overall}{All in all}, the highest sensitivity is concentrated on the four conductivity values, followed by the hyperparameters corresponding to the east well and the discharge of the west well. The remaining hyperparameters are non-sensitive.

The comparison between Morris and ASM is useful to confirm the results presented by \citep{ziliotto_2025}. In fact, it is good practice to compare different approaches in sensitivity analysis, since there can be a difference in the conclusion. From a physical point of view, we conclude that in the quadrupole experiment, the values of the hydraulic conductivities are the most important parameters, followed by the operational parameters of the wells activated at the beginning of the process. 
While the efficiency depends on both the design parameters of the system (configuration of the pumping wells) and the physical parameters of the aquifer (hydraulic conductivity), we can only vary the design parameters, e.g., pumping rates and well locations. The control of the hydraulic conductivity on the system is expected and was also identified in previous studies \citep{bertran_2023}. On the contrary, the dependence on the design parameters is very relevant, although they are not the most sensitive ones, since this information can be used in the optimization of the design of the system.

\section{Conclusion and Outlook}\label{sec:Conclusion}

In our work, we use three different sensitivity analysis methods, Sobol indices, Morris scores, and activity scores of the active subspace method (ASM), to analyze the mixing enhancement of two chaotic flow fields depending on their design parameters. We use a time-varying approach to capture the temporal evolution of the sensitivity.
To do so, we introduce a scaling to normalize the sensitivity metrics over time, making all sensitivity metrics quantitative. For the comparison of the values of the sensitivity metrics, we consider their units and introduce a transformation to make the units of the Morris scores the same as those of the other two metrics. 
Our analysis supports the decision-making on the best method for sensitivity analysis, depending on the input dimension, the computational budget, and the goals of the analysis. We show that the sensitivities are largely consistent over the three methods while having different computational costs and giving different insights into interactions between hyperparameters. 
% interactions
With Sobol indices, we can directly compute second-order interactions at the cost of $N \cdot n$ additional samples.
The Morris method can indicate the existence of interactions by looking at the ratio $\frac{\hat{\sigma}}{\hat{\mu}^*}$, and the first eigenvector of ASM gives insights into the important linear directions of the input hyperparameter space. 
% computational costs
Although the second-order Sobol indices come with increased computational costs, the effort might be reasonable, especially for low-dimensional applications, as the results give a precise and easy-to-understand estimation of second-order interactions.
In contrast, when using high-dimensional or computationally expensive methods, we demonstrate the strength of the Morris method and, subsequently, of the ASM activity scores. Using our low-dimensional experiment, we demonstrate that Morris yields the same relative error as the total Sobol index with only $\frac{1}{4}$ of the model evaluations. Additionally, as we reuse the Morris samples to compute the ASM activity scores, we can analyze the model's sensitivity with two approaches at the cost of one approach. Beyond sensitivity analysis, the ASM method offers the possibility to develop a surrogate model that approximates the original model's behavior at a low cost. These surrogates can be used for a broad range of tasks, such as uncertainty quantification, optimization, or real-time decision-making, causing ASM to be a good starting tool for deeper analysis.

The results of our sensitivity analysis show that mixing in chaotic systems relies sensitively on their flow parametrization. \added{The sensitivity rankings can be used to optimize the most influential design parameters of a chaotic flow field to enhance mixing. Additionally, when implementing the flow field, the more influential hyperparameters must be implemented precisely, as low aberration can lead to significant differences in the model response. }The time-varying analysis reveals that the sensitivity ranking changes over time. For applications that generate effective chaotic mixing, this means that, depending on the duration of the process, different hyperparameters need special consideration. For the randomized RPM flow, for example, the sensitivity of the deviation parameters $\Theta_r$ and $\tau_r$ increases over time. The longer the mixing lasts, the more important the deviation parameters are to the final mixing enhancement\added{, while for short times, they play a minor role}. Additionally, we find that the sensitivity also depends on the hyperparameter intervals. To generate meaningful results, we recommend using hyperparameter intervals of comparable sizes.
\added{We emphasize that the mixing enhancement can also depend on physical parameters that can not be controlled to enhance mixing. If a physical parameter is highly sensitive, however, it may be beneficial to estimate its value through measurements before optimizing the design parameters to enhance mixing.} 

Especially for the quadrupole flow, the computational cost of the flow and transport model limits our analysis in terms of model realizations. In general, the sensitivity values require more samples for convergence than a ranking or screening of the hyperparameters \citep{sarrazin_2016}. The authors of \citep{campolongo_1997} show that $10$ trajectories are insufficient for their $35$-dimensional model. Other authors use more trajectories like \citep{richieri_2024} with $500$ trajectories on a $ 17$-dimensional model. We reinforce the hypothesis that the optimal number of trajectories is model-specific and hard to predict.
Regarding the ASM activity scores, we extend the original method to enable the usage of Morris samples. Consequently, the theorems presented in \citep{constantine_2017} do not hold anymore. However, the connection between our activity scores and the Morris methods motivates that we yield valid results. If $m = n$, like in our experiments, the Morris and ASM activity scores are based on the same data and use a similar formula, which means that the results of these two methods will always be correlated.

For future work, we envision comparing our adapted version of the activity scores to the original activity scores. 
As a robust sensitivity analysis often requires multiple methods, a good agreement of our adapted activity scores and the original version would support our claim that we can get this sensitivity method ``for free" when using Morris. 
Further, our analysis only features one method that can explicitly indicate second-order interaction. Including another method to analyze these effects and comparing the results to our second-order Sobol indices would be interesting. The approach in \citep{campolongo_1999,cropp_2002} is an extension to Morris that requires a total of $N \cdot(\frac{n^2+n}{2} + 1)$ model evaluations to compute these interactions (including $\hat{\mu}^*$ and $\hat{\sigma}$). As this number grows quadratically with the input dimension $n$, it becomes infeasible for most high-dimensional models.
We would also like to investigate the influence of randomization on the mixing enhancement of the RPM flow. Noticing that our non-randomized RPM flow is equivalent to our randomized RPM flow with $\Theta_r = 0$ and $\tau_r = 0$, we found the sensitivity of the mean parameters $\bar{\Theta}$ and $\bar{\tau}$ to be comparable to $\Theta$ and $\tau$ in the non-randomized case. At late times, the sensitivity of the deviation parameters $\Theta_r$ and $\tau_r$ increases. The higher sensitivity of the deviation parameters indicates that the randomization influences the mixing enhancement and could be an important factor in optimizing chaotic flow fields to enhance mixing.
The mixing efficiency depends on design parameters, like well locations and pumping rates. Hydrological parameters, like the hydraulic conductivity, are fixed and are, in general, subject to a large uncertainty. By applying risk-aware optimization, it may be possible to optimize the design parameters by considering the uncertainty of the hydrological parameters.

\backmatter

\section*{Statements \& Declarations}

\bmhead{Funding}

This research is a result of the ChaosAD project (Chaotic ADvection and Mixing Enhancement in Porous Media: The Quest for Experimental Evidence), which is supported by the Deutsche Forschungsgemeinschaft (DFG) under the grant numbers CH 981/8-1 and SI 2853/2-1. Financial support for Francesca Ziliotto was provided by TUM International Graduate School of Science and Engineering (IGSSE), within the INE2 project (Innovative Engineering Injection Extraction systems for in-situ groundwater remediation. From model- and laboratory-based evidence to stakeholder involvement). The authors gratefully acknowledge the scientific support and HPC resources provided by the Erlangen National High Performance Computing Center (NHR@FAU) of the Friedrich-Alexander-Universität Erlangen-Nürnberg (FAU). The hardware is funded by the German Research Foundation (DFG).

\bmhead{Competing Interests}

The authors have no relevant financial or non-financial interests to disclose.

\bmhead{Author Contributions}

Conceptualization was performed by Carla Feistner and Gabriele Chiogna. Carla Feistner also worked on the formal analysis and methodology. Software, investigation, and visualization of the RPM flow were performed by Carla Feistner. For the Quadrupole flow, the implementation of the flow and transport model and the sensitivity analysis with Morris was done by Francesca Ziliotto, and the analysis with the active subspace method was added by Carla Feistner. The supervision was done by Gabriele Chiogna and Barbara Wohlmuth. 
The original draft was written by Carla Feistner, and all authors contributed to the review and editing of the manuscript. All authors read and approved the final manuscript. The project administration and funding acquisition were done by Gabriele Chiogna. We thank Prof. David Mays for the fruitful discussion on the manuscript.

\bmhead{Supplementary information}

We created a PDF file that contains supplementary figures and tables that support the results presented in this manuscript. The file contains a detailed table describing the parametrization of the quadrupole flow, a figure motivating the choice of our quantity of interest to describe mixing, an analysis of the output variability of both chaotic flow fields, all convergence figures for the different experiments of the randomized RPM flow, and the eigenvectors of the covariance matrices for all experiments.


\bibliography{references_bibtex}% common bib file

\begin{appendices}



\section{Sobol indices}\label{app:Meth:Sobol}

The Sobol index is a variance-based approach to assess the sensitivity of the flow parameters \citep{sobol_2001,saltelli_2010,sobol_1993}. The theory is based on the ANOVA representation \citep{archer_1997,jansen_1999,sobol_2001,sobol_2003,sobol_2005} of $f$
\begin{align}\label{app:eq:ANOVA}
	\begin{split}
 		f(x) = f_0 &+ \sum_i f_i(x_i) + \sum_{i<j} f_{i,j}(x_i, x_j) \\&+ ... + f_{1,2,...,n}(x_1, x_2, ..., x_n).
	\end{split}
\end{align}
The $f_{i_1, ..., i_s}$ terms in \eqref{app:eq:ANOVA} are orthogonal and can be explicitly computed as integrals of $f(x)$ \citep{sobol_2001, homma_1996}. Thereby,
\begin{align}\label{app:eq:f_i}
	\begin{array}{lcll}
 		f_0 &=& \E[f(x)] & \text{grand mean} \\
 		f_k(x_i) &=& \E[f(x)|x_i] - f_0 & \text{$i$'th main effect} \\
 		f_{i,j}(x_i, x_j) &=& \E[f(x)|x_i, x_j] & \text{$ij$'th interaction} \\
 		~&~& - \E[f(x)|x_i] & ~ \\
 		~&~& - \E[f(x)|x_j] - f_0 & ~
	\end{array}
\end{align}
and so on \citep{archer_1997, saltelli_2002}.
The variance of the model $f$ is given by 
\begin{align*}
 V &= \int f(x)^2 \, dx - f_0^2\\
	&= \sum_{s=1}^n \sum_{i_1<...<i_s} V_{i_1, ..., i_s}
\end{align*}
where 
\begin{align}\label{app:eq:V_i_continuous}
 V_{i_1, ..., i_s} := \int f_{i_1, ..., i_s}^2 dx_{i_1} ... dx_{i_s}
\end{align}
represents the shared contribution of the parameters $x_{i_1}, ..., x_{i_s}$ to the variance of the model $V$ \citep{saltelli_2010,sobol_1993,homma_1996}. Using this, we can define the Sobol indices as
\begin{align}\label{app:eq:S_i}
 S_{i_1, ..., i_s} := \frac{V_{i_1, ..., i_s}}{V}
\end{align}
where $s$ is called the order/dimension of the index \citep{sobol_2001, sobol_1993,homma_1996}. It directly follows 
\begin{align}\label{app:eq:Sobol_sum}
	\sum_i S_i + \sum_i \sum_{j>i} S_{i,j} + ... + S_{1, 2, ..., k} = 1.
\end{align}
%Using \eqref{app:eq:f_i} and \eqref{app:eq:V_i_continuous} we find for the first-order Sobol indices \citep{saltelli_2002}
%\begin{align*}
% S_i = \frac{\var(\E[f(x)|x_i])}{V},
%\end{align*}
%which represents the output variance that is due to the input factor $x_i$ \citep{pianosi_2016}. 

\end{appendices}

\pagebreak

\clearpage
\onecolumn
\begin{center}
  {\LARGE Supplementary material}
\end{center}

\renewcommand{\thesection}{S\arabic{section}}
% Prefix supplementary floats with "S" and reset counters
\renewcommand{\thefigure}{S\arabic{figure}}
\renewcommand{\thetable}{S\arabic{table}}
\setcounter{table}{0}
\setcounter{figure}{0}
\setcounter{section}{0}

\section{Introduction}

This supplementary material contains additional information on the sensitivity analysis presented in the paper. We start by presenting the parameters of the quadrupole model in \cref{sec:parameters_quad} and motivate the usage of the peak concentration as quantity of interest for the quadrupole flow in \cref{sec:quantity_of_interest}. We then show the output variability of the RPM flow and the quadrupole model in \cref{sec:output_variability}. Afterward, we give more detailed plots of our convergence analysis in \cref{sec:convergence}. Finally, we show the first eigenvector of the covariance matrix for the RPM flow \cref{sec:first_eigenvec_RPM} and the quadrupole flow in \cref{sec:first_eigenvec_quad}, which we use to analyze the correlation of the hyperparameters.

\section{Parameters for the Quadrupole model}\label{sec:parameters_quad}

To complete the description of the quadrupole model, initially introduced in \citep{mays_2012}, we give the model geometry in \cref{tab:Quad_model_geometry} and the details of the EIE sequence in \cref{tab:Quad_EIE_sequence}. The model geometry is the same used in \citep{ziliotto_2025}.

\begin{table*}[tb!]
	\renewcommand{\arraystretch}{1.25}
	\centering
	\begin{subtable}[t]{\textwidth}
		\centering
		\begin{tabular}{|l|l|}
			\hline
			\rowcolor[gray]{0.9}\multicolumn{2}{|c|}{Geometry of the model} \\
			\hline
			Aquifer size & $201 \times 201 \times 10$ $\text{m}^3$ \\
			Discretization & $\Delta x = \Delta y = 0.5$ m \\
			Aquifer thickness $b$ & $10$ m \\
			Boundary conditions & \begin{tabular}{@{}l@{}} west \& east: constant head,\\[-0.2cm] north \& south: no flow \end{tabular} \\
			Distance from the origin to the ideal wells $L$ & $25$ m \\
			\hline
			\rowcolor[gray]{0.9}\multicolumn{2}{|c|}{Random hydraulic conductivity field} \\
			\hline
			mean $\mu_{\log(K)}$ & $0.7$ \\
			variance $\sigma_{\log(K)}$ & $0.25$ \\
			correlation length $\lambda$ & $10$ m \\
			\hline
			\rowcolor[gray]{0.9}\multicolumn{2}{|c|}{Flow model} \\
			\hline
			Porosity $\epsilon$ & $0.25$ \\
			Storage coefficient $S$ & $10^{-5}$ \\
			\hline 
			\rowcolor[gray]{0.9}\multicolumn{2}{|c|}{EIE system} \\
			\hline
			Total duration of the EIE & $75$ days \\
			Number of EIE steps & $12$ \\
			Duration of EIE step $T$ & $6.25$ days \\
			Injection rate $Q$ of the wells & $\Lambda^2 = \frac{QT}{\pi \epsilon b L^2}$, $\Lambda^2$ from \cref{tab:Quad_EIE_sequence} \\
			\hline 
			\rowcolor[gray]{0.9}\multicolumn{2}{|c|}{Transport model} \\
			\hline
			Size of contamination & $11 \times 11$ $\text{m}^2$ \\
			Initial concentration & $1$ $\frac{\text{kg}}{\text{m}^3}$ \\
			Isotropic hydrodynamic dispersion coefficient $D$ & $4 \cdot 10^{-2}$ $\frac{\text{m}^2}{\text{day}}$ \\
			\hline
		\end{tabular}
		\caption{Summary of the model geometry and flow and transport input parameters from \citep{ziliotto_2025}.}
		\label{tab:Quad_model_geometry}

	\end{subtable}

	\vspace{1ex}
	\begin{subtable}[t]{\textwidth}
		\renewcommand{\arraystretch}{1.4}
		\centering
		\begin{tabular}{|c||c|c|c|c|c|c|c|c|c|c|c|c|}
			\hline
			Step & $1$ & $2$ & $3$ & $4$ & $5$ & $6$ & $7$ & $8$ & $9$ & $10$ & $11$ & $12$ \\
			\hline
			Well & W & E & W & E & W & E & S & N & S & N & S & N \\
			\hline 
			$\Lambda^2$ & $\frac{3.5}{\pi}$ & $\frac{3.5}{\pi}$ & $-\frac{1}{\pi}$ & $-\frac{3}{\pi}$ & $-\frac{1.6}{\pi}$ & $-\frac{1.4}{\pi}$ & $\frac{3.5}{\pi}$ & $\frac{3.5}{\pi}$ & $-\frac{1}{\pi}$ & $-\frac{3}{\pi}$ & $-\frac{1.6}{\pi}$ & $-\frac{1.4}{\pi}$ \\
			\hline
		\end{tabular}
		\caption{Details of the EIE sequence or the quadrupole flow. The first row gives the step number, the second row indicates the well that is operated at this step (W: west, E: east, S: south, N: north), and the last row shows the value of $\Lambda^2$ that controls the injection rate $Q$. $\Lambda^2 > 0$ indicates injection \citep{mays_2012}.}
		\label{tab:Quad_EIE_sequence}
	\end{subtable}
	
	\caption{Summary of the model geometry, flow and transport input parameters and details about the EIE sequence.}
	\label{tab:Quad_model}
\end{table*}

\section{Quantity of interest to analyze mixing}\label{sec:quantity_of_interest}

\begin{figure}[b!]
	\centering
	\begin{subfigure}[t]{13.1cm}
		\includegraphics[width=13.1cm]{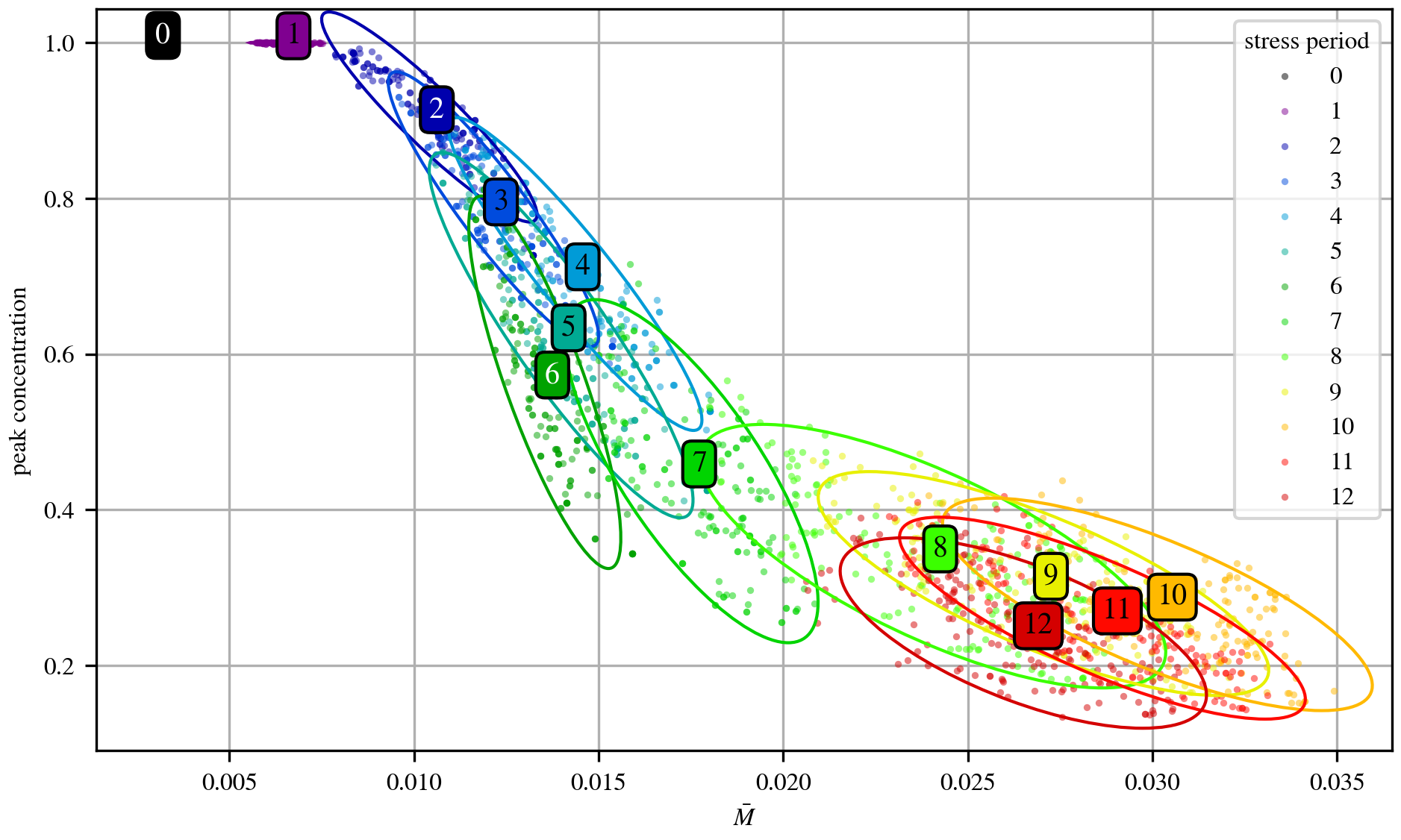}
	\end{subfigure}

	\caption{Relation between $\M$ and the peak concentration for all $170$ samples considering all $12$ stress periods. The colored ellipses highlight the preferential region of the datapoints for each stress period that was computed using the covariance matrix of the data.}
	\label{fig:peak_over_M}
\end{figure}

For the quadrupole flow, we claim in our paper that $\M$ is not a good quantity of interest to analyze mixing, as we experience solute extraction due to our modification to the ideal flow introduced in \citep{mays_2012}. Considering all $10$ sample paths that were used in our analysis, we show the relation between $\M$ and the peak concentration in \cref{fig:peak_over_M}. We observe that while the peak concentration decreases monotonically, we observe an occasional decrease of $\M$ due to mass extraction.

\section{Output variability}\label{sec:output_variability}

\begin{figure}[tb!]
	\centering
	\begin{subfigure}[t]{0.864cm}
		\includegraphics[width=0.864cm]{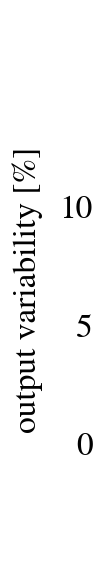}
	\end{subfigure}
	\begin{subfigure}[t]{4.536cm}
		\includegraphics[width=4.536cm]{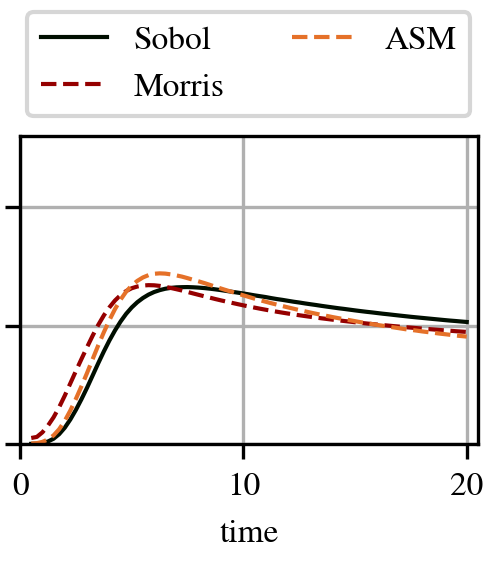}
		\caption{Large intervals experiment}
		\label{fig:sSens_output_variability_big}
	\end{subfigure}
	\begin{subfigure}[t]{4.536cm}
		\includegraphics[width=4.536cm]{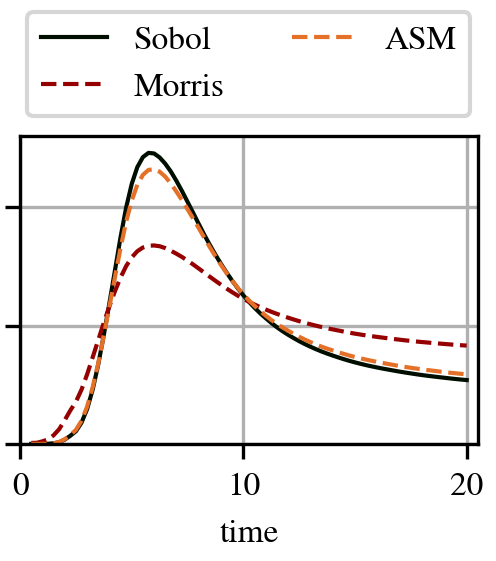}
		\caption{Small intervals experiment}
		\label{fig:sSens_output_variability_small}
	\end{subfigure}

	\caption{Comparison of the output variability measured using the three sensitivity metrices for the non-randomized RPM flow with $\Theta \in [0, \pi]$ and $\tau \in [0, 1]$ in \cref{fig:sSens_output_variability_big} and $\Theta \in [0.4\pi, 0.6\pi]$ and $\tau \in [0.4, 0.6]$ in \cref{fig:sSens_output_variability_small} over time. The output variability is given as a percentage of the total variability integrated over all times.}
	\label{fig:sSens_output_variability}
\end{figure}

\begin{figure}[tb!]
	\centering
	\begin{subfigure}[t]{0.864cm}
		\includegraphics[width=0.864cm]{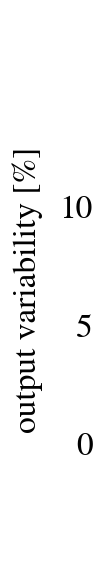}
	\end{subfigure}
	\begin{subfigure}[t]{4.538cm}
		\includegraphics[width=4.538cm]{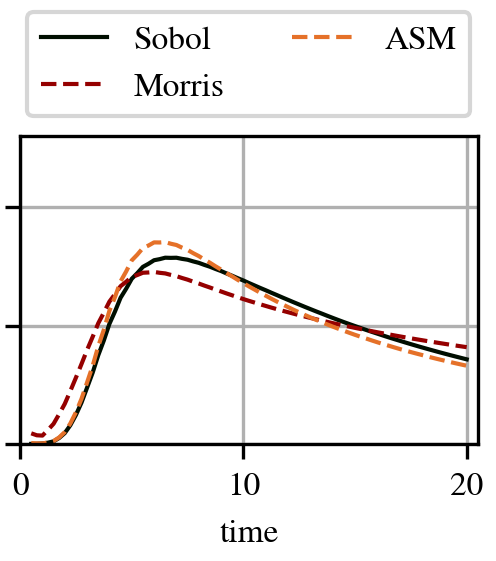}
		\caption{Large intervals experiment}
		\label{fig:Sens_output_variability_big}
	\end{subfigure}
	\begin{subfigure}[t]{4.538cm}
		\includegraphics[width=4.538cm]{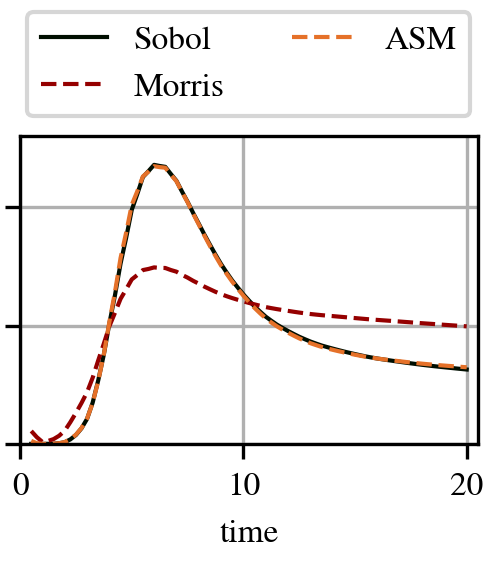}
		\caption{Small intervals experiment}
		\label{fig:Sens_output_variability_small}
	\end{subfigure}

	\caption{Comparison of the output variability measured using the three sensitivity metrics for the randomized RPM flow with $\bar{\Theta} \in [0, \pi]$, $\Theta_r \in [0, 0.2\pi]$, $\bar{\tau} \in [0.1, 1]$ and $\tau_r \in [0, 0.2]$ in \cref{fig:Sens_output_variability_big} and $\bar{\Theta} \in [0.45, 0.55\pi]$, $\Theta_r \in [0, 0.1\pi]$, $\bar{\tau} \in [0.45, 0.55]$ and $\tau_r \in [0, 0.1]$ in \cref{fig:Sens_output_variability_small} over time. The output variability is given as a percentage of the total variability integrated over all times.}
	\label{fig:Sens_output_variability}
\end{figure}

In the paper, we mention the output variability of the model, which we can also measure using the three sensitivity methods. For Sobol, we use the output variance 
\begin{align*}
 V = \frac{1}{2N} \sum_{j=1}^{2N} \left(f(AB)_j - \frac{1}{2N}\sum_{k=1}^{2N} f(AB)_k \right)^2.
\end{align*}
For ASM and Morris, we use the sum of all sensitivity scores before normalization
\begin{align*}
    \mu^* = \sum_{i=1}^n \mu_i^*, \qquad \alpha = \sum_{i=1}^n \alpha_i.
\end{align*}
We show the resulting output variabilities for our two experiments on the non-randomized RPM flow in \cref{fig:sSens_output_variability} and for the randomized RPM flow in \cref{fig:Sens_output_variability}. We represent the output variability in terms of percentage to the integral of the variability over $t \in [0,20]$ to maintain comparability between the three methods.

\begin{figure}[tb!]
	\centering
	\begin{subfigure}[t]{6.876cm}
		\includegraphics[width=6.876cm]{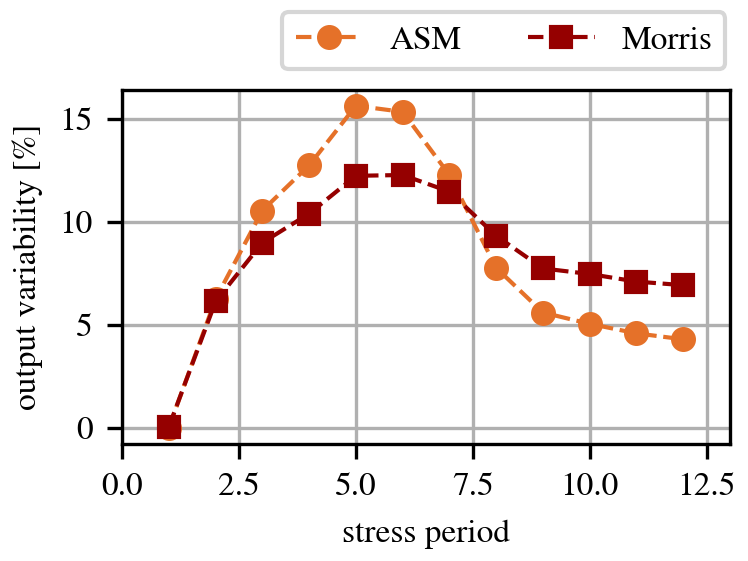}
	\end{subfigure}

	\caption{Comparison of the output variability measured using The Morris method and the ASM activity scores over time. The output variability is given as a percentage of the total variability integrated over all stress periods.}
	\label{fig:Quad_output_variability}
\end{figure}

We observe similar output variabilities for the randomized and the non-randomized RPM flow. We find very low variabilities for small times, especially in the small intervals experiment. The low output variability is likely to be the reason for the significant changes in the sensitivities at early times. As there is only very little overall sensitivity, all hyperparameters must be insensitive to the output.
We also show the output variability of the Quadrupole model in \cref{fig:Quad_output_variability}. The output variability at stress period one is very low, so we can assume again that although we identify sensitive and insensitive hyperparameters, all hyperparameters need to be considered insensitive.

\section{Convergence of sensitivity methods}\label{sec:convergence}

\begin{figure*}[tb!]
	\centering
	\begin{subfigure}[t]{0.996cm}
		\includegraphics[width=0.996cm]{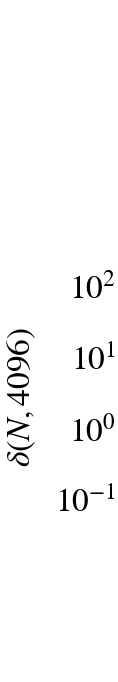}
	\end{subfigure}
	\begin{subfigure}[t]{3.486cm}
		\includegraphics[width=3.486cm]{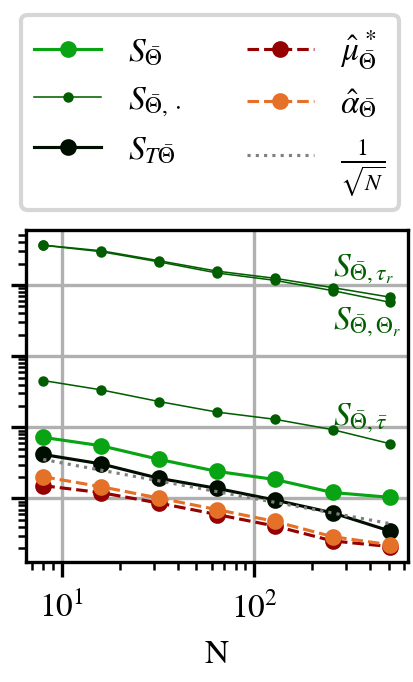}
		\caption{$\bar{\Theta}$}
	\end{subfigure}
	\begin{subfigure}[t]{3.486cm}
		\includegraphics[width=3.486cm]{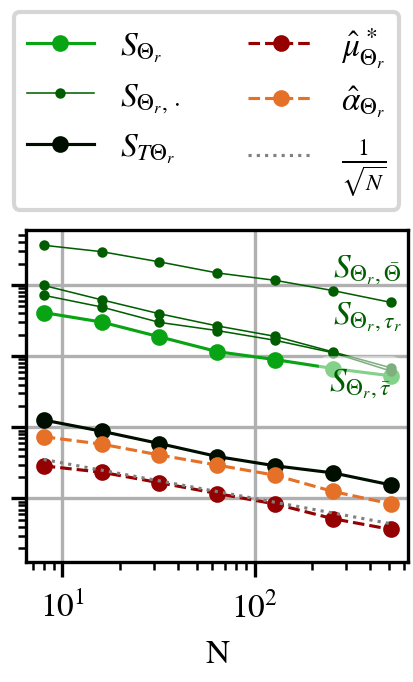}
		\caption{$\Theta_r$}
	\end{subfigure}
	\begin{subfigure}[t]{3.486cm}
		\includegraphics[width=3.486cm]{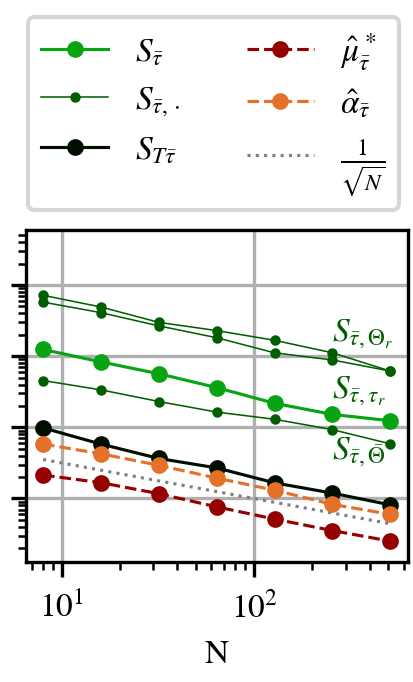}
		\caption{$\bar{\tau}$}
	\end{subfigure}
	\begin{subfigure}[t]{3.486cm}
		\includegraphics[width=3.486cm]{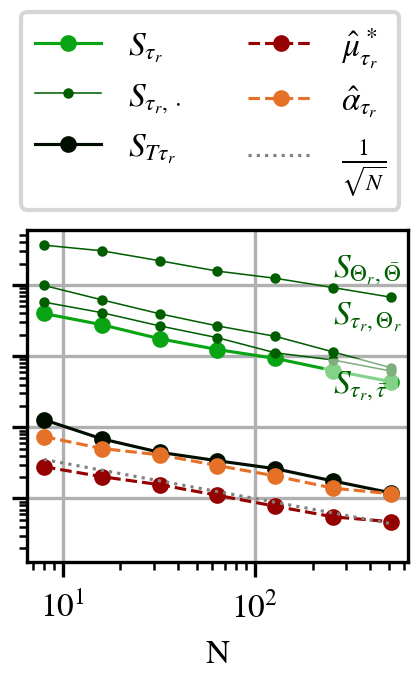}
		\caption{$\tau_r$}
	\end{subfigure}
	\hspace{0.72cm}

	\caption{Mean relative error estimate $\delta(N, 4096)$ over batch size $N$ for the different sensitivity scores for the randomized RPM flow with $\bar{\Theta} \in [0, \pi]$, $\Theta_r \in [0, 0.2\pi]$, $\bar{\tau} \in [0.1, 1]$ and $\tau_r \in [0, 0.2]$. }
	\label{fig:app:Sens_convergence_big}
\end{figure*}

\begin{figure*}[tb!]
	\centering
	\begin{subfigure}[t]{0.996cm}
		\includegraphics[width=0.996cm]{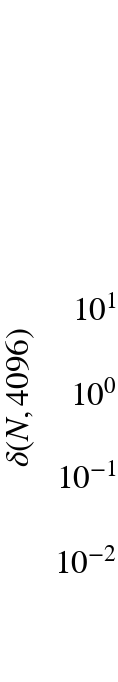}
	\end{subfigure}
	\begin{subfigure}[t]{3.486cm}
		\includegraphics[width=3.486cm]{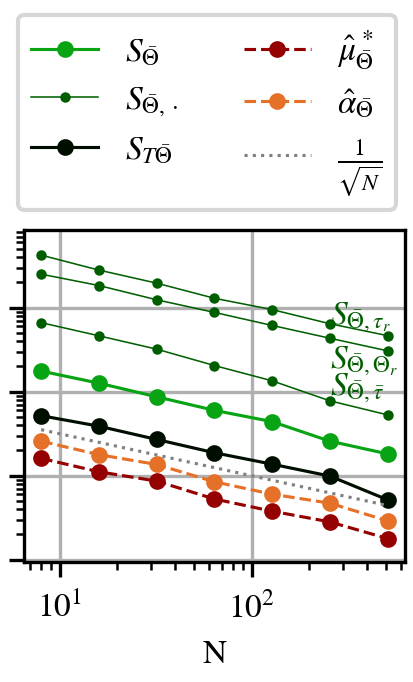}
		\caption{$\bar{\Theta}$}
	\end{subfigure}
	\begin{subfigure}[t]{3.486cm}
		\includegraphics[width=3.486cm]{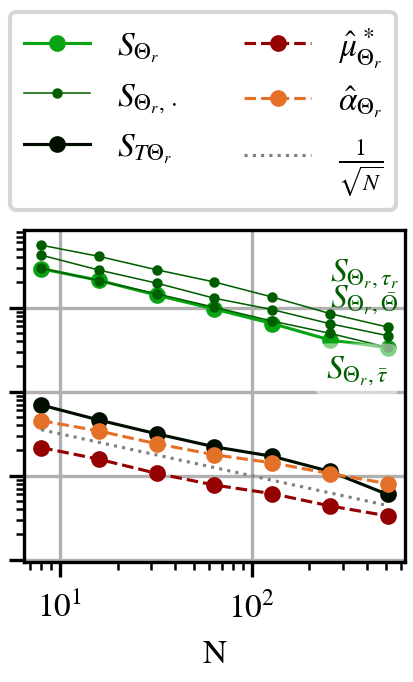}
		\caption{$\Theta_r$}
	\end{subfigure}
	\begin{subfigure}[t]{3.486cm}
		\includegraphics[width=3.486cm]{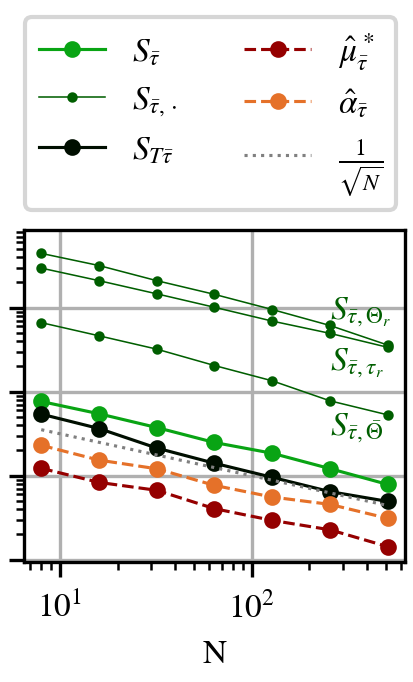}
		\caption{$\bar{\tau}$}
	\end{subfigure}
	\begin{subfigure}[t]{3.486cm}
		\includegraphics[width=3.486cm]{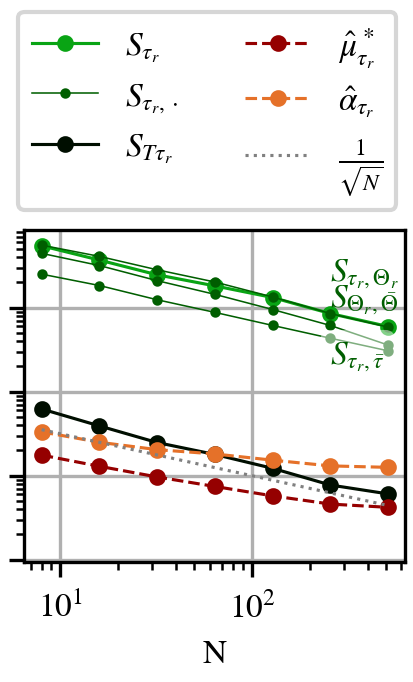}
		\caption{$\tau_r$}
	\end{subfigure}
	\hspace{0.72cm}

	\caption{Mean relative error estimate $\delta(N, 4096)$ over batch size $N$ for the different sensitivity scores for the randomized RPM flow with $\bar{\Theta} \in [0.45, 0.55\pi]$, $\Theta_r \in [0, 0.1\pi]$, $\bar{\tau} \in [0.45, 0.55]$ and $\tau_r \in [0, 0.1]$. }
	\label{fig:app:Sens_convergence_small}
\end{figure*}

The paper only contains the convergence of the total Sobol index $S_{T}$, the Morris scores $\hat{\mu}^*$ and the activity score of ASM $\hat{\alpha}$ for the randomized RPM flow with $\bar{\Theta} \in [0, \pi]$, $\Theta_r \in [0, 0.2\pi]$, $\bar{\tau} \in [0.1, 1]$ and $\tau_r \in [0, 0.2]$. For completeness, we show the results for all sensitivity measures and both randomized configurations in \cref{fig:app:Sens_convergence_big} and \cref{fig:app:Sens_convergence_small}. Notice that $\delta(N 4096)$ is a relative error estimate. This choice results in large errors for the first and second-order Sobol indices that yield values close to zero. 

\section{First eigenvector of the RPM flow models}\label{sec:first_eigenvec_RPM}

\begin{figure}[tb!]
	\centering
	\begin{subfigure}[t]{6.5cm}
	\centering
		\includegraphics[width=5.16cm]{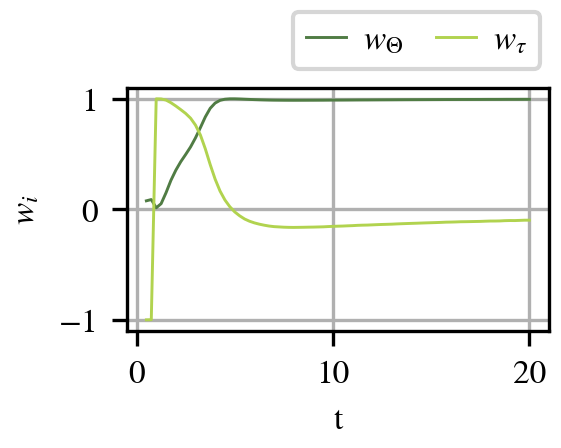}
		\caption{large hyperparameter intervals experiment}
		\label{fig:app:vec_non-rand_big}
	\end{subfigure}
	~
	\begin{subfigure}[t]{6.5cm}
	\centering
		\includegraphics[width=5.16cm]{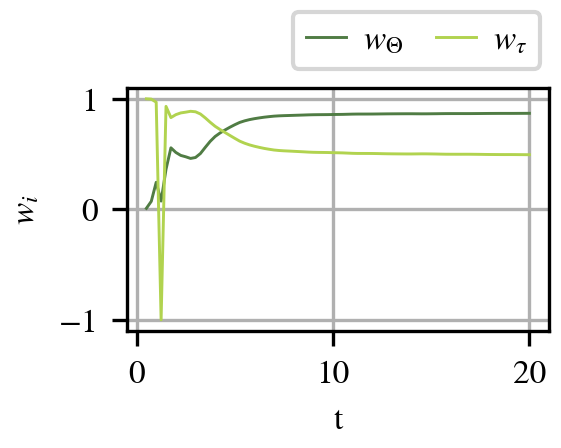}
		\caption{small hyperparameter intervals experiment}
		\label{fig:app:vec_non-rand_small}
	\end{subfigure}

	\caption{Entries of the first eigenvector of the covariance matrix build for the non-randomized RPM flow with $\Theta \in [0, \pi]$, $\tau \in [0, 1]$ in \cref{fig:app:vec_non-rand_big} and $\Theta \in [0.45, 0.55\pi]$, $\tau \in [0.45, 0.55]$ in \cref{fig:app:vec_non-rand_small}.}
	\label{fig:app:vec_non-rand}
\end{figure}

\begin{figure}[tb!]
	\centering
	\begin{subfigure}[t]{6.5cm}
	\centering
		\includegraphics[width=5.16cm]{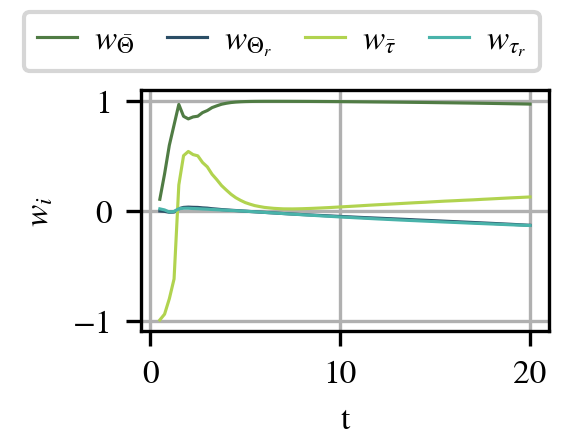}
		\caption{large hyperparameter intervals experiment}
		\label{fig:app:vec_rand_big}
	\end{subfigure}
	~
	\begin{subfigure}[t]{6.5cm}
	\centering
		\includegraphics[width=5.16cm]{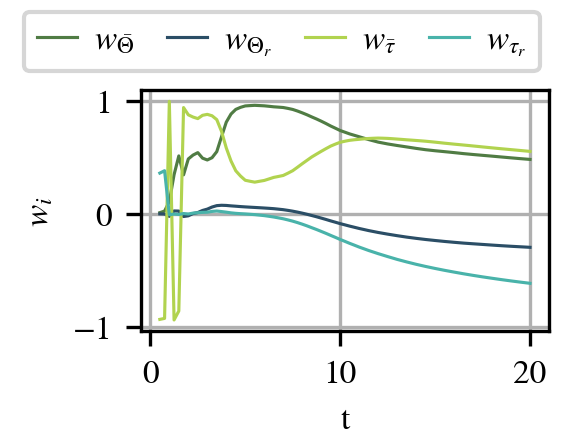}
		\caption{small hyperparameter intervals experiment}
		\label{fig:app:vec_rand_small}
	\end{subfigure}

	\caption{Entries of the first eigenvector of the covariance matrix build for the randomized RPM flow with $\bar{\Theta} \in [0, \pi]$, $\Theta_r \in [0, 0.2\pi]$, $\bar{\tau} \in [0.1, 1]$ and $\tau_r \in [0, 0.2]$ in \cref{fig:app:vec_rand_big} and $\bar{\Theta} \in [0.45, 0.55\pi]$, $\Theta_r \in [0, 0.1\pi]$, $\bar{\tau} \in [0.45, 0.55]$ and $\tau_r \in [0, 0.1]$ in \cref{fig:app:vec_rand_small}.}
	\label{fig:app:vec_rand}
\end{figure}

We use the first eigenvector of the covariance matrix to analyze if parameters are positively or negatively correlated. We show this first eigenvector over time in \cref{fig:app:vec_non-rand} for the non-randomized RPM flow and in \cref{fig:app:vec_rand} for the randomized RPM flow. 

\section{First eigenvectors of the Quadrupole model}\label{sec:first_eigenvec_quad}

\begin{figure}[tb!]
	\centering
	\begin{subfigure}[t]{16cm}
		\includegraphics[width=16cm]{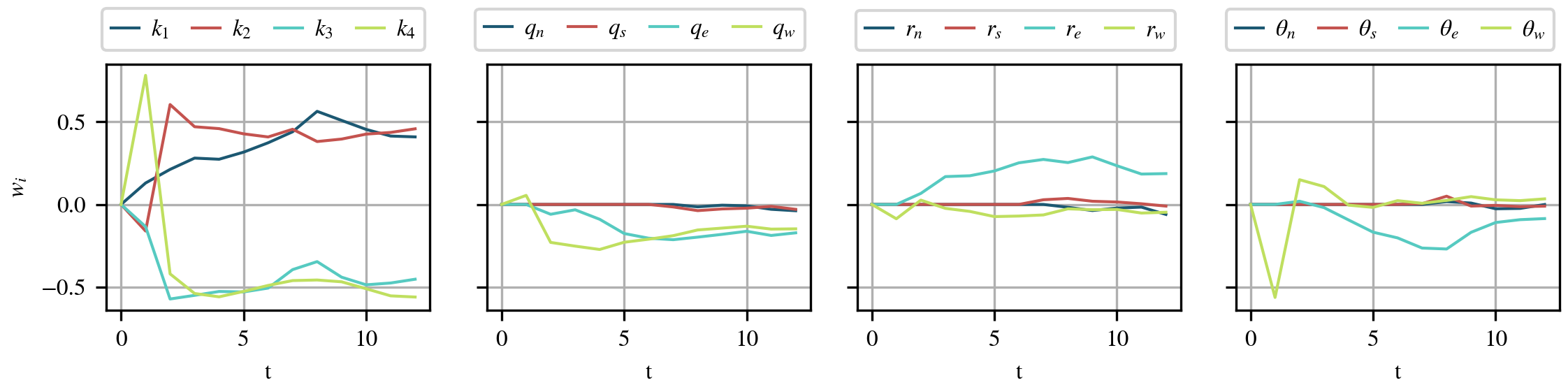}
		\caption{First eigenvector}
		\label{fig:app:quad_vec1}
	\end{subfigure}
	
	\begin{subfigure}[t]{16cm}
		\includegraphics[width=16cm]{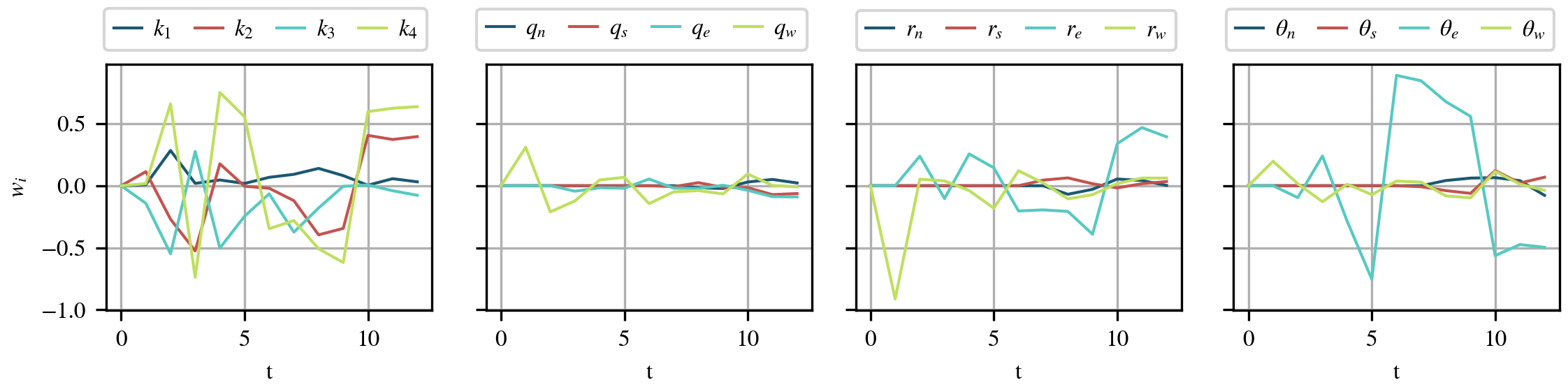}
		\caption{Second eigenvector}
		\label{fig:app:quad_vec2}
	\end{subfigure}

	\caption{Entries of the first and second eigenvector of the covariance matrix build for the quadrupole flow.}
	\label{fig:app:vec_quad}
\end{figure}

We use the first eigenvector of the covariance matrix corresponding to the Quadrupole flow to analyze the correlation of the hyperparameter. We show the entries of the first eigenvector in \cref{fig:app:quad_vec1}. We also include the second eigenvector in \cref{fig:app:quad_vec2}; however, the second eigenvector does not show a clear trend for the correlation. Therefore, we do not use it in our analysis in the paper.

\section{Validation of the simulation time for the RPM flow}\label{sec:validation_tmax}

\begin{figure}[tb!]
	\centering
	\begin{subfigure}[t]{6.4cm}
		\includegraphics[width=6.33cm]{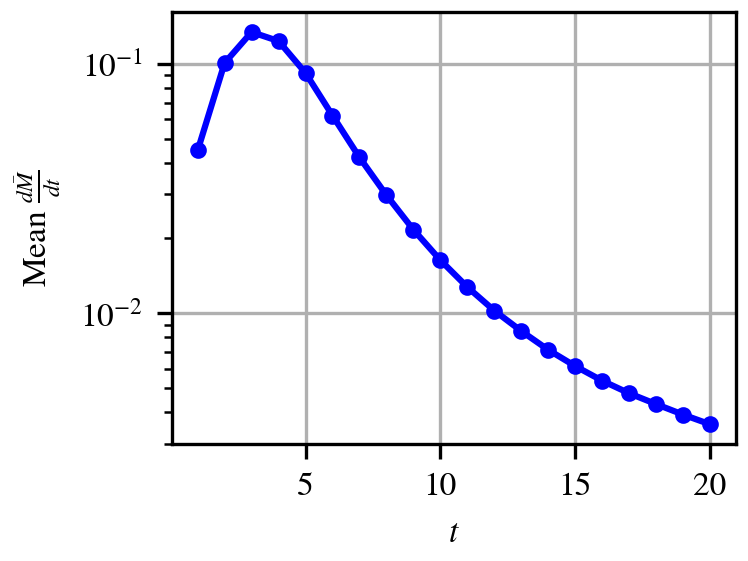}
		\caption{non-randomized configuration with large intervals}
		\label{fig:validation_tmax_nrand_br_evolution}
	\end{subfigure}
	~
	\begin{subfigure}[t]{6.4cm}
		\includegraphics[width=6.33cm]{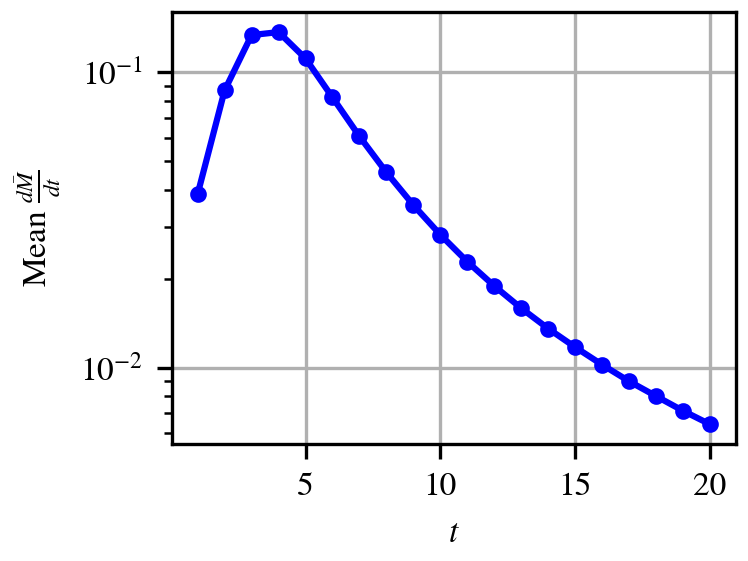}
		\caption{randomized configuration with large intervals}
		\label{fig:validation_tmax_nrand_sr_evolution}
	\end{subfigure}

	\caption{Mean derivative $\frac{d\M}{dt}$ over time for all samples of the non-randomized RPM flow with $\Theta \in [0, \pi]$ and $\tau \in [0, 1]$ in \cref{fig:validation_tmax_nrand_br_evolution} and the randomized RPM flow with $\bar{\Theta} \in [0, \pi]$, $\Theta_r \in [0, 0.2\pi]$, $\bar{\tau} \in [0.1, 1]$ and $\tau_r \in [0, 0.2]$ in \cref{fig:validation_tmax_nrand_sr_evolution}.}
	\label{fig:validation_tmax_nrand_evolution}
\end{figure}

In the paper, we choose a simulation time of $t_{\max} = 20$ for the RPM flow and only compute the time-varying sensitivity upon this time interval. This approach is sufficient as the mixing ($\M$) reaches a steady state after sufficient long time, hence the results of the sensitivity analysis will converge towards a long term stable result. For the randomized configurations the steady state corresponds to all grid cells containing at least one particle ($\M \approx 1$) and in the non-randomized configuration the chaotic region is filled with particles while the KAM-islands remain empty. To validate that the configurations converge towards the steady state, we approximate the derivative $\frac{d\M}{dt}(t^*)$ for all samples using backward finite differences between $t^*$ and $t^* - 0.25$. The results for the mean derivative $\frac{d\M}{dt}(t^*)$ for the non-randomized and randomized configurations are shown in \cref{fig:validation_tmax_nrand_evolution}. We observe that the derivatives decrease over time for both configurations, indicating convergence towards a steady state.

\begin{figure}[tb!]
	\centering
	\begin{subfigure}[t]{7cm}
		\includegraphics[width=7cm]{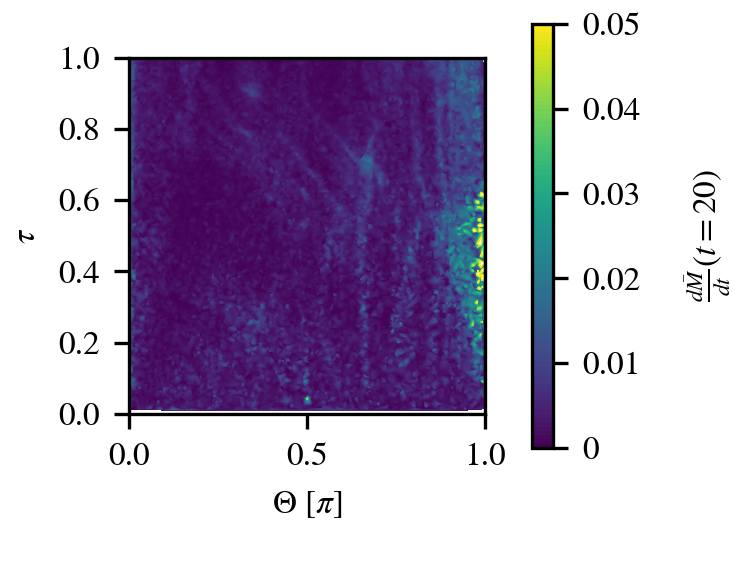}
		\caption{Large intervals experiment}
		\label{fig:validation_tmax_nrand_br}
	\end{subfigure}
	~
	\begin{subfigure}[t]{7cm}
		\includegraphics[width=7cm]{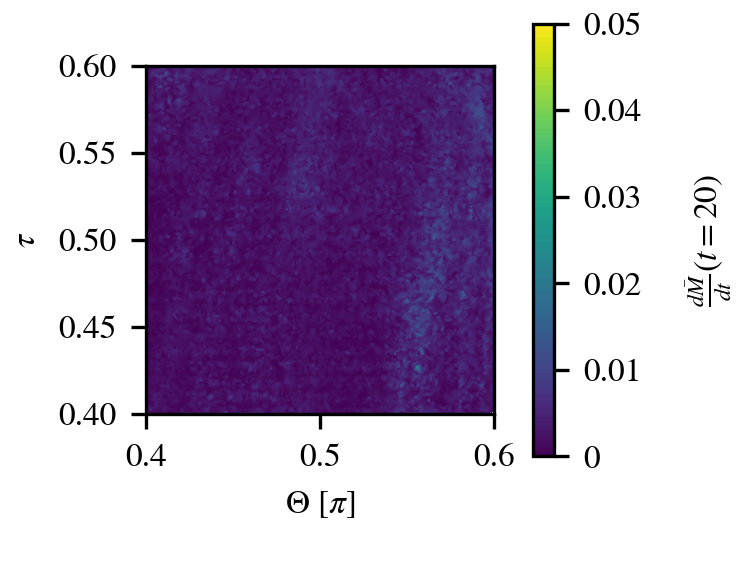}
		\caption{Small intervals experiment}
		\label{fig:validation_tmax_nrand_sr}
	\end{subfigure}

	\caption{Derivative $\frac{d\M}{dt}$ at time $t=20$ for all samples of the non-randomized RPM flow with $\Theta \in [0, \pi]$ and $\tau \in [0, 1]$ in \cref{fig:validation_tmax_nrand_br} and $\Theta \in [0.4\pi, 0.6\pi]$ and $\tau \in [0.4, 0.6]$ in \cref{fig:validation_tmax_nrand_sr}.}
	\label{fig:validation_tmax_nrand}
\end{figure}

\begin{figure}[tb!]
		\centering
	\begin{subfigure}[t]{13cm}
	\centering
		\includegraphics[width=9.5cm]{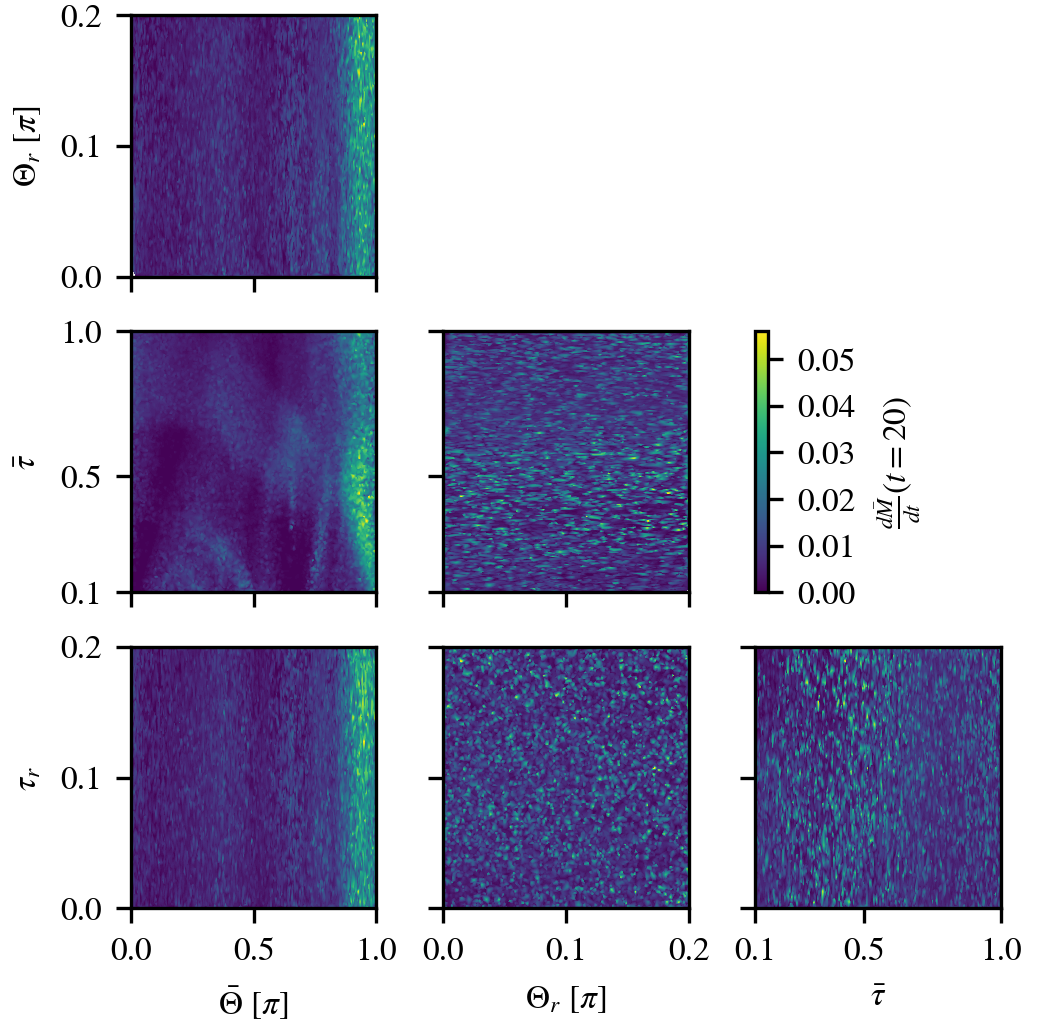}
		\caption{Large intervals experiment}
		\label{fig:validation_tmax_rand_br}
	\end{subfigure}

	\begin{subfigure}[t]{13cm}
		\centering
		\includegraphics[width=9.5cm]{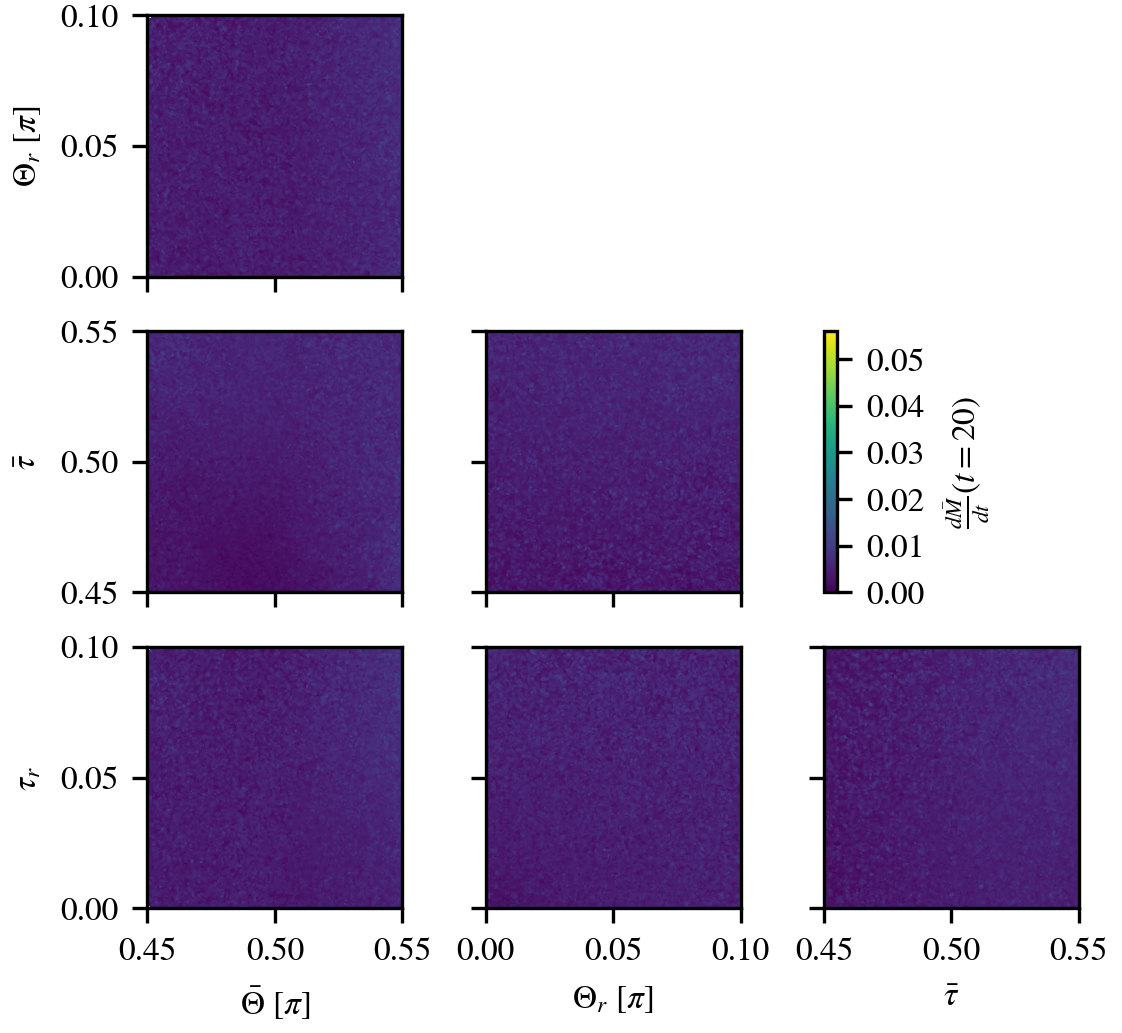}
		\caption{Small intervals experiment}
		\label{fig:validation_tmax_rand_sr}
	\end{subfigure}

	\caption{Derivative $\frac{d\M}{dt}$ at time $t=20$ for all samples of the randomized RPM flow with $\bar{\Theta} \in [0, \pi]$, $\Theta_r \in [0, 0.2\pi]$, $\bar{\tau} \in [0.1, 1]$ and $\tau_r \in [0, 0.2]$ in \cref{fig:validation_tmax_rand_br} and $\bar{\Theta} \in [0.45, 0.55\pi]$, $\Theta_r \in [0, 0.1\pi]$, $\bar{\tau} \in [0.45, 0.55]$ and $\tau_r \in [0, 0.1]$ in \cref{fig:validation_tmax_rand_sr}.}
	\label{fig:validation_tmax_rand}
\end{figure}

In \cref{fig:validation_tmax_nrand,fig:validation_tmax_rand}, we further show the derivative $\frac{d\M}{dt}(t=20)$ . 
We observe very low derivatives (mostly smaller than $0.05$), hence we can assume that the mixing state of many configurations is already close to a steady state by $t_{\max} = 20$ and we don't expect large changes of the sensitivities for $t > t_{\max}$.

\end{document}